\documentclass[onecolumn,trackchanges]{aastex631}

\shorttitle{Calibrating the WSA model in EUHFORIA}
\shortauthors{Samara et al.}
\usepackage{graphicx}
\usepackage{subfigure}
\usepackage[figuresleft]{rotating}

\hypersetup{colorlinks=true,citecolor=cyan,urlcolor=cyan,linkcolor=blue}
\begin{document}

\title{Calibrating the WSA model in EUHFORIA based on PSP observations}

\correspondingauthor{Evangelia Samara}
\email{evangelia.samara@nasa.gov}

\author[0000-0002-7676-9364]{E. Samara}
\affiliation{NASA Goddard Space Flight Center, Greenbelt, MD, USA}

\author{C. N. Arge}
\affiliation{NASA Goddard Space Flight Center, Greenbelt, MD, USA}

\author{R. F. Pinto}
\affiliation{IRAP, Université de Toulouse, CNRS, UPS, CNES, Toulouse, France}

\author{J. Magdaleni\'{c}}
\affiliation{Solar-Terrestrial Centre of Excellence -- SIDC, Royal Observatory of Belgium, Brussels, Belgium}
\affiliation{CmPA/Dept.\ of Mathematics, KU Leuven, Leuven, Belgium}

\author{N. Wijsen}
\affiliation{CmPA/Dept.\ of Mathematics, KU Leuven, Leuven, Belgium} 

\author{M. L. Stevens}
\affiliation{Smithsonian Astrophysical Observatory, Cambridge, MA, USA}

\author{L. Rodriguez}
\affiliation{Solar-Terrestrial Centre of Excellence -- SIDC, Royal Observatory of Belgium, Brussels, Belgium\\}

\author{S. Poedts}
\affiliation{CmPA/Dept.\ of Mathematics, KU Leuven, Leuven, Belgium}
\affiliation{Institute of Physics, University of Maria Curie-Sk{\l}odowska, Lublin, Poland}

\begin{abstract}
{We employ Parker Solar Probe (PSP) observations during the latest solar minimum period (years 2018 - 2021) to calibrate the version of the Wang-Sheeley-Arge (WSA) coronal model used in the European space weather forecasting tool EUHFORIA. WSA provides a set of boundary conditions at 0.1\;au necessary to initiate the heliospheric part of EUHFORIA, namely, the domain extending beyond the solar Alfv\'enic point. To calibrate WSA, we observationally constrain four constants in the WSA semi-empirical formula based on PSP observations. We show how the updated (after the calibration) WSA boundary conditions at 0.1\;au are compared to PSP observations at similar distances, and we further propagate these conditions in the heliosphere according to EUHFORIA's magnetohydrodynamic (MHD) approach. We assess the predictions at Earth based on the Dynamic Time Warping technique. Our findings suggest that, for the period of interest, the WSA configurations which resembled optimally the PSP observations close to the Sun, were different than the ones needed to provide better predictions at Earth. One reason for this discrepancy can be attributed to the scarcity of fast solar wind velocities recorded by PSP. The calibration of the model was performed based on unexpectedly slow velocities that did not allow us to achieve generally and globally improved solar wind predictions, compared to older studies. Other reasons can be attributed to missing physical processes from the heliospheric part of EUHFORIA but also the fact that the currently employed WSA relationship, as coupled to the heliospheric MHD domain, may need a global reformulation beyond that of just updating the four constant factors that were taken into account in this study.} 

\end{abstract}


\section{Introduction} \label{Sec:Introduction}

Validation of space weather forecasting models is crucial for accurately predicting space weather conditions in the heliosphere. This is necessary to mitigate possibly destructive consequences of extreme space weather on technological infrastructure at Earth or at any point in the heliosphere where man-made technologies exist \citep[][]{hapgood2011towards, green2015coronal, schrijver15}{}{}. However, the pace at which space weather forecasting tools evolve has outstripped the pace of improvements in the quality of the input data they employ \citep[][]{MacNeice2018}{}{}. Minimization of uncertainties in those data is one way to advance our prediction capabilities. Calibration of models with observations is another. Both directions require the advent of novel missions with cutting-edge instrumentation that will not only provide data we never had before, but also help improve the existing observations that are used to drive and validate models.

Parker Solar Probe (PSP) is one of such novel missions launched only a few years ago \citep[][]{fox2016, PSPmission2, PSPmission1}. In an effort to address a number of physical questions that the solar community is struggling to answer for the past decades, e.g., how the solar corona is heated, or how the fast solar wind and coronal mass ejections (CMEs) are accelerated, PSP approaches the Sun unprecedentedly closely (eventually as close as $\approx9.9\;$R$_{\odot}$). This provides a unique opportunity to study the in situ conditions at close proximity from the Sun, as well as take wide-field images of solar structures and outflows. In situ observations at this region can be very useful for the calibration of models employed to reconstruct solar wind conditions close to the Sun. Such models, usually called coronal models, reconstruct the global 3D structure of the magnetic corona up to the Alfv\'enic point %radius of 30 Rsun\footnote{This radius differs depending on the model but is usually lying either at 21.5 Rs or 30 Rs.} and provide a set of boundary conditions for the 
and provide the boundary conditions that heliospheric models need as input to initiate and propagate the solar wind in the inner heliosphere.

In this paper, we aim to calibrate the implementation of Wang-Sheeley-Arge coronal model \citep[WSA;][]{ Arge04, arge03} used in the European Heliospheric Forecasting Information Asset \citep[EUHFORIA;][]{pomoell18}{}{} by employing PSP observations. We note that the version of WSA in EUHFORIA is different from the original WSA model (see differences between \citealt[][]{pomoell18}{}{} and \citealt[][]{arge03, Arge04}{}{}). WSA has been routinely used for many years to reconstruct the solar wind conditions at close-to-the-Sun distances and predict solar wind velocities and magnetic field polarities in the heliosphere based on a 1D kinematic code. In the frame of this study, we only use the first (or coronal) part of WSA. The boundary conditions it generates are used to initiate the heliospheric part of EUHFORIA at 0.1~au \citep[][]{pomoell18}{}{}. The degree of EUHFORIA's success to reliably forecast space weather conditions throughout the heliosphere depends on many factors, one of the most important being the quality of the boundary conditions from WSA. Until recently, it was not possible to validate them by employing in situ observations. However, this is currently possible thanks to the unique observations from PSP which is the first spacecraft to cross the boundary of 0.1~au. Therefore, the aim is to investigate how much we can improve solar wind modeling in the heliosphere by undertaking the task to calibrate the WSA version used in EUHFORIA based on PSP data.

\section{The WSA model in EUHFORIA}
\label{section:Section_2}

To initiate the heliospheric part of EUHFORIA, a set of solar wind boundary conditions are needed at the inner boundary of the heliospheric domain (0.1~au). The first of them, the radial velocity (v$_{r}$), is provided by the WSA model based on the following equation:

\begin{equation}
 \text{v}_{r}(f,d)= V_{0} +\frac{V_{1}}{(1+f)^{\alpha}}\left[1-0.8\exp\left(-\left(\frac{d}{w}\right)^{\beta}\right)\right]^3.
\label{WSA_vr}
\end{equation}

\noindent The constants used by default in EUHFORIA are $V_{0} =\;$240 km/s, $V_{1} = 675\;$km/s, $\alpha$ = 0.222, $\beta$ = 1.25 and $w = 0.02\;$rad \citep[see][for more details]{pomoell18}{}{}. The parameter $f$ denotes the flux tube expansion factor and is defined as: 

\begin{equation}
f= \left(\frac{R_{\odot}}{R_{b}}\right)^{2} \frac{B_{r}(R_{\odot}, \theta,\phi)}{B_{b}(R_{b}, \theta_{b}, \phi_{b})},
\label{fte_factor}
\end{equation}

\noindent where $R_{b}=0.1\;$au, while $B_{r}$ and $B_{b}$ are the radial magnetic fields at the photosphere and at $0.1$~au, respectively. For the purposes of this study, we revised this definition by taking $R_{b}=2.5\;$R$_{\odot}$ and $B_{b}$ at 2.5R$_{\odot}$, according to the original Wang-Sheeley prescription \citep[see, e.g.,][]{wang1997}{}{}. The parameter $d$ is the minimum angular distance of a magnetic flux-tube footpoint to the closest coronal hole (CH) boundary, otherwise known as the ``distance to the CH boundary''. Both $f$ and $d$ are calculated based on the geometry of the magnetic flux tubes after the reconstruction of the 3D solar corona based on the Potential Field Source Surface \citep[PFSS;][]{altschuler69, schatten69, Wiegelmann2017}{}{} and the Schatten Current Sheet \citep[SCS;][]{schatten1971current_npp}{}{} models.

With eq.~\ref{WSA_vr} used to specify solar wind velocities at 0.1~au, the remaining parameters needed to initiate the heliospheric part of EUHFORIA (i.e., number density; $n$, temperature; $T$, radial magnetic field; $B_{r}$), are calculated as follows:

\begin{equation}
n=n_\textup{fsw}\cdot(\text{v}_\textup{fsw}/\text{v}_{r})^{2},
\label{WSA_n}
\end{equation}

\begin{equation}
    T=T_\textup{fsw}\cdot(\rho_\textup{fsw}/\rho),
\label{WSA_T}
\end{equation}
and

\begin{equation}
    B_{r}=\textup{sgn}(B_\textup{corona})\cdot B_\textup{fsw}\cdot(\text{v}_{r}/\text{v}_\textup{fsw}).
\label{WSA_Br}
\end{equation}

\noindent where v$_\textup{fsw}$~=~675~km/s is the velocity of the fast solar wind that carries a magnetic field of $B_\textup{fsw}$~=~300~nT at 0.1~au.
The plasma number density of the fast solar wind at the same radius is $n_\textup{fsw}$~=~300~cm$^{-3}$, while \textup{sgn}($B_\textup{corona}$) is the sign of the magnetic field as given by the coronal model. Also, the plasma thermal pressure is constant at the boundary and equal to P~=~3.3~nPa corresponding to a temperature of $T_\textup{fsw}$ = 0.8 MK in the fast solar wind (see \citet[][]{pomoell18} for more details). The parameter $\rho$ denotes the mass density with $\rho_\textup{fsw}$~=~0.5$n_\textup{fsw}$m$_{p}$, where m$_{p}$ is the proton mass. 

Equation~\ref{WSA_vr} is of utmost importance since it provides velocities at $0.1\;$au based on which all other plasma and magnetic parameters are calculated. However, it is interesting to understand how this semi-empirical formula was first derived. If we go back to the literature \citep[][]{arge03, Arge04, vanderHolst10}, we see that eq.~\ref{WSA_vr} was formulated after ballistically mapping the solar wind from L1 back to 0.1~au and searching for patterns between speeds and a number of photospheric/coronal parameters. A large number of test functions and values of constants were tried in eq.~\ref{WSA_vr} until a good comparison with observations at \mbox{1\;au} was achieved, after propagating the WSA velocities ballistically to Earth using a simple 1D kinematic model and a specific type of magnetograms (originally Mount Wilson Solar observatory maps and more recently, GONG and VSM mpas). %Besides the fact that the procedure used to map the solar wind from L1 back to 0.1~au in order to formulate Eq.~\ref{WSA_vr} contains lots of uncertainties and can result in big errors, 
However, a kinematic model is different from an MHD polytropic model (as EUHFORIA) in the following ways. First, the velocity of the solar wind in the kinematic model is radial and does not change with distance except to conserve mass and mass flux at the stream interaction regions \citep[][]{kim2014}{}{}. No acceleration of the solar wind takes place as it propagates outwards, only decelerations at the stream interaction regions. On the other hand, an MHD polytropic model accelerates the supersonic and super-Alvfenic solar wind due to polytropic expansion and it takes into account much more complex stream interaction dynamics. Therefore, it is apparent that eq.~\ref{WSA_vr} does not seem to be consistent when used to produce boundary conditions for 3D MHD polytropic models.

% Therefore, we notice some differences with the way we use Eq.~\ref{WSA_vr} currently in EUHFORIA. First, we drive our 0.1~au boundary conditions at Earth based on an MHD and not kinetic approach. Second, we do not only use GONG synoptic magnetograms, but a variety of other maps (e.g., GONG ADAPT, HMI etc) depending on the needs of each particular study. We also need to take into account that the procedure used to map the solar wind from L1 back to 0.1~au in order to formulate Eq.~\ref{WSA_vr} contains lots of uncertainties and can result in big errors. As a result, calibration of Eq.~\ref{WSA_vr} at the boundary based on observations is imperative to improve our solar wind predictions. 

To account for this inconsistency, a number of ad hoc assumptions were adopted in EUHFORIA \citep[and in other models similar to it, such as Enlil, see][for more details]{mcgregor08, owens08}. First, $50\;$km/s are subtracted from the velocities at $0.1\;$au to account for the acceleration caused by the MHD heliospheric model when it propagates velocities in the interplanetary space \citep[][]{pomoell18}. Secondly, velocities at $0.1\;$au are capped between a range, usually that is $[275, 625]\;$km/s. The reason of these constraints has not been clearly justified throughout the literature. The main argument behind the upper limit, is that velocities above $625\;$km/s tend to overestimate the fast solar wind at Earth. On the other hand, the reasoning behind the lower limit, is that velocities below 275 km/s could be sub-Alfv\'enic, something that it is not handled correctly by the MHD code. Even though these assumptions help us reconstruct the large-scale structure of the solar wind at Earth, we often miss the precise prediction of e.g., well-defined high-speed streams \citep[HSSs; see, for example, ][]{Hinterreiter19, Samara2021MVP}{}. Moreover, the dynamics between the slow and the fast solar wind are generally poorly reproduced \citep[see e.g.,][]{Samara2022DTW}{}{}. However, it will be unfair to claim that the not fully accurate performance of EUHFORIA (and similar models) is solely caused because of this discrepancy. The problem of improving solar wind modeling is a multi-variable problem that depends on a number of parameters, such as the magnetograms used, the kind of model used and assumptions made in it, the phase of the solar cycle etc. In the framework of this paper, we will only focus on dealing with the discrepancy discussed above.

This work is inspired by the study of \citet[][]{mcgregor11} who calibrated the WSA equation in Enlil. Their study relied on Helios data to perform the calibration since Helios was the only mission, before PSP, to have approached the Sun closer than any other mission had ever done before (up to \mbox{0.3\;au}). Their calibrated solar wind results proved to better reconstruct the solar wind velocities at Earth. They also proved to better reconstruct the Ulysses measurements. Now, the proximity of PSP close to the Sun gives us the unprecedented chance to use more accurate data to calibrate our model and improve solar wind predictions by applying the method of \citet[][]{mcgregor11}.

\section{Calibrating the WSA equation}
\label{section:Section_3}

\subsection{Datasets}

In EUHFORIA, the WSA velocities at the boundary are typically produced with 2$^{o}$ resolution in both longitude and latitude on a 3D sphere of a radius equal to $0.1\;$au. For a proper calibration of those velocities, we would ideally need observations at exactly $0.1\;$au for all the latitudes and longitudes of that sphere. This is, of course, not possible since (a)~the orbit of PSP is elliptical in the ecliptic plane, so it only records data within a very restricted latitudinal zone ($\approx \pm 5^{o}$ around the solar equator), and (b)~PSP only stays shortly close to $0.1\;$au due to Kepler's second law. 

To compare the WSA velocities with PSP data and perform the calibration, we first removed the influence of CMEs from the PSP observations based on the HELIO4CAST interplanetary coronal mass ejection (ICME) catalog \citep[][]{mostl2017, mostl2020}{}{}. Then, we made the assumption that WSA velocities within the latitudinal zone of $\pm 5^{o}$, should resemble PSP velocities between \mbox{0.1 - 0.4\;au}. The reason we chose to compare our modeled data with measurements from an extended radial zone is twofold: first, because PSP measurements very close to $0.1\;$au are statistically poor (not so many data points to perform a proper comparison and calibration). Second, because \citet[][]{mcgregor11} had relied on Helios measurements between \mbox{0.3 - 0.4\;au} to perform the calibration of the WSA velocities in Enlil. Due to lack of observations at close-to-the-Sun distances, this was the best they could do. Therefore, we decided to retain McGregor's radial range and extend it downward to $0.1\;$au with PSP data. The longitudes of the WSA velocities taken into account were also in accordance with the longitudinal PSP passage for the periods of interest. The calibration is performed based on level-3 10-min cadence data from SPC \citep[][]{case2020}{}{} and SPAN \citep[][]{whittlesey2020}{}{} instruments onboard PSP for the first eight encounters that took place during 2018 - 2021. These years are around the minimum phase of solar cycle 24/25 and the data selection is restricted to those in order to have the minimal influence from transient solar phenomena (such as CMEs) that contaminate the solar wind measurements.

\subsection{Calibration of $V_{0}$}

\begin{figure}
\centering
\gridline{\fig{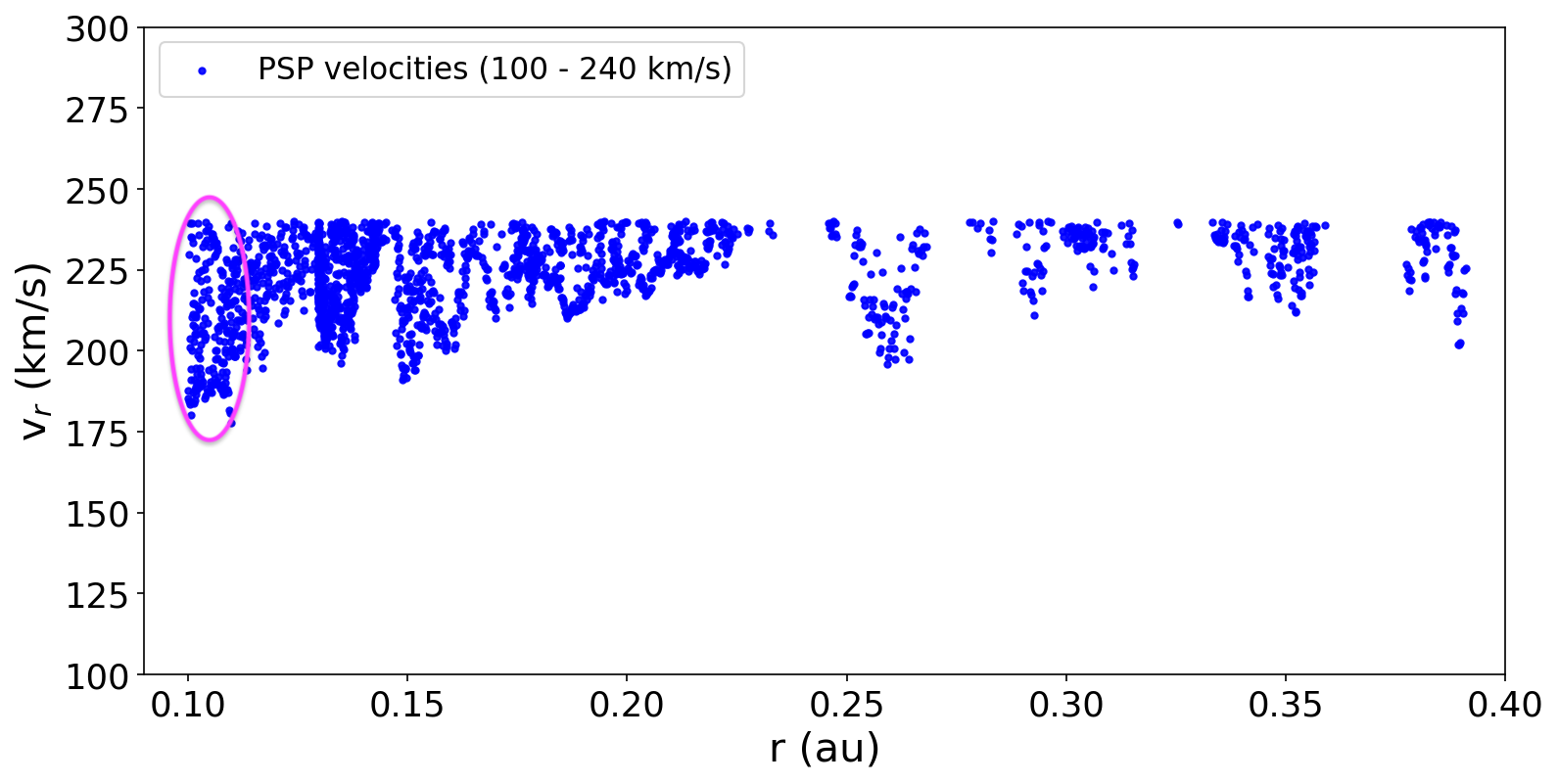}{0.8\textwidth}{}}     
\caption{The lowest velocities observed by PSP between $0.1 - 0.4\;$au during the first eight encounters. The velocity range is restricted between $100 - 240\;$km/s. The points that were averaged for the calculation of the lowest velocity that PSP recorded ($V_{0}$), are circled in magenta (see text for more details).}
\label{Fig:V0_psp}
\end{figure}

The terms subject to calibration in eq.~\ref{WSA_vr} are four: $V_{0}$, $V_{1}$, $\beta$ and $w$. For a global calibration of the WSA formula, all constant parameters should be calibrated. Nevertheless, we chose to focus on the aforementioned four because they can be fine-tuned based on PSP observations and past studies \citep[][]{mcgregor11}. We first discuss $V_{0}$ which should reflect the lowest velocity close to the 0.1~au boundary. Its default value is currently $240\;$km/s (see eq.~\ref{WSA_vr}) and it represents the lowest solar wind velocity typically seen at the vicinity of Earth. To understand what is the best value to update $V_{0}$, we plot in Fig.~\ref{Fig:V0_psp} the velocities up to 240 km/s recorded by PSP between 0.1 - 0.4\;au during the first eight encounters. Based on this figure, $V_{0}$ should be replaced by $\approx 177\;$km/s since this is the lowest velocity recorded at $\approx 0.11\;$au. However, this value is not representative of the velocities seen close to the boundary, i.e., it only characterizes 2-3 PSP measurements. To be more consistent with the overall velocities that the spacecraft measured, we took their average in the range 100~km/s\footnote{Velocities recorded by PSP below 100~km/s are not assumed reliable for science and should be ignored.} to 240~km/s between $0.10 - 0.11\;$au. This resulted to the value of $207$~km/s, which will be the updated $V_{0}$ from now on. 

\subsection{Calibration of $V_{1}$}

\begin{figure}
\centering
\gridline{\fig{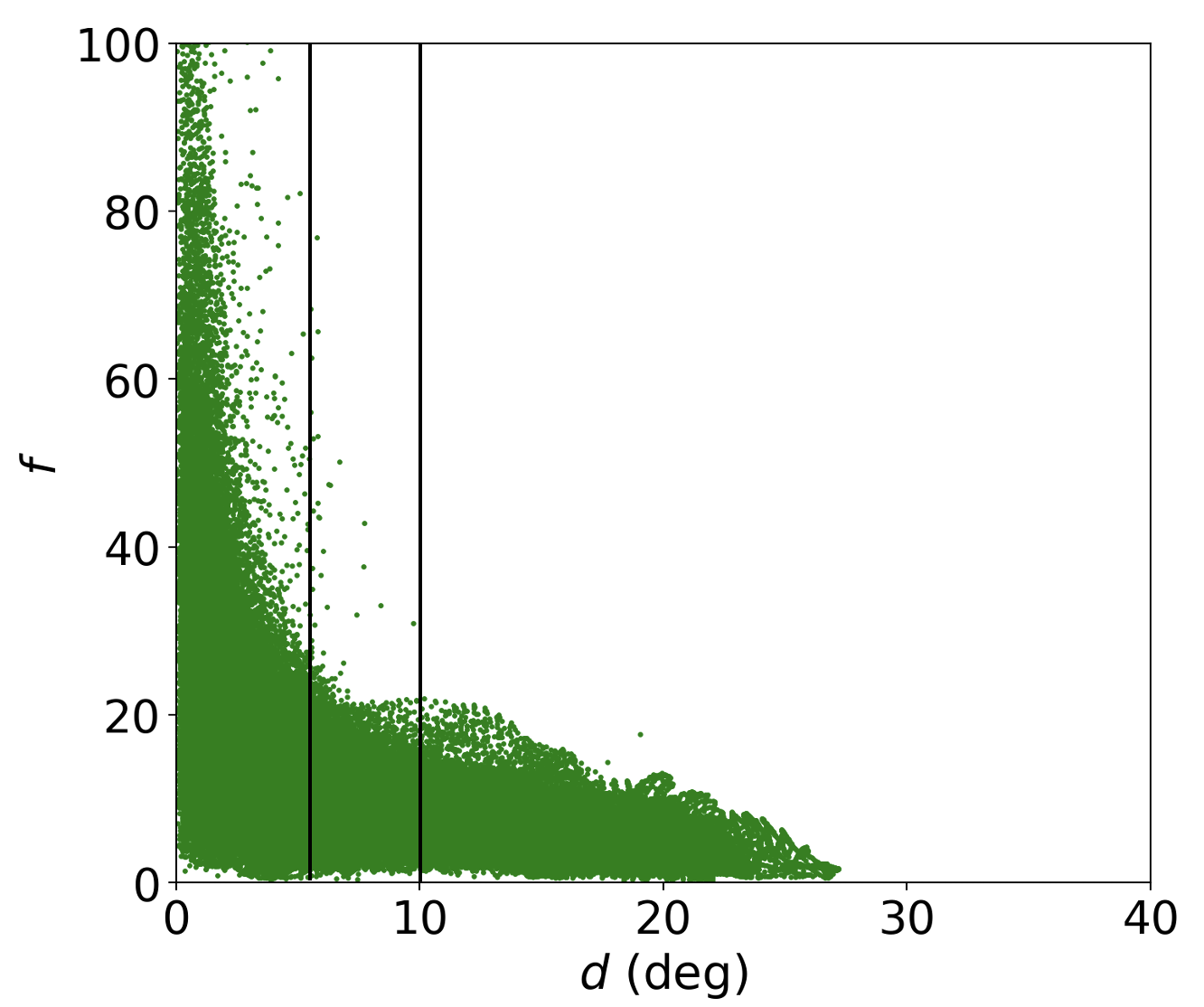}{0.6\textwidth}{}}     
\caption{Flux tube expansion factor ($f$) as a function of the distance to the CH boundary ($d$) during the first eight PSP encounters around the Sun. The points have been produced based on the default WSA model in EUHFORIA. The two vertical black lines signify the regime that velocities between 600 - 750~km/s originate from, based on the study of \citet[][]{schwadron2005}.}
\label{Fig:FTEvsDCHB}
\end{figure}

$V_{1}$ should reflect the maximum speed range that solar wind can vary above $V_{0}$. As suggested by \citet[][]{mcgregor11}{}{}, to calibrate this value we rely on (a) the work of \citet[][]{riley2003}, (b) the work of \citet[][]{schwadron2005}, and (c) Fig.~\ref{Fig:FTEvsDCHB}. In their study, \citet[][]{riley2003} ballistically mapped Ulysses solar wind measurements from twelve different Carrington rotations to the outer boundary of a coronal MHD model. Then, they mapped the velocities to their source regions in the photosphere and showed that velocities below $600\;$km/s come from regions up to $6^{o}$ from the CH boundary. Two years later, \citet[][]{schwadron2005} showed that speeds in the range $600 - 740\;$km/s originate from a region between $6^{o} - 10^{o}$ from the CH boundary. This is the range that is restricted by the two vertical black lines in Fig.~\ref{Fig:FTEvsDCHB}. More specifically, Fig.~\ref{Fig:FTEvsDCHB} shows the flux tube expansion factor ($f$) as a function of the distance to the CH boundary ($d$) based on the default WSA model in EUHFORIA, for the first eight encounters of PSP. Following \citet[][]{mcgregor11}{}{}, we adjusted $V_{1}$ so that v$_{r}$ in eq.~\ref{WSA_vr} gives the value $600\;$km/s for $f = 15$ assuming an asymptotically large $d$ and the updated value of $V_{0}$. Based on the aforementioned assumptions $V_{1}$ is 725~km/s, namely, 50~km/s higher than the default $V_{1}$.

\begin{figure}[h!]
\centering
\gridline{\fig{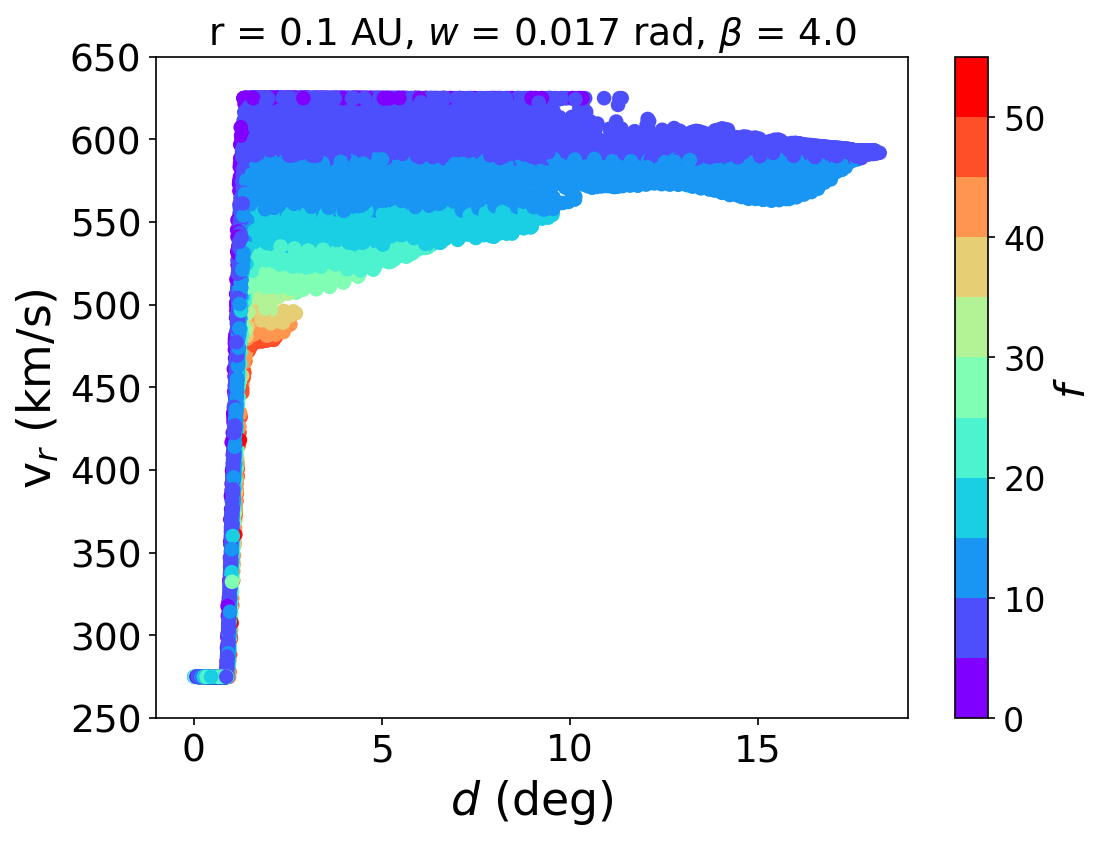}{0.3\textwidth}{(a)}
        \fig{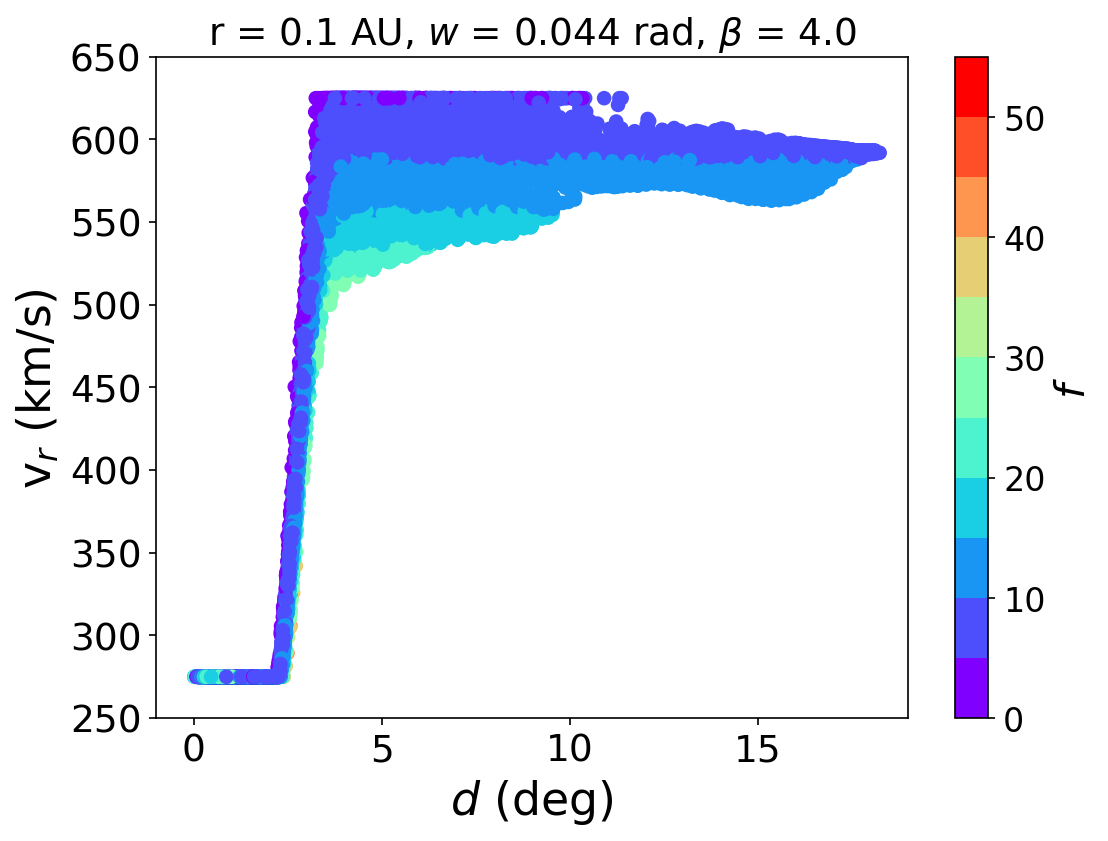}{0.3\textwidth}{(b)}
        \fig{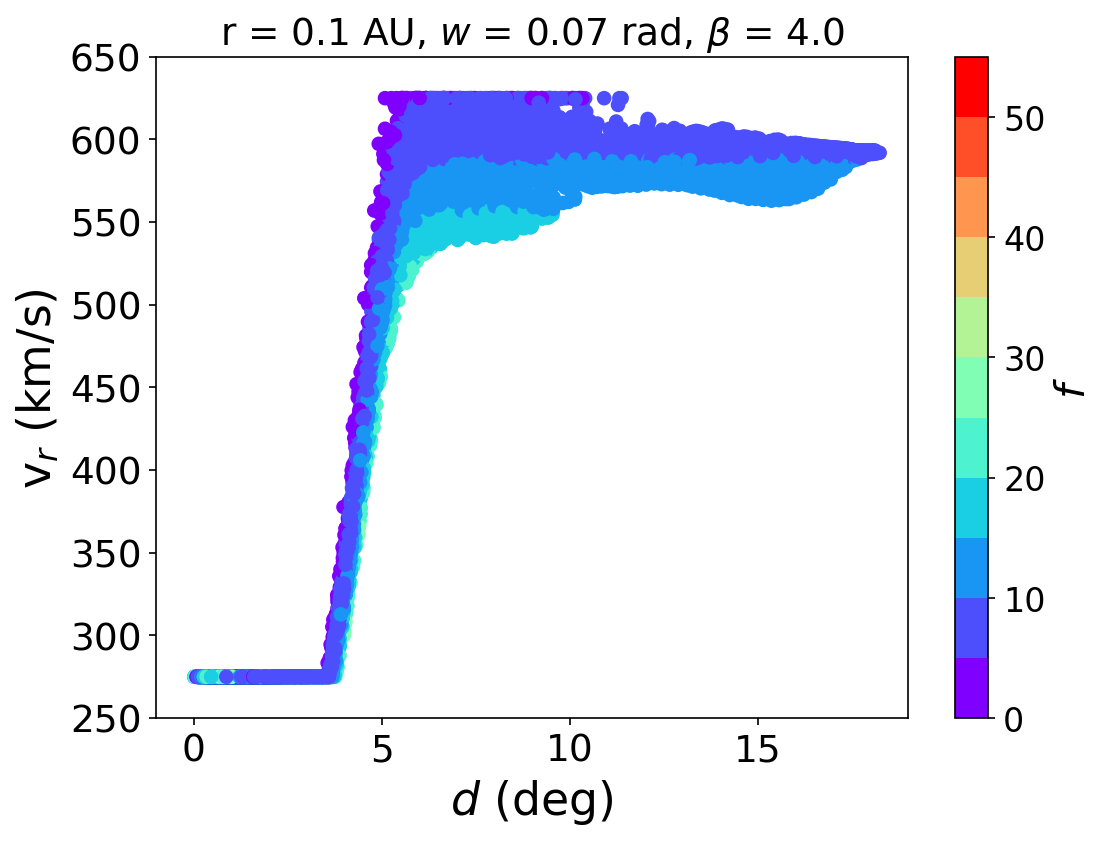}{0.3\textwidth}{(c)}} 

\gridline{\fig{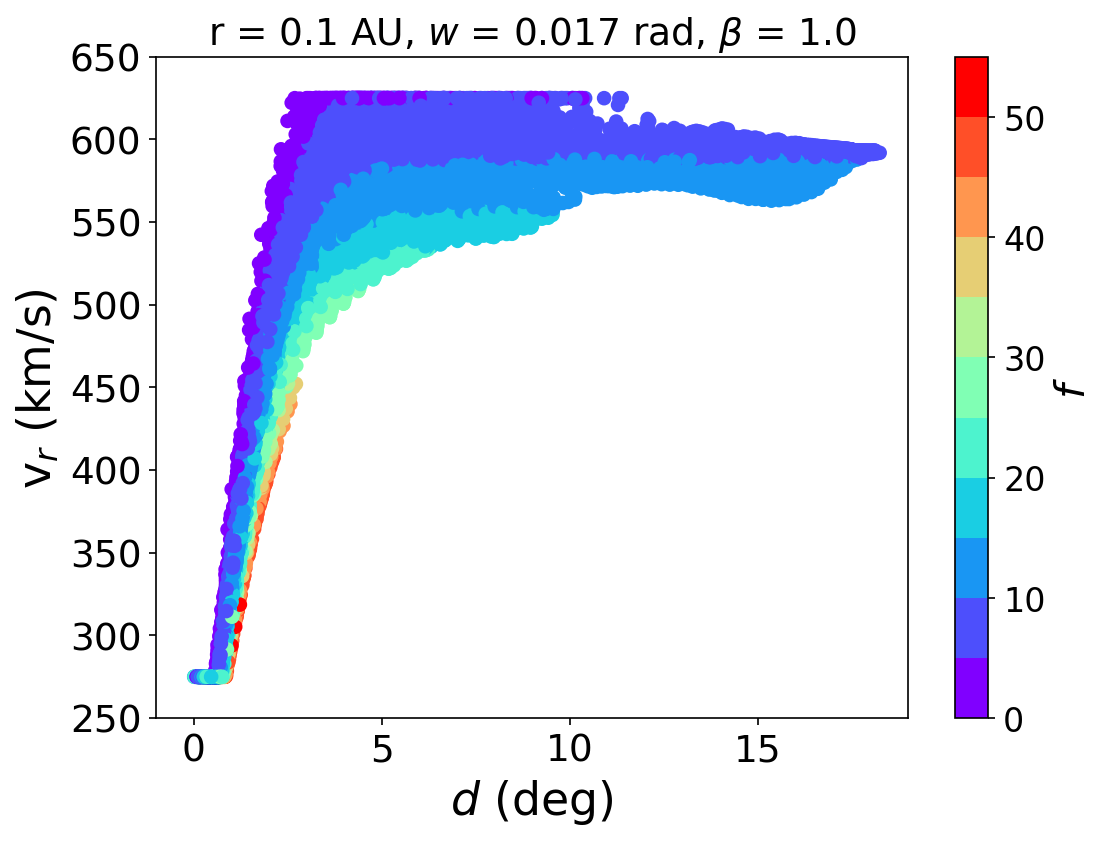}{0.3\textwidth}{(d)}
        \fig{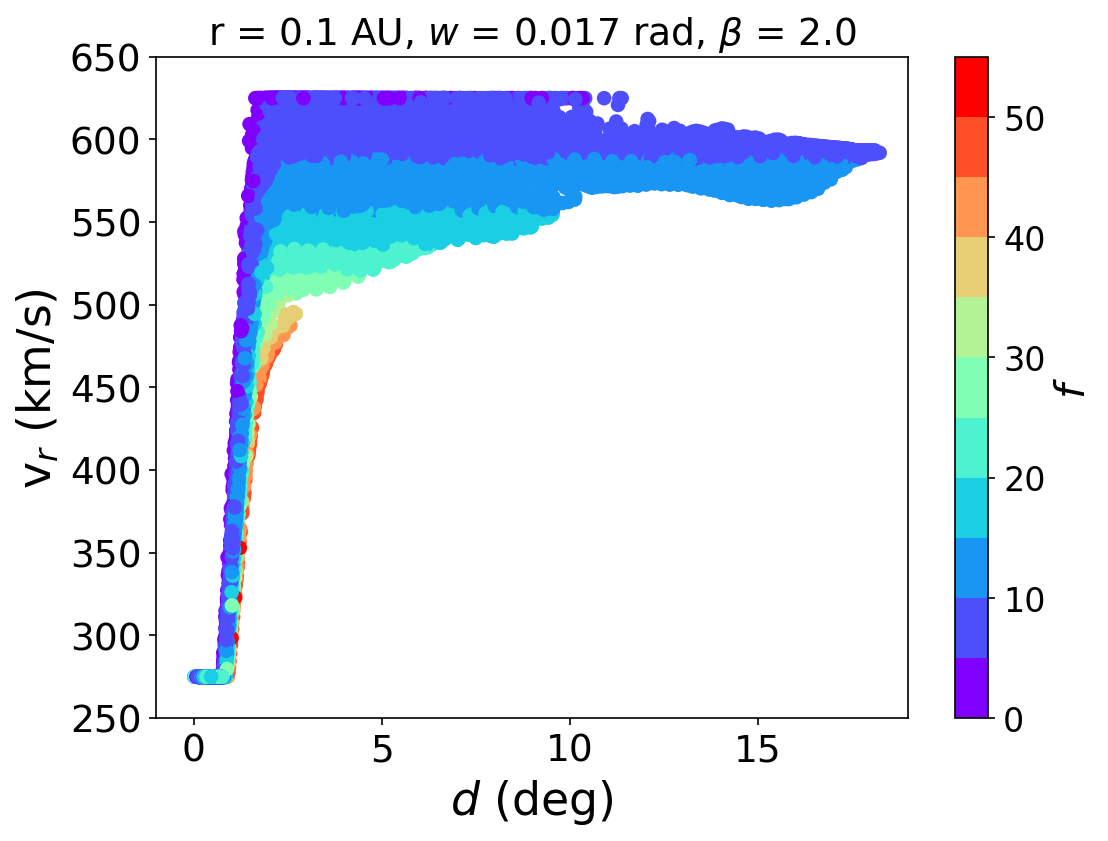}{0.3\textwidth}{(e)}
        \fig{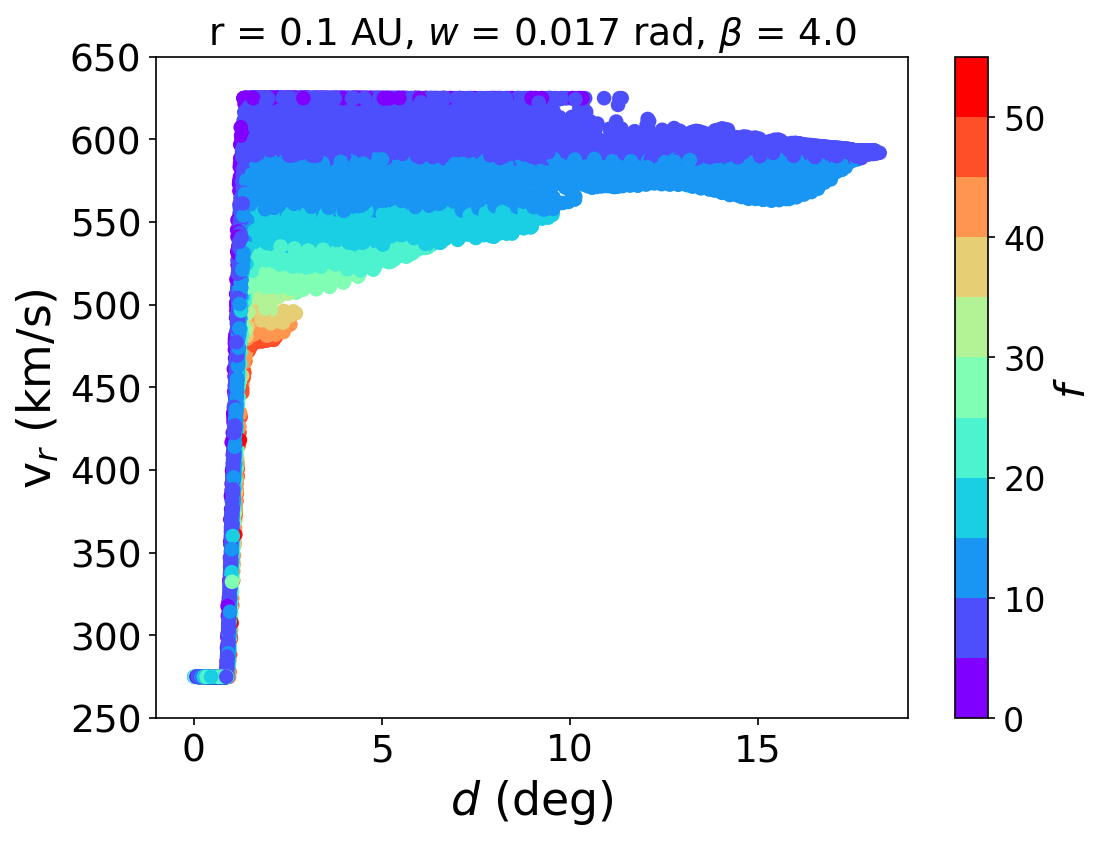}{0.3\textwidth}{(f)}}
           
\caption{Radial velocity at the boundary (v$_{r}$) as a function of the distance to the
CH boundary ($d$). From panel (a) to (c), $w$ increases while $\beta$ stays fixed. From panel (d) to panel (e), $\beta$ increases as $w$ stays fixed.
The flux tube expansion factor ($f$) is color-coded.}
\label{Fig:same_b_different_w}
\end{figure}

\subsection{Calibration of $w$ and $\beta$} 
\label{subsection:Calibration_w_beta}

The parameters $w$ and $\beta$ define the influence of the distance to the CH boundary ($d$) in eq.~\ref{WSA_vr}. They reflect \textit{where} the transition from slow to fast solar wind takes place from the CH boundary, and \textit{how abrupt} this transition is, respectively. The best choice of $w$ and $\beta$ defines the dynamics between the fast and slow solar wind in the predicted solar wind time series, something that we want to improve in the model. Figure~\ref{Fig:same_b_different_w} depicts how the WSA radial velocity (v$_{r}$) at 0.1~au changes for different $w$ values (when keeping $\beta$ fixed), and for different $\beta$ values (when keeping $w$ fixed), respectively. The flux tube expansion factor ($f$) is color-coded. It is important to notice that the difference between Fig.~\ref{Fig:same_b_different_w}a, b, c is the angular distance ($d$) from the CH boundary at which the transition between the slow and fast solar wind takes place. In other words, how quickly from (or, how far inside in) the CH boundary the transition occurs. As $w$ grows bigger, the transition migrates to larger values of $d$, namely, it takes place deeper into the CH. The deeper it is, the more slow solar wind we get before the transition, with the transition to intermediate solar wind speeds occurring deeper in the CHs.
%meaning that intermediate solar wind velocities gradually disappear and turn to slow solar wind. 
That leaves a more restricted range of fast solar wind velocities after the transition zone, which can only be obtained if the CH is big enough. On the other hand, in Fig.~\ref{Fig:same_b_different_w}d, e, f the distance from the CH boundary stays approximately the same but the slope of the transition changes. For lower $\beta$ values, the transition is more gradual, while for larger $\beta$ values, slow wind changes abruptly to fast wind within only a fraction of the degree.

In order to find the best $w, \beta$ combination, we performed a parametric study of these values, keeping $V_{0}$ and $V_{1}$ fixed at their updated values. For each PSP encounter we selected one GONG ADAPT global photospheric magnetic field map. The date of each map was taken at the time that PSP was the closest to the solar central meridian as seen from Earth at the Heliocentric Earth Equatorial frame. For each encounter, we tested the following values for $w$: 0.017, 0.020, 0.026, 0.031 and 0.035 $rad$. This range was decided after conducting many runs to understand the extreme cases of the modelled output. Then, for every value of $w$, we varied $\beta$ in the range [0.50, 3] with a step of 0.5. Additional values of 0.25, 0.75 and 1.25 were considered. Therefore, for each $w$ value, we ran the WSA model 9 times. Taking into account all $w$ and $\beta$ values and all encounters, the total number of WSA runs for this parametric study was 360. 

\begin{sidewaysfigure}[ht]
   \centering
   \includegraphics [width=\textwidth, keepaspectratio, angle=0]{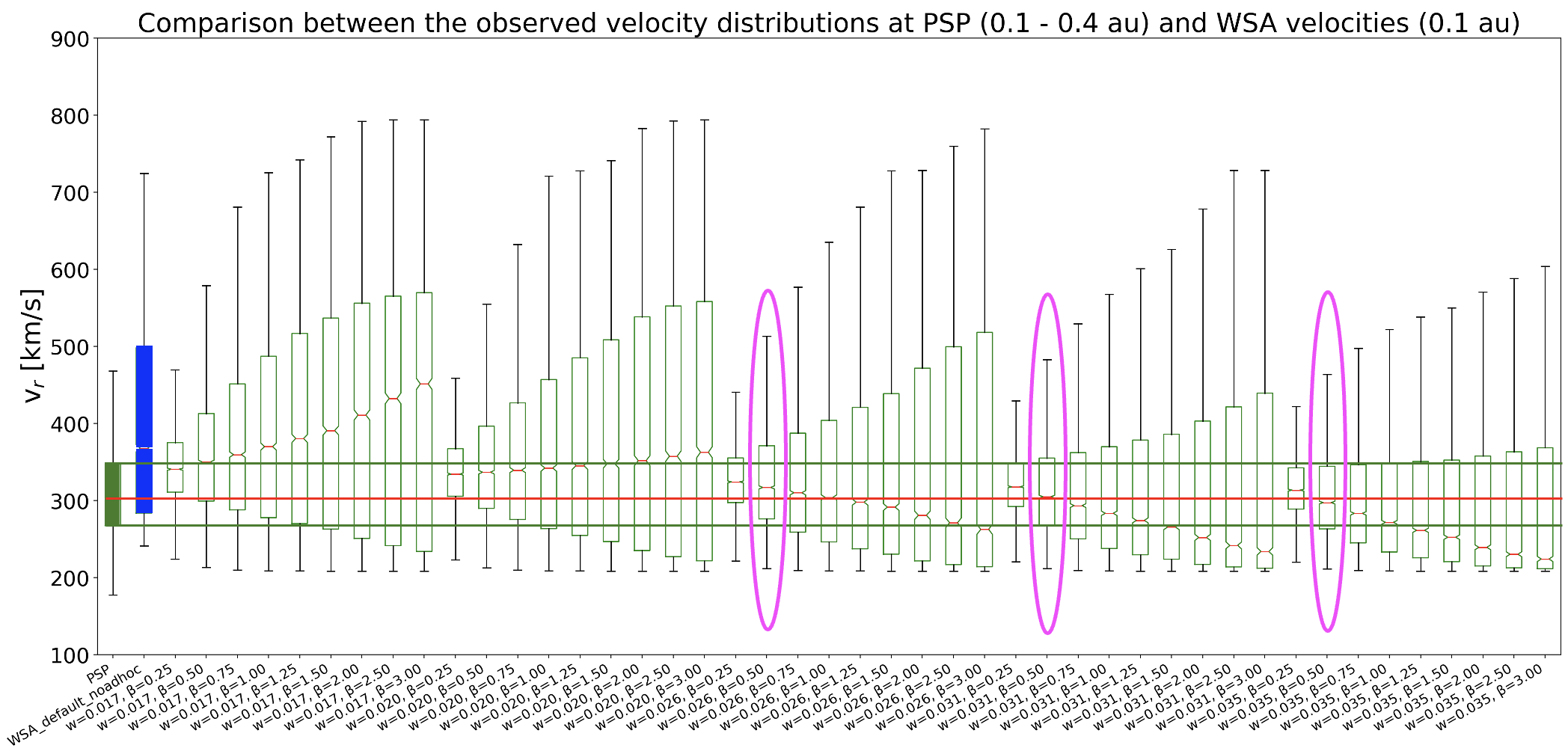}
  \caption{Box-whiskers plot of velocities distributions. Green box: velocity values recorded by PSP between $0.1 - 0.4\;$au for all first eight encounters. Blue box: WSA velocities at 0.1 au without the ad hoc modifications. All other boxes reflect the WSA velocity distributions resulted out of the parametric study of $w, \beta$. The median value of each distribution is shown in red, while the first and third quartiles are represented by the lower and upper box edges, respectively. The black legs extending above and below the boxes show the maximum and minimum velocity values of those distributions.}
  \label{Fig:Comp_of_all_distributions_1stway}
\end{sidewaysfigure}

The results of the parametric study are summarized in the box-whisker plot of Fig.~\ref{Fig:Comp_of_all_distributions_1stway}. Every box represents a velocity distribution from all eight PSP encounters with a median value shown in red, and the first and third quartiles shown by the lower and upper box edges, respectively. The black legs extending above and below the boxes show the maximum and minimum velocity values of those distributions, respectively. The first (green) box in Fig.~\ref{Fig:Comp_of_all_distributions_1stway} corresponds to the distribution of the PSP velocities between \mbox{0.1 - 0.4 au} during the first eight encounters. The second (blue) box depicts the velocities from the default WSA model at $0.1\;$au without the ad hoc assumptions of subtracting $50\;$km/s and capping the velocities between $[275, 625]\;$km/s at the boundary (from now on, stated as ``noadhoc'' in the plots). All other boxes reflect the WSA velocity distributions obtained after the parametric study of $w$ and $\beta$, having included the updated values \mbox{for $V_{0}$, $V_{1}$.}

Fig.~\ref{Fig:Comp_of_all_distributions_1stway} shows that the default WSA distribution (blue box) does not resemble the PSP distribution (green box). It seems that the default WSA set-up generally overestimates the measured PSP velocities. To understand if any other box approximates better the green distribution, we extended the lines of the median (red), first and third quartiles (green) of the PSP box throughout the whole figure for a better comparison with the rest of the distributions. The WSA configuration that resembles the green PSP distribution the closest is circled in magenta and corresponds to [$w, \beta$] combination $[0.031 \text{ rad}, 0.50]$. As a second and third optimal choice, we further select the combinations $[0.035 \text{ rad}, 0.50]$ and $[0.026 \text{ rad}, 0.50]$, respectively.

\begin{figure}
\centering
\gridline{\fig{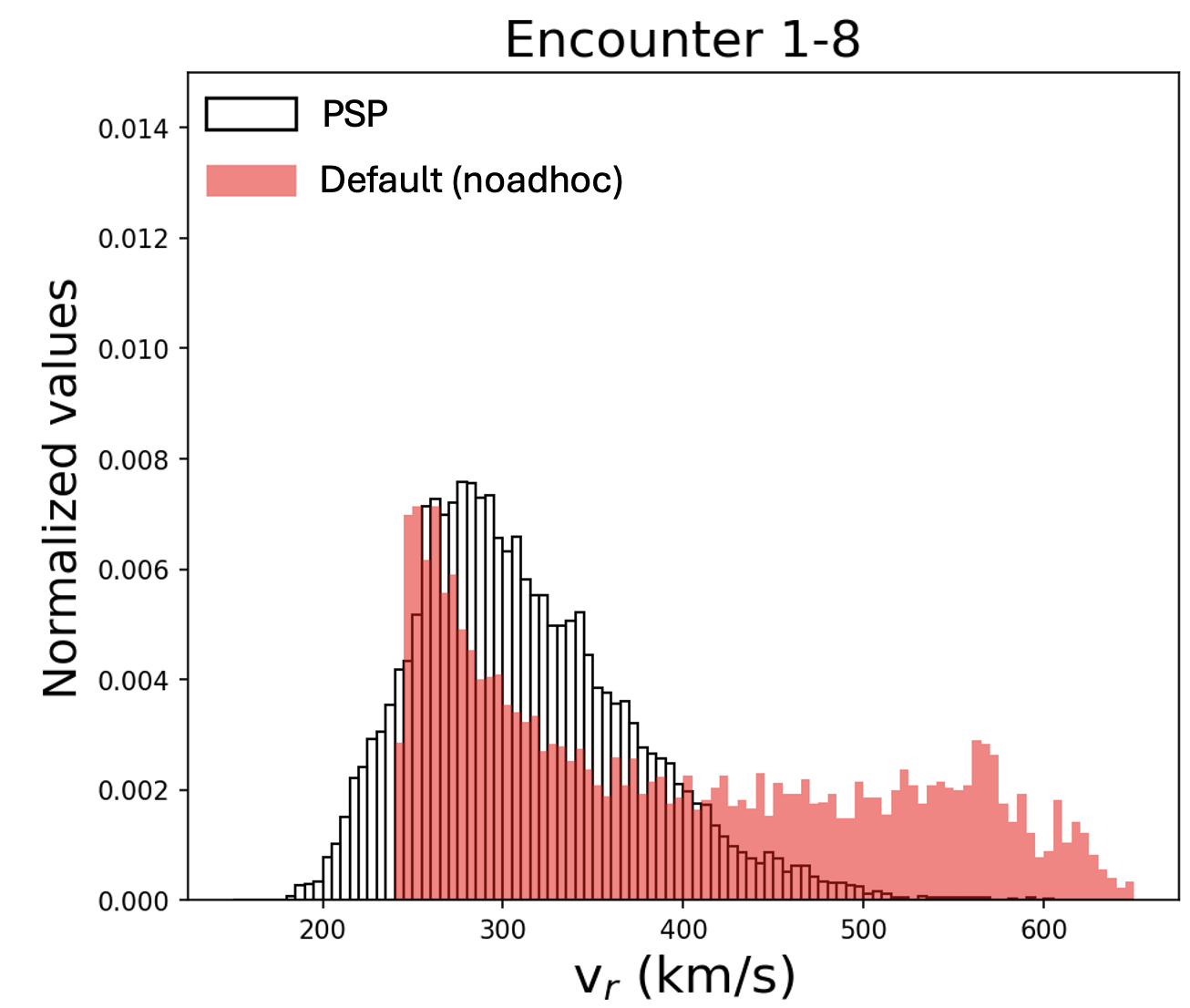}{0.5\textwidth}{(a)}
            \fig{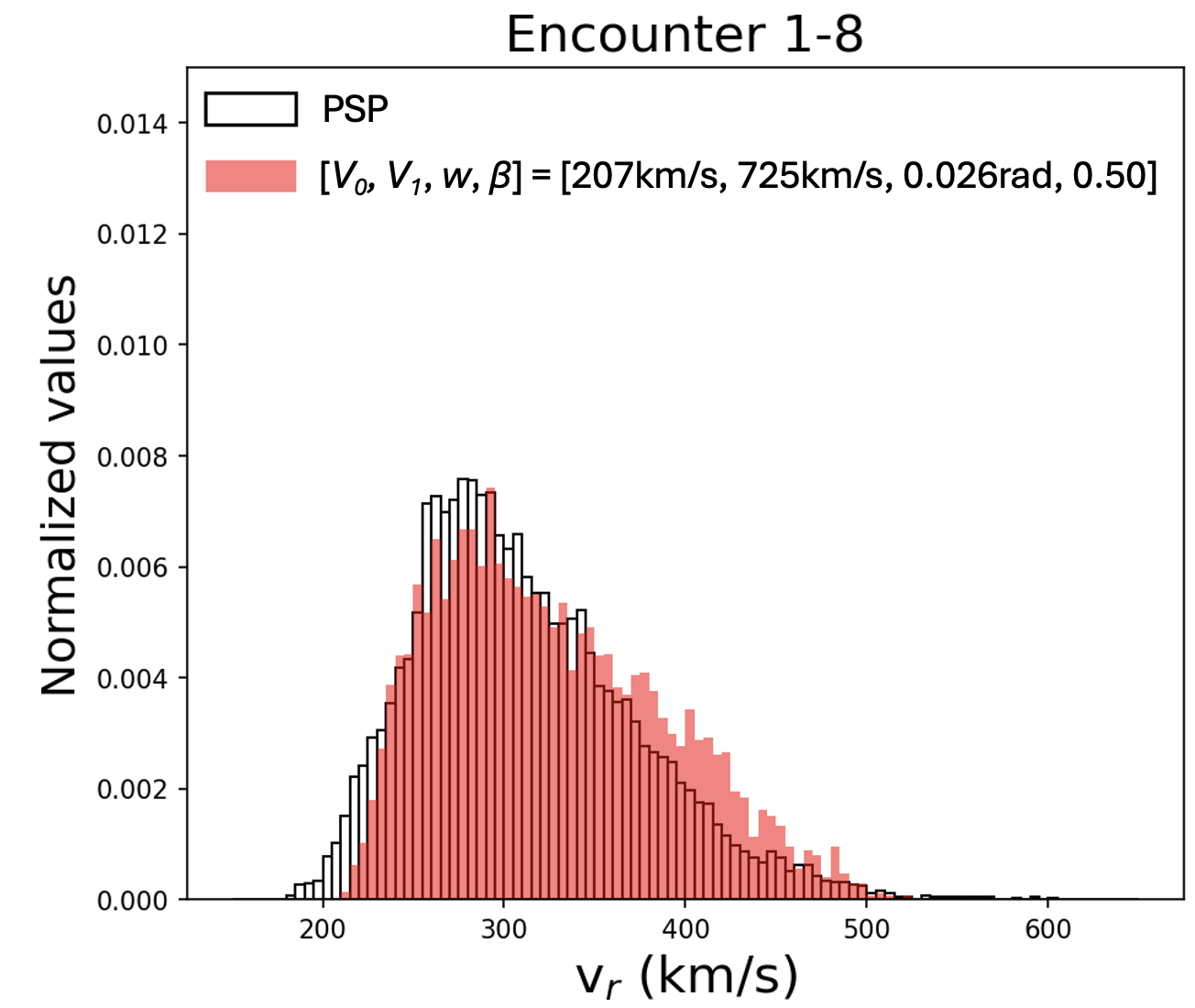}{0.51\textwidth}{(b)}}

\gridline{\fig{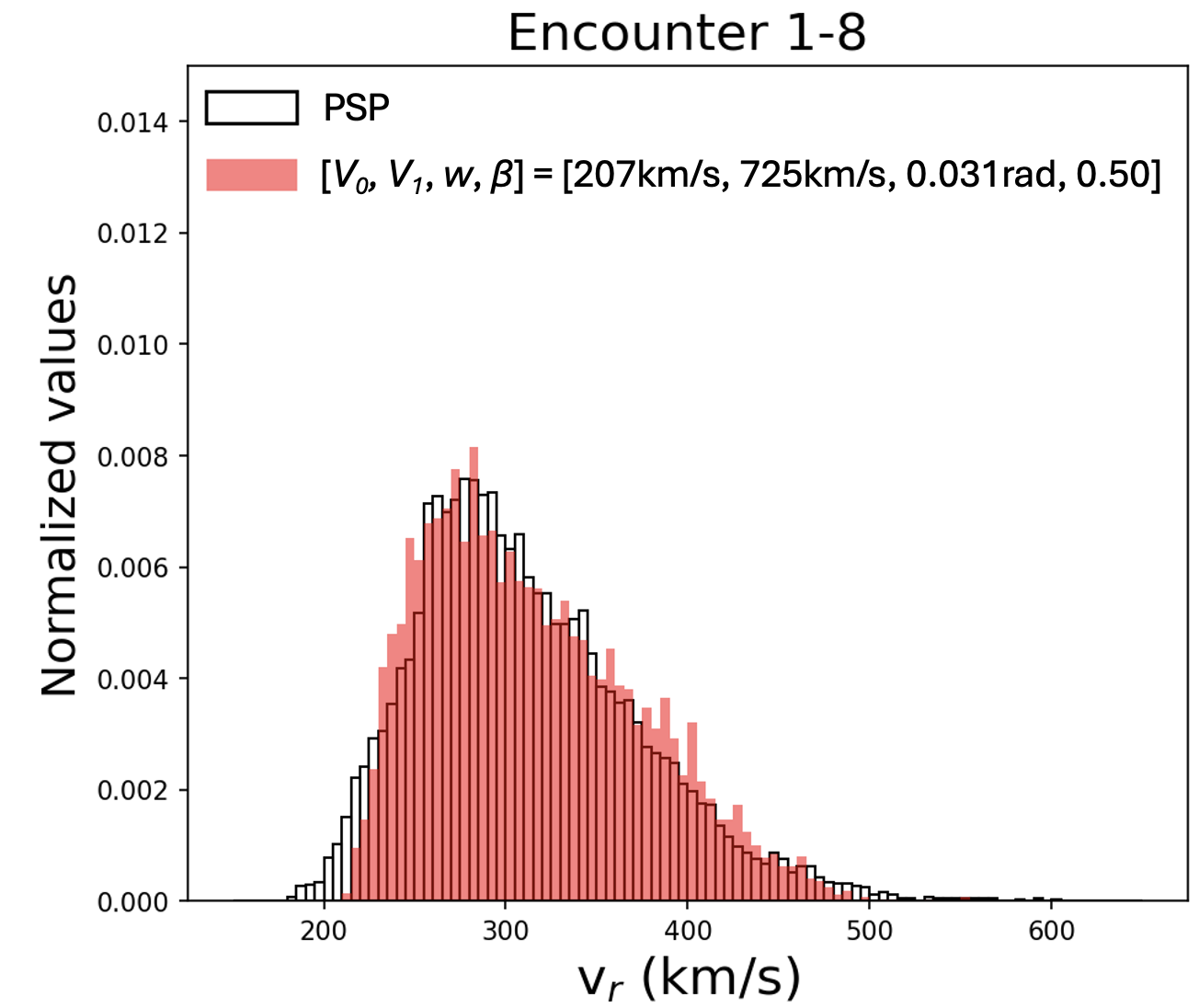}{0.5\textwidth}{(a)}
            \fig{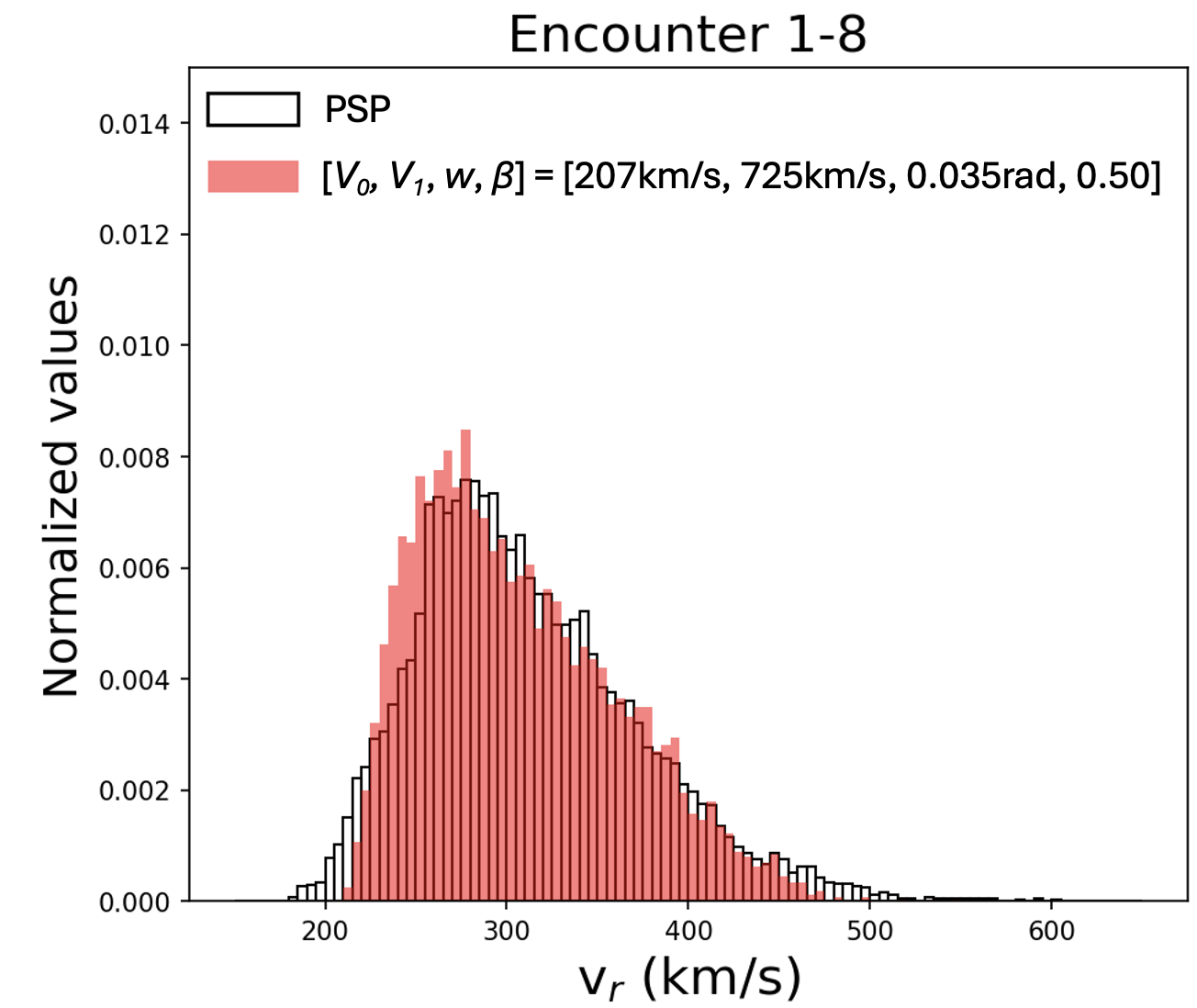}{0.507\textwidth}{(b)}}
            
\caption{Histograms of PSP (white) and WSA (red) velocities for the first eight PSP encounters while the spacecraft was between 0.1-0.4 au. Panel (a): comparison between PSP observations and the default WSA model without the ad hoc modifications. Panel (b): comparison between PSP observations and WSA model with $[w, \beta]=[0.026 \text{ rad}, 0.50]$ corresponding to the first circled distribution of Fig. \ref{Fig:Comp_of_all_distributions_1stway}. Panel (c): same as (b) but for WSA modeling setup of $[w, \beta]=[0.031 \text{ rad}, 0.50]$ corresponding to the second circled distribution of Fig.~\ref{Fig:Comp_of_all_distributions_1stway}. Panel (d): same as (b) but for WSA modeling setup of $[w, \beta]=[0.035 \text{ rad}, 0.50]$ corresponding to the third circled distribution of Fig.~\ref{Fig:Comp_of_all_distributions_1stway}.}
\label{Fig:Distributions}
\end{figure}

% To obtain a better idea how the $w, \beta$ variables change the WSA velocity distributions each time, we present in Fig.~\ref{Fig:Distributions} the histograms of PSP observations and WSA modeled output for all values of $w$ and selected values of $\beta$ (i.e., $\beta$ = [0.25, 0.50, 2]. 

To obtain a better idea how the selected distributions look and compare to the PSP one, we present in Fig.~\ref{Fig:Distributions} the histograms of velocities from PSP and WSA modeled output close to the Sun. The red distributions of Fig.~\ref{Fig:Distributions}a, b, c, d correspond to the blue box and the three optimal WSA configurations (circled in magenta) of Fig.~\ref{Fig:Comp_of_all_distributions_1stway}, respectively. We notice that the optimal WSA configurations selected in Fig.~\ref{Fig:Comp_of_all_distributions_1stway} resemble better the PSP distribution compared to the default WSA set-up (see, e.g., comparison between Fig.~\ref{Fig:Distributions}b, c, d and Fig.~\ref{Fig:Distributions}a). The question now is, does this better agreement between the WSA boundary conditions and the PSP data close to the Sun lead to better solar wind forecasts in the inner heliosphere? To see if this is the case, we ran the heliospheric part of EUHFORIA. The heliospheric predictions arising from the optimal WSA configurations are presented and discussed in section \ref{section:Section_4}.

\section{Solar wind predictions at PSP and Earth}
\label{section:Section_4}

\begin{figure}
\centering
\gridline{\fig{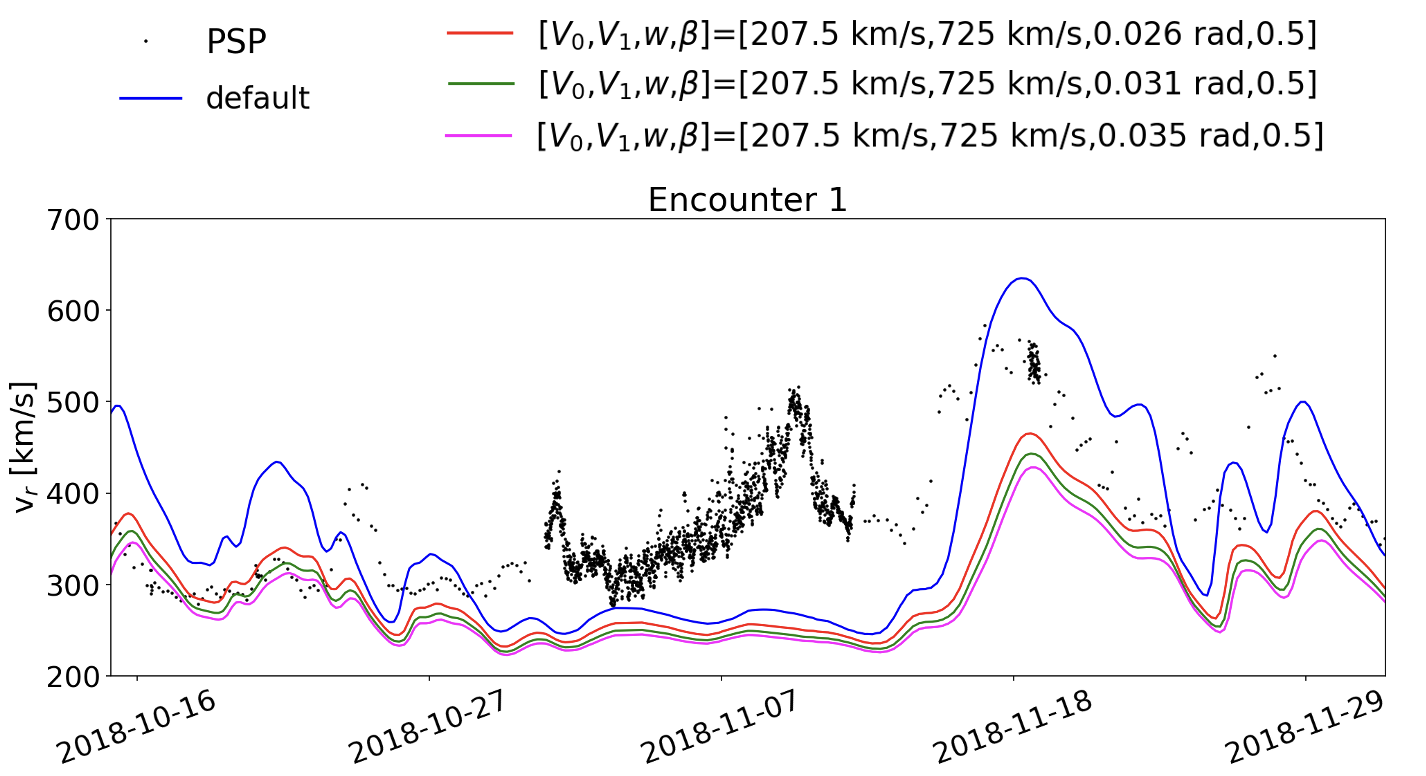}{0.48\textwidth}{(a)}
            \fig{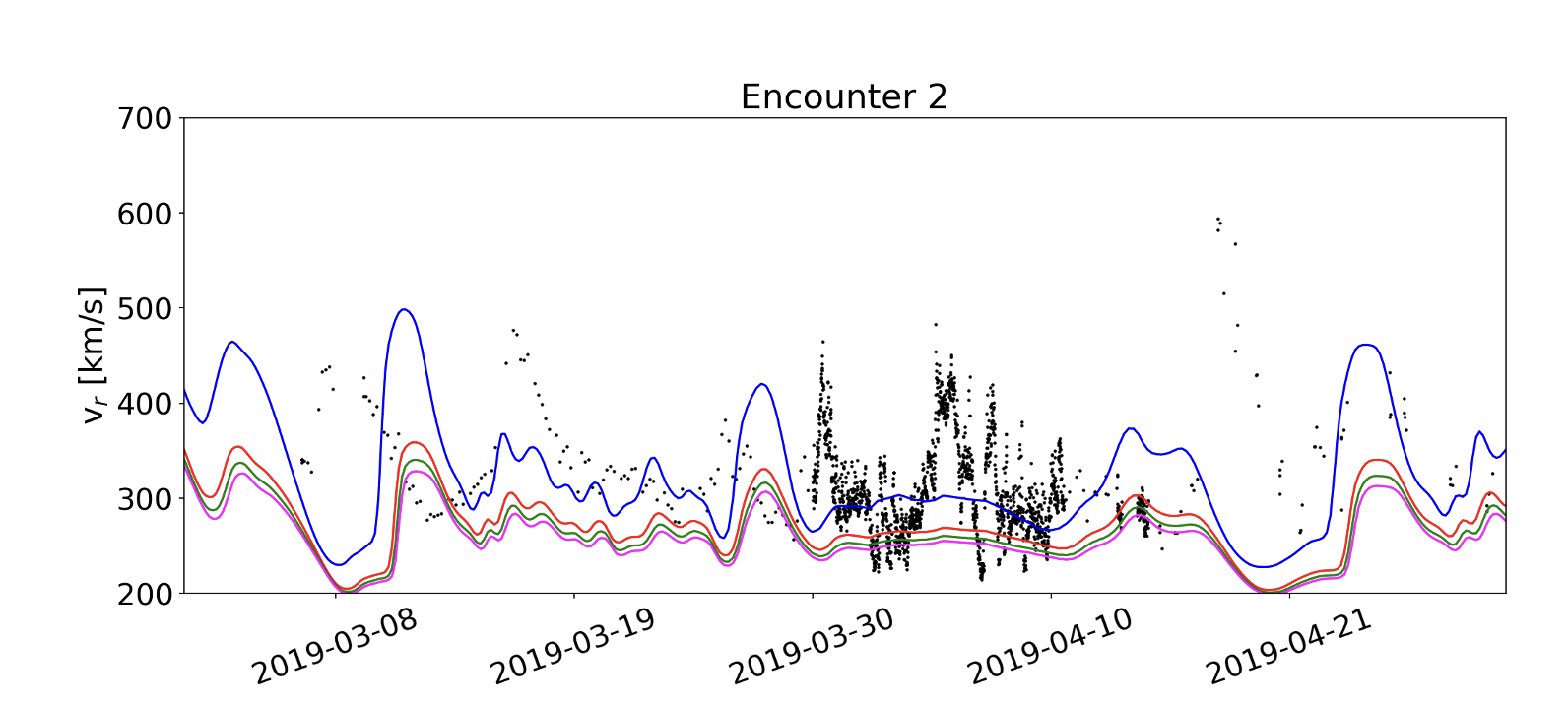}{0.50\textwidth}{(b)}}

\gridline{\fig{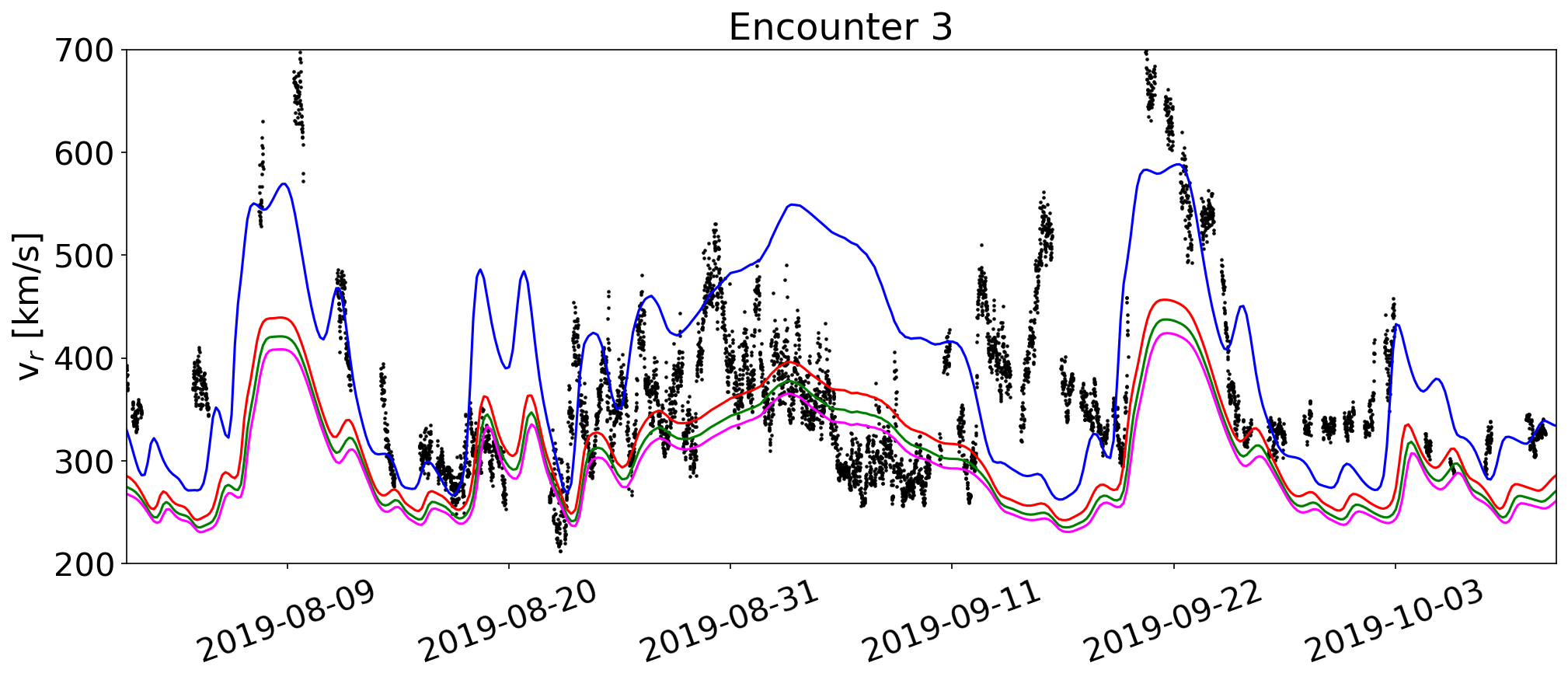}{0.465\textwidth}{(c)}
            \fig{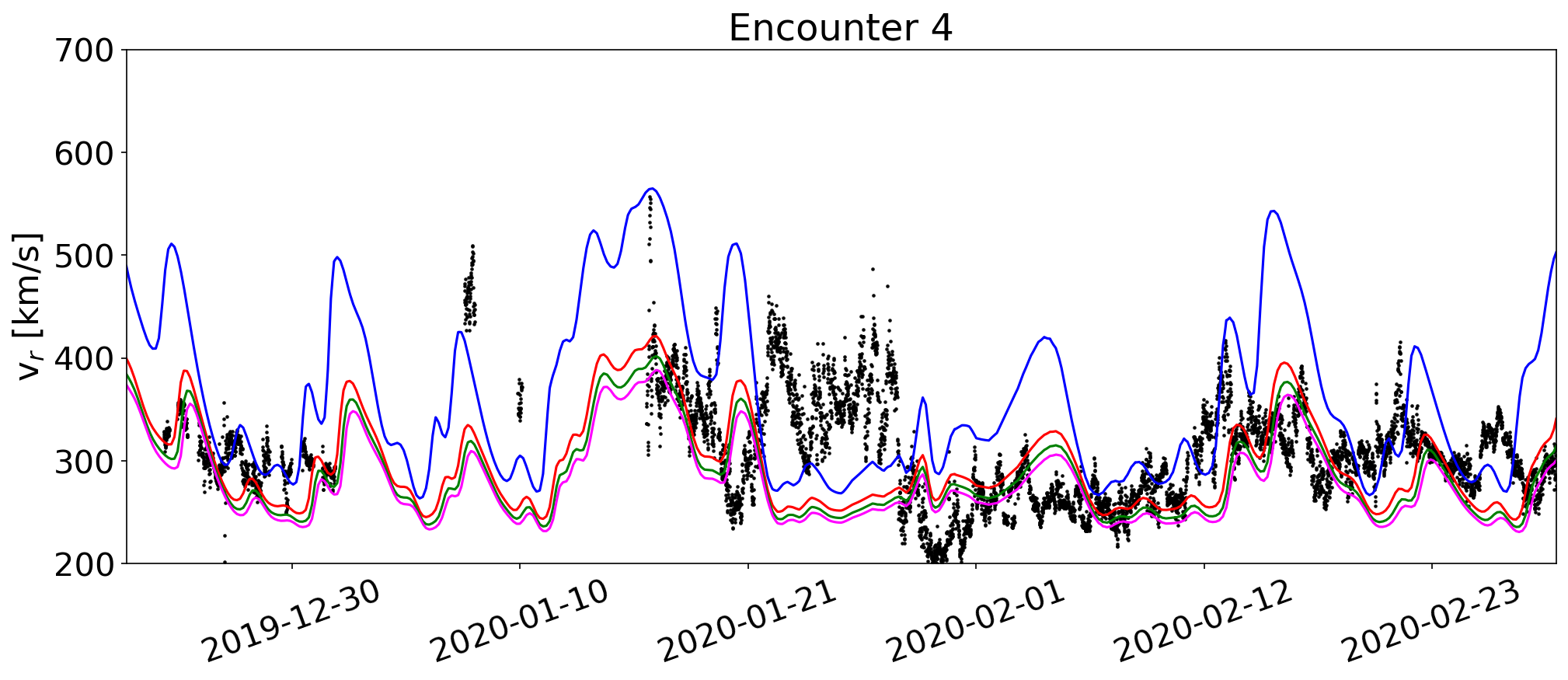}{0.47\textwidth}{(d)}}

\gridline{\fig{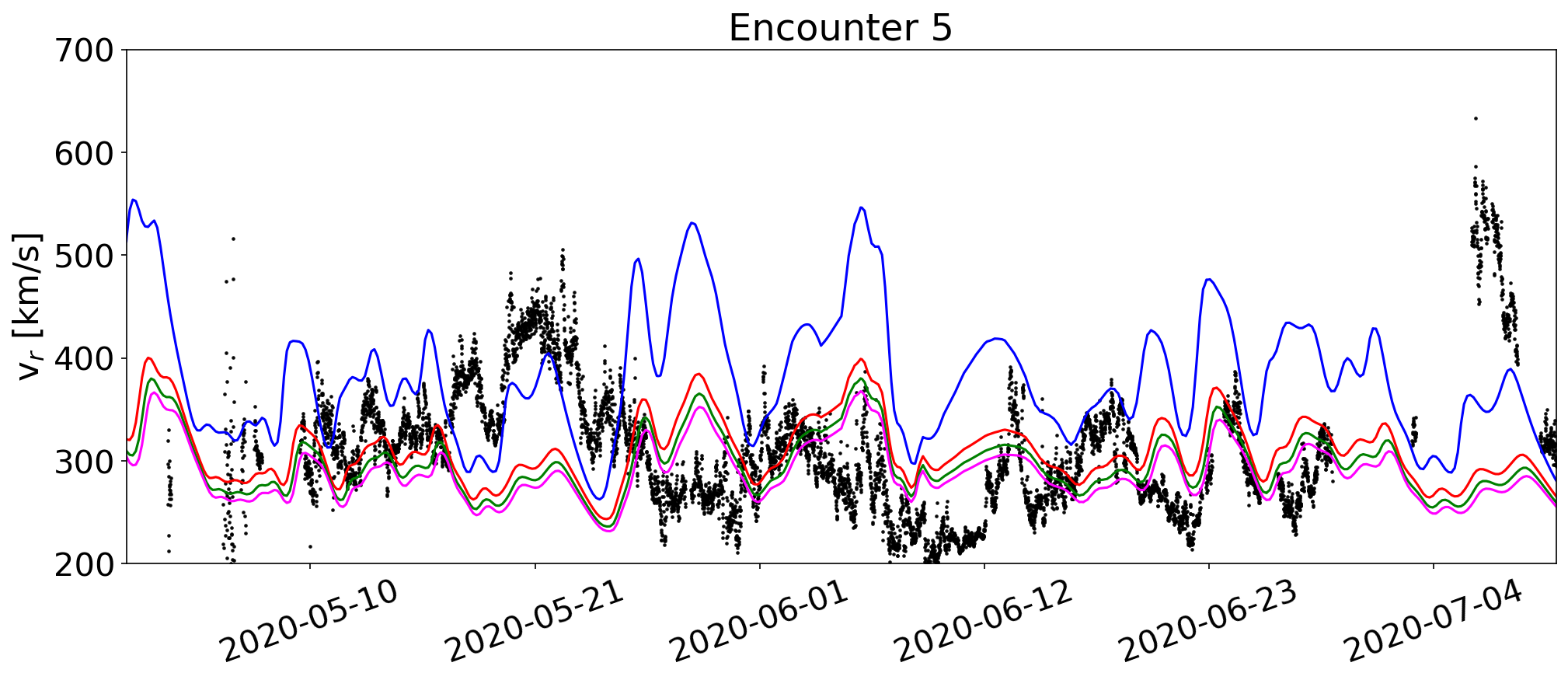}{0.47\textwidth}{(e)}
            \fig{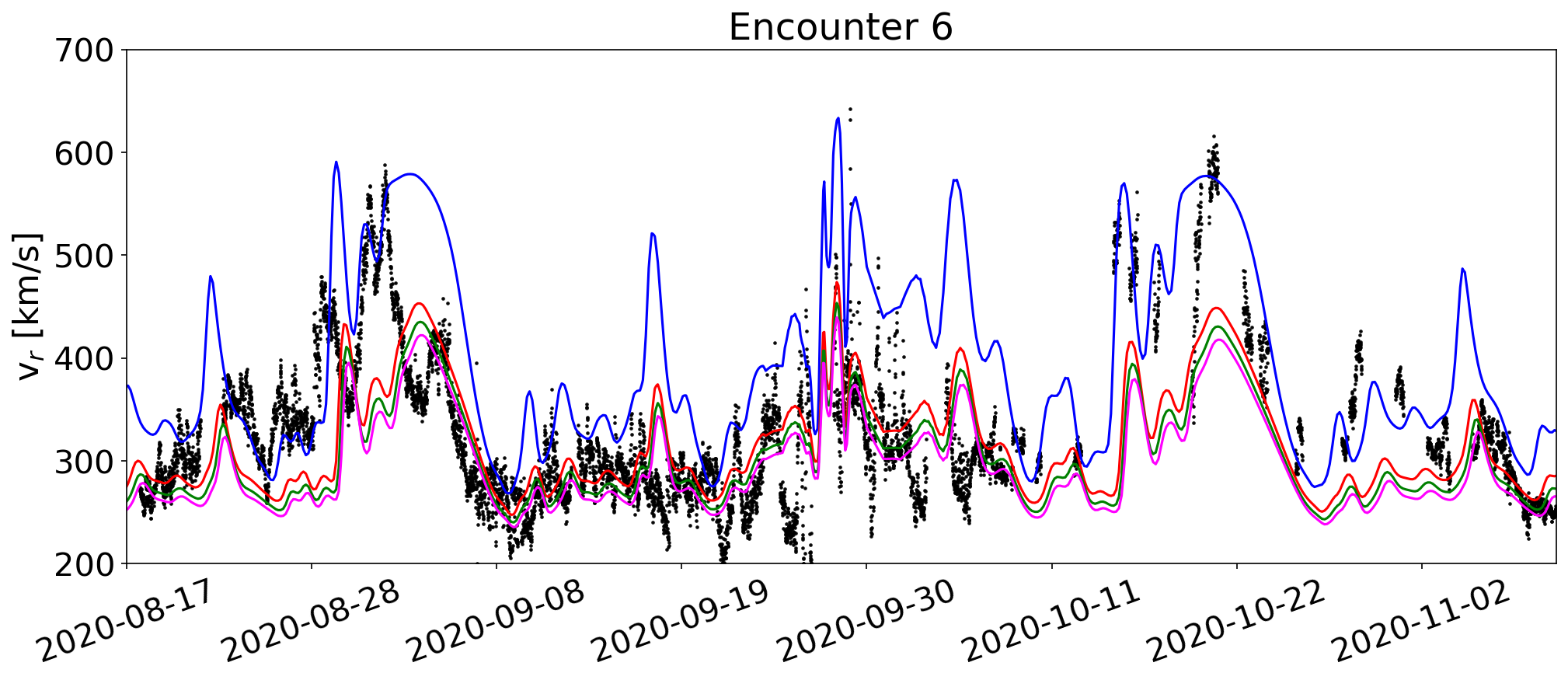}{0.465\textwidth}{(f)}}

\gridline{\fig{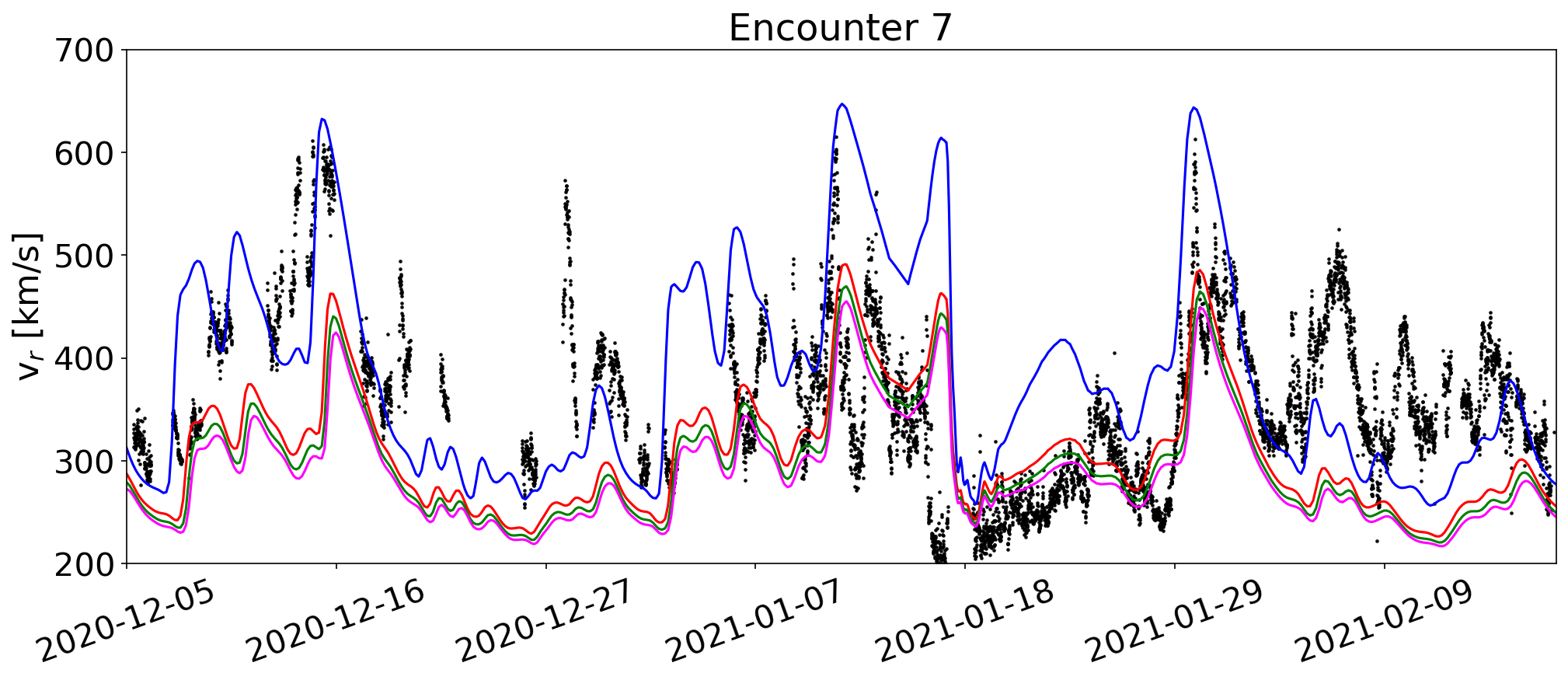}{0.475\textwidth}{(g)}
            \fig{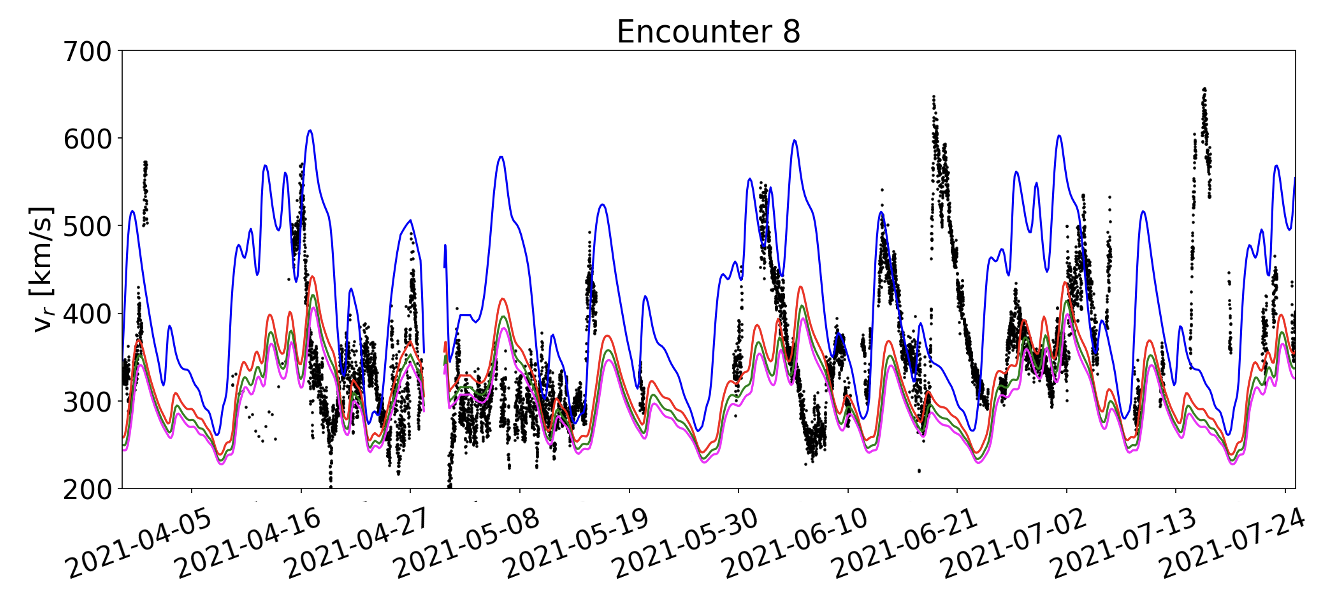}{0.48\textwidth}{(h)}}
                     
\caption{Comparison between PSP time series (black) and modeled output
from WSA+EUHFORIA (colorful lines). Blue time series correspond to runs made with the 
the default WSA configuration. The magenta and green time series correspond to runs made with the optimized configurations of Fig.~\ref{Fig:Comp_of_all_distributions_1stway}. ICMEs have been removed from observations according to the HELIO4CAST ICME catalogue \citep[see][for more details]{mostl2017, mostl2020}{}{}}.
\label{Fig:TimeseriesPSP_1stway}
\end{figure}

\begin{figure}
\centering
\gridline{\fig{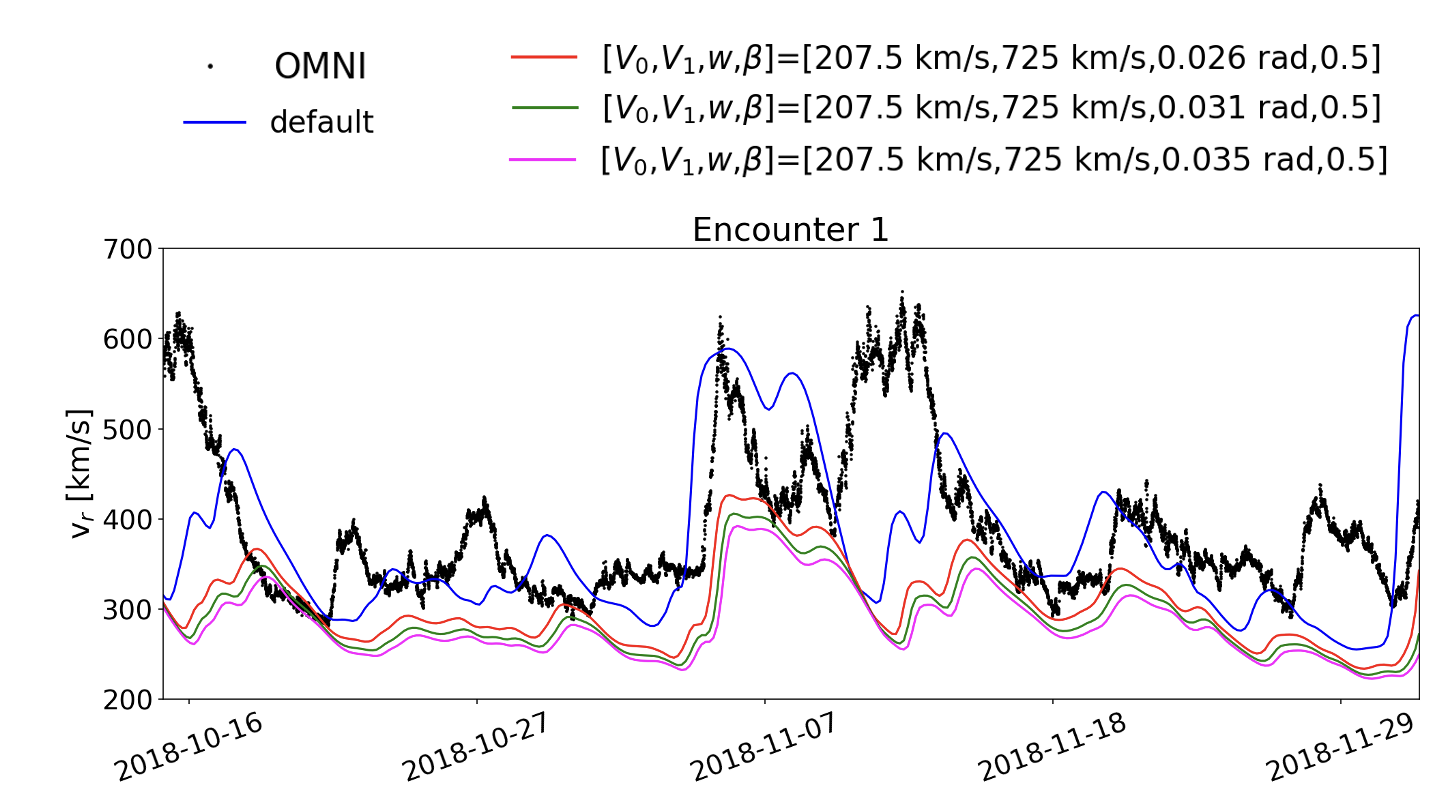}{0.49\textwidth}{(a)}
            \fig{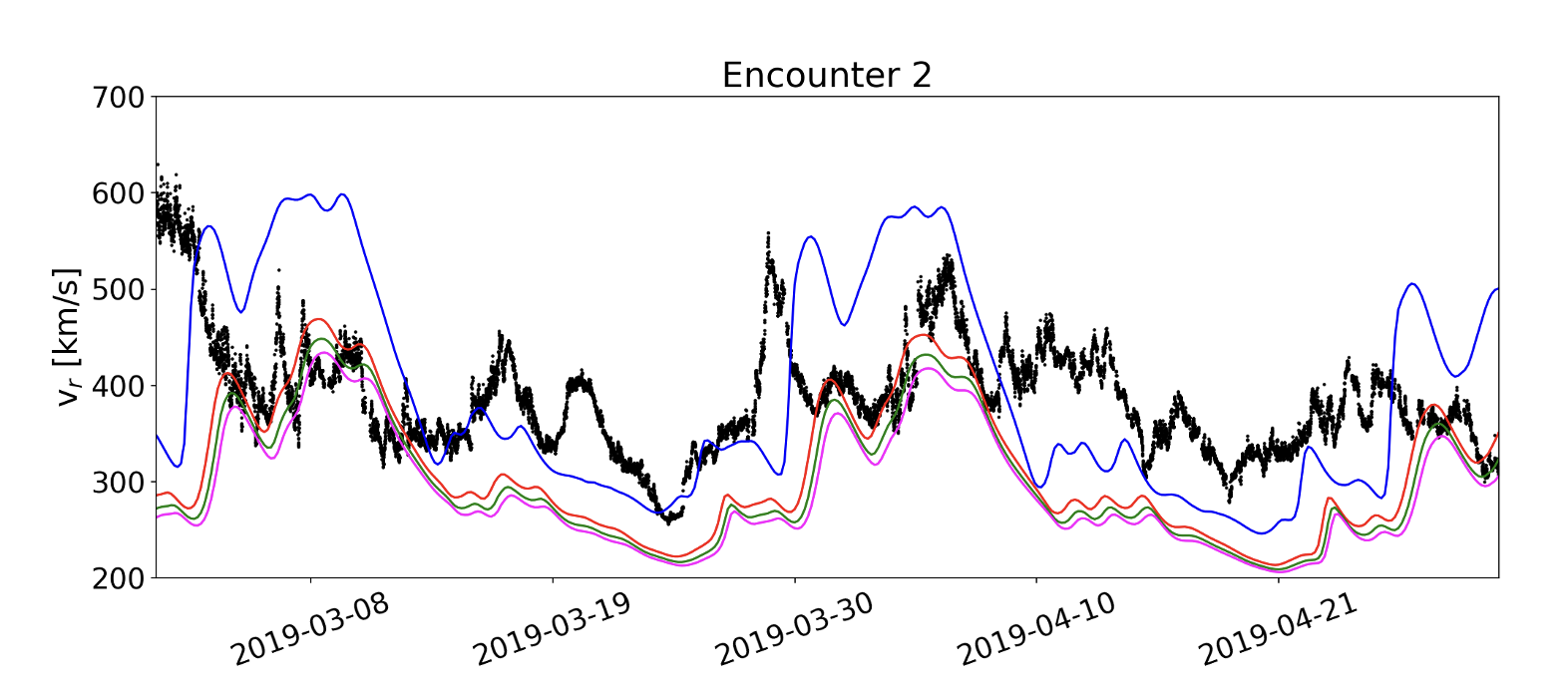}{0.505\textwidth}{(b)}}

\gridline{\fig{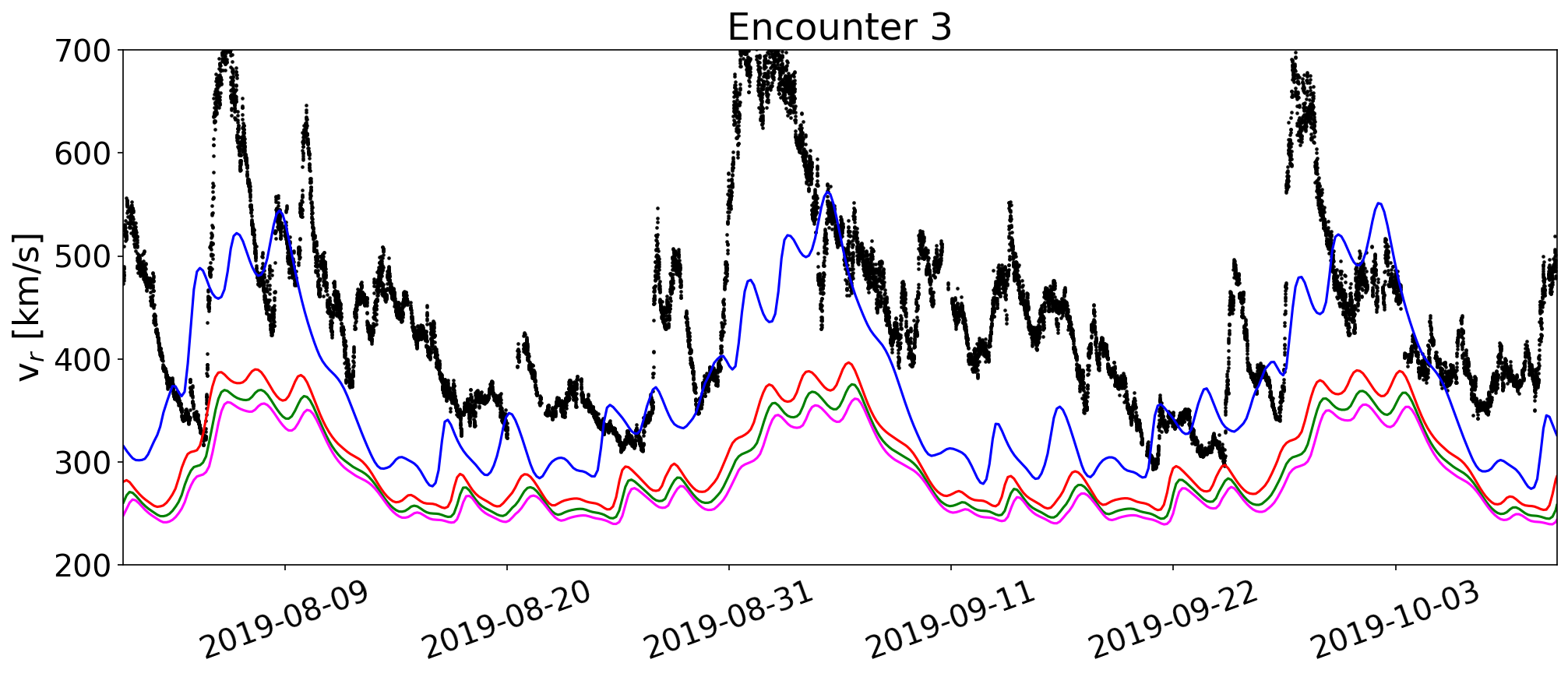}{0.465\textwidth}{(c)}
            \fig{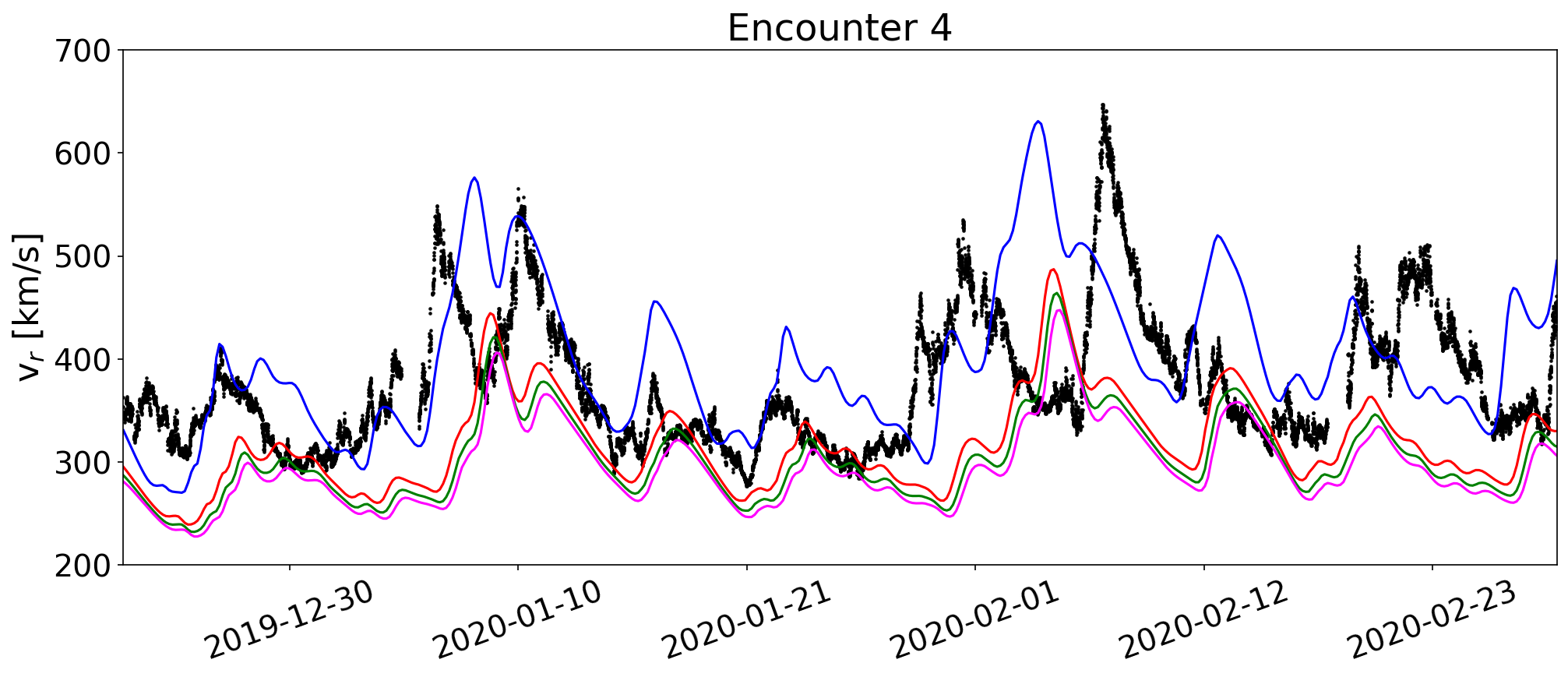}{0.465\textwidth}{(d)}}

\gridline{\fig{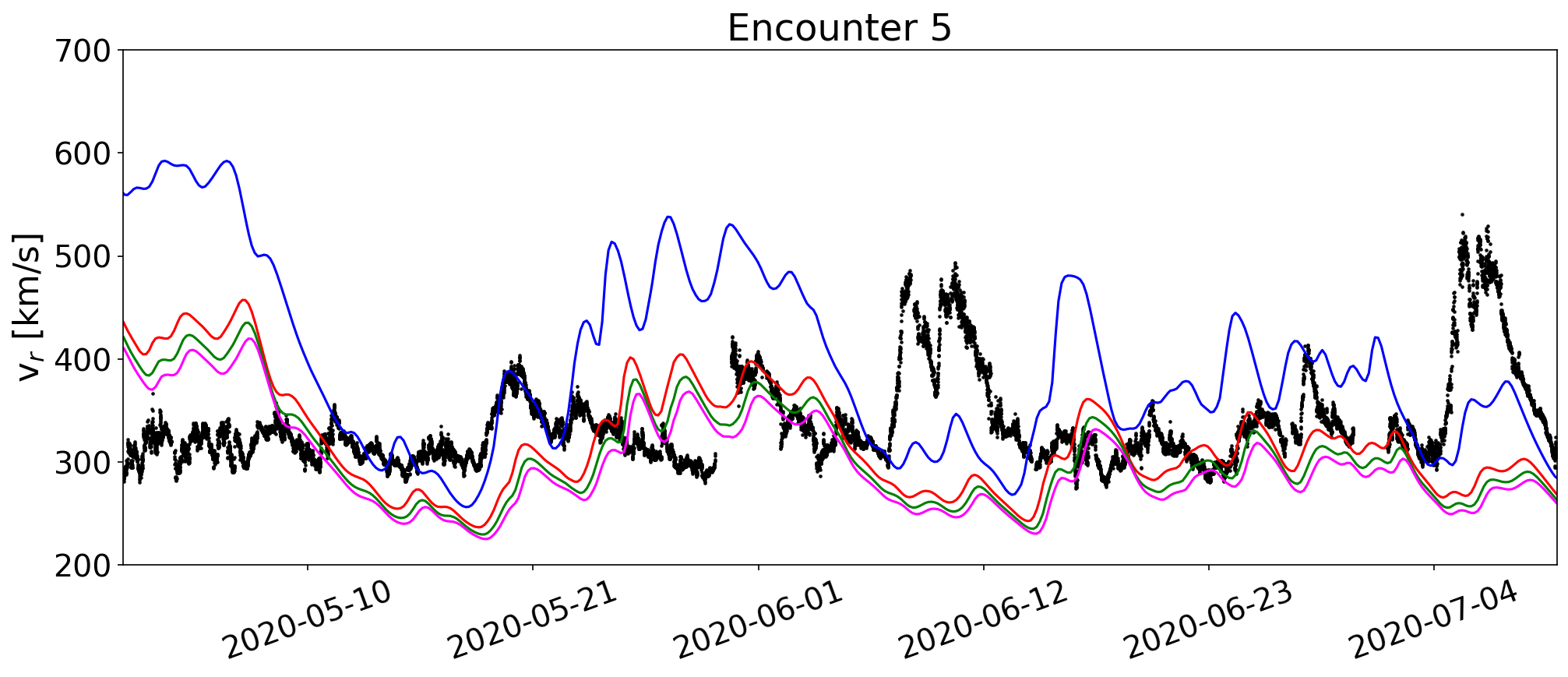}{0.465\textwidth}{(e)}
            \fig{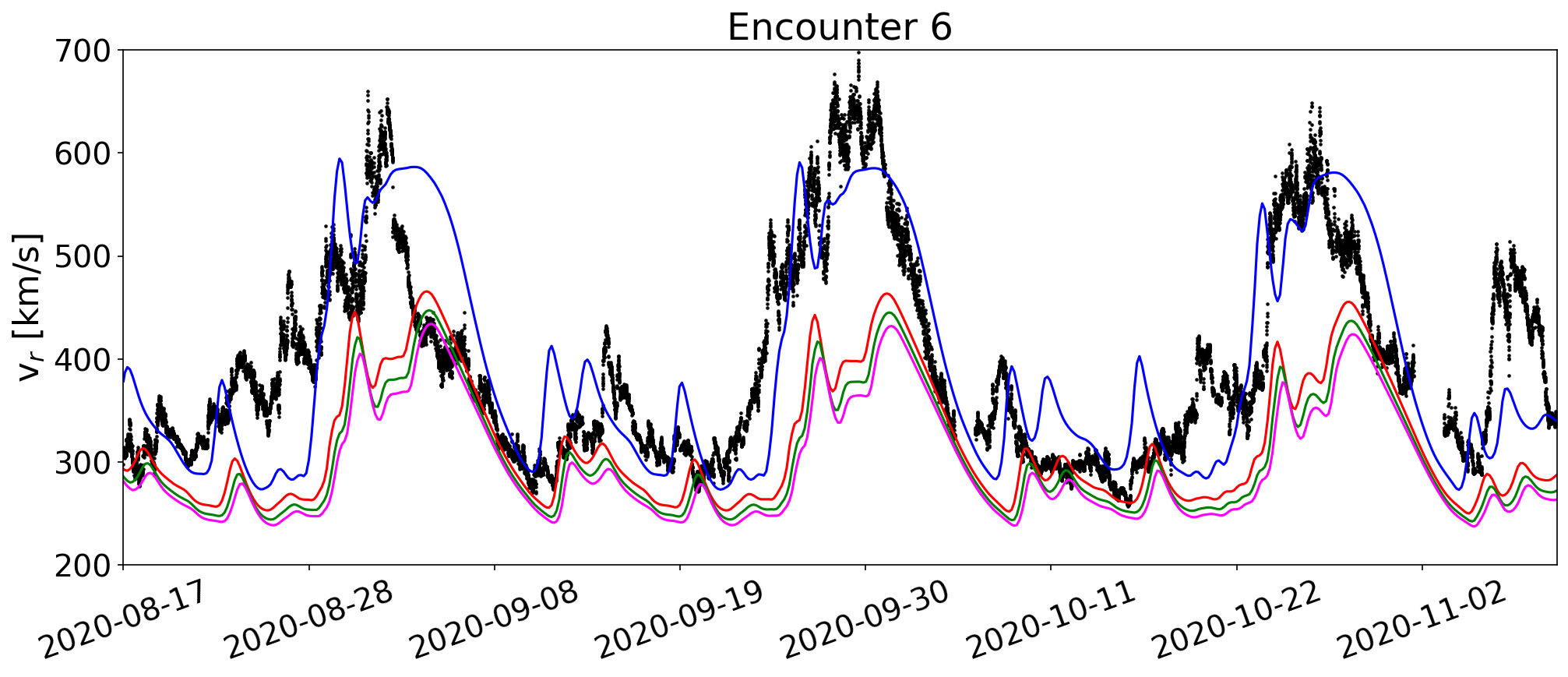}{0.465\textwidth}{(f)}}

\gridline{\fig{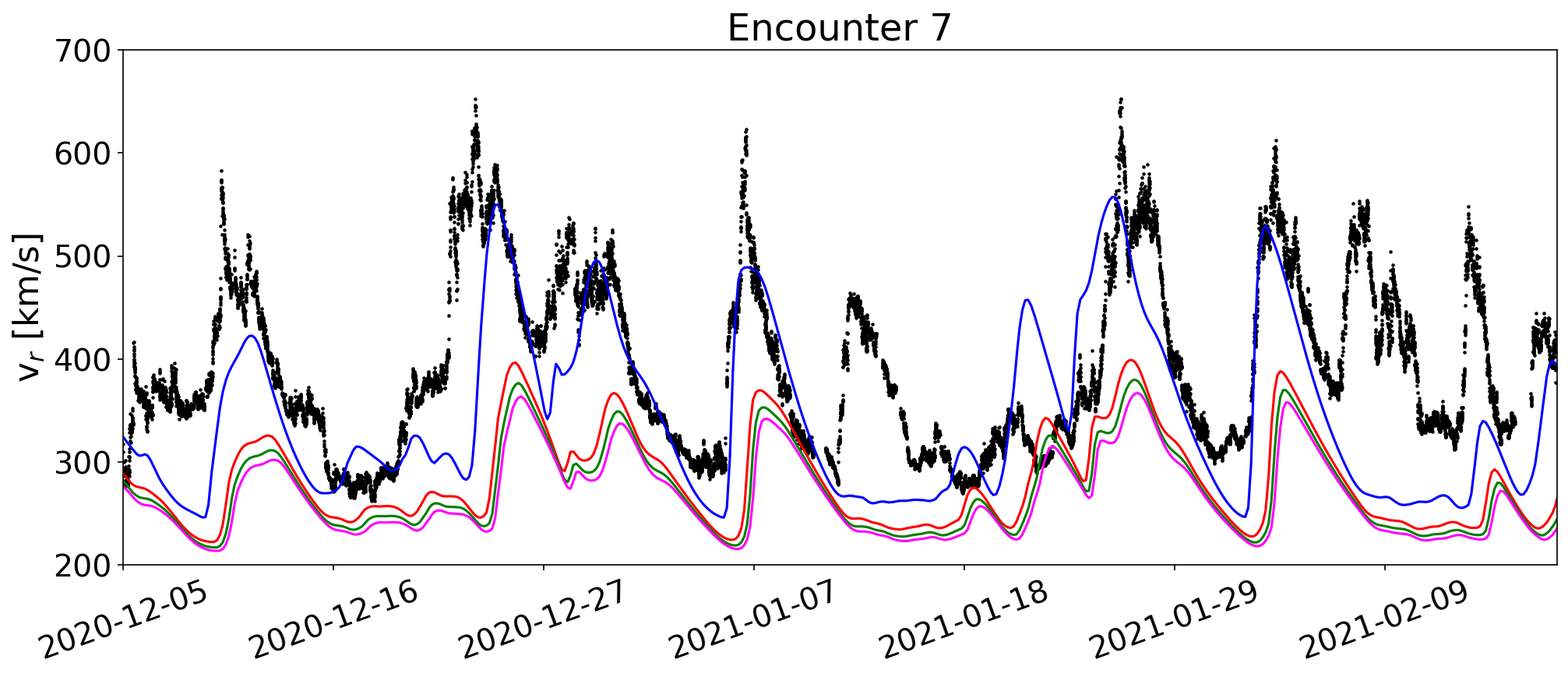}{0.465\textwidth}{(g)}
            \fig{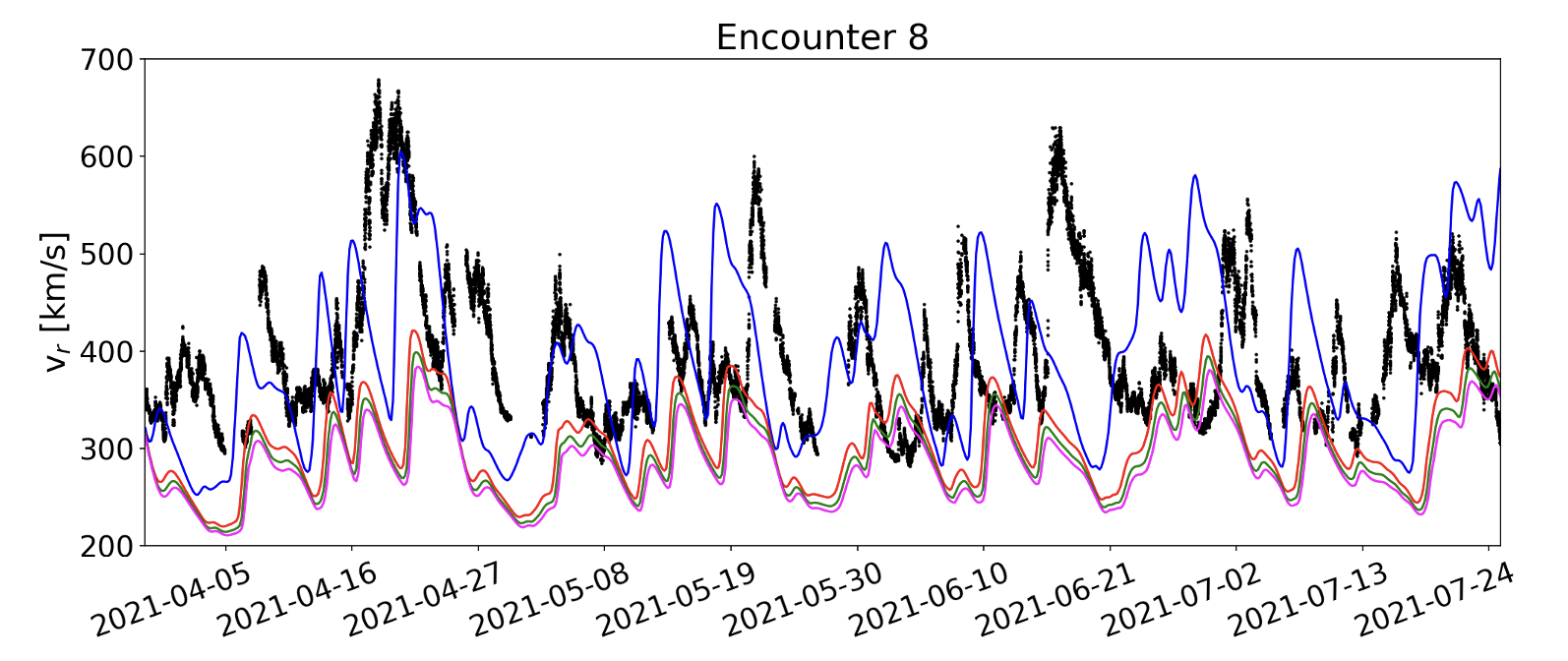}{0.49\textwidth}{(h)}}
                     
\caption{Same as Fig.~\ref{Fig:TimeseriesPSP_1stway} but for time series at Earth.}
\label{Fig:TimeseriesEarth_1stway}
\end{figure}

The solar wind velocity predictions resulting from the three optimal WSA configurations of Fig.~\ref{Fig:Comp_of_all_distributions_1stway} are presented in Fig.~\ref{Fig:TimeseriesPSP_1stway} and \ref{Fig:TimeseriesEarth_1stway} for PSP and Earth position, respectively. Comparisons with in situ data and predictions based on the default WSA configuration, are also shown. In Fig.~\ref{Fig:TimeseriesPSP_1stway} we notice that for some time intervals solar wind velocity is reconstructed better from the default WSA setup (blue line; see e.g., the last HSSs at the end of encounter 1, the HSSs at the beginning of encounter 3, or the HSSs at the beginning and end of encounter 7). On the other hand, other time intervals seem to be modeled better when using one of the three optimized configurations of Fig.~\ref{Fig:Comp_of_all_distributions_1stway} (red, green or magenta line; see e.g., the slow solar wind at the beginning of encounter 1, or the dynamics between slow and fast solar wind during parts of the encounter 5, 6 and 7). From Fig.~\ref{Fig:TimeseriesEarth_1stway} we conclude that the solar wind forecasts from the optimized configurations of Fig.~\ref{Fig:Comp_of_all_distributions_1stway} (red, green and magenta lines) generally under predict observations at Earth. It is, overall, the default WSA set up that provides the best predictions, even though these predictions have still a lot of room for improvement. 

To accurately quantify the performance of our modeled output compared to observations and see if the aforementioned ``by-eye'' conclusions are true, we applied the Dynamic Time Warping method \citep[DTW;][]{Samara2022DTW}{}{}. This method was applied only for the time series at Earth as PSP observations are characterized by big data gaps. Based on DTW, we can define a skill-score called Sequence Similarity Factor (SSF):

\begin{equation}
\centering
    \textrm{SSF} = \frac{\textrm{DTW}_{score}(O,M)}{\textrm{DTW}_{score}(O,R)}, \quad \textrm{SSF}\in [0,\infty),
\label{eq:SSF}
\end{equation}

\noindent a number that provides a quantification of how good the forecast is, compared to the ideal and a reference prediction scenarios. In Eq.~\ref{eq:SSF}, the DTW$_{score}$ is a distance measure between modelled and observed series. More specifically, the numerator is the DTW score between the observed ($O$) and modelled ($M$) time series. The denominator is the DTW score between the observed time series and a reference-case scenario which, for the purposes of this study, corresponds to the average value of observations per encounter (mean model). For DTW$_{score}$ = 0, we have the perfect agreement between predictions and observations. Therefore, SSF equal to zero indicates the perfect forecast while SSF equal to one shows that the forecast performs the same as our reference model. For SSF values higher than one, the model’s predictions are worse than the predictions from the reference model. The closer the SSF is to zero, the better the forecast is. 

\begin{table*}[ht]
\centering
\caption{Evaluation of solar wind velocity time series at Earth based on the DTW method for the dates of each PSP encounter. Column~1 indicates the number of the encounter while columns 2 to 5 show the SSF skill score for predictions performed with the default WSA configuration and the three optimal WSA configurations of Fig.~\ref{Fig:Comp_of_all_distributions_1stway}, respectively. The closer the SSF is to zero, the better the forecast is.}
\begin{center}
  %\resizebox{\textwidth}{!}{%
 \begin{tabular}{|c|c|c|c|c|} 
 \hline
  Encounter &  SSF WSA default &  SSF WSA [0.026 rad, 0.50] &  SSF WSA [0.031 rad, 0.50] &  SSF WSA [0.035 rad, 0.50]  \\
 \hline\hline
 \hline

1 & 0.63 & 1.19  & 1.42 & 1.56 \\
\hline
2 & 1.44 & 1.49  & 1.72 & 1.92\\
\hline
3 & 0.88 & 1.79  & 2.01 & 2.16\\
\hline
4 & 0.68 & 0.98  & 1.25 & 1.44\\
\hline
5 & 2.50 & 1.65  & 1.75 & 1.84\\
\hline
6 & 0.30 & 0.75  & 0.91 & 1.02\\
\hline
7 & 0.58 & 1.33  & 1.52 & 1.65\\
\hline
8 & 0.87 & 1.44  & 1.69 & 1.86\\
\hline

\end{tabular}
%}
\end{center}
\label{Table:DTW_Earth}
\end{table*}

\section{Results}
\label{section:Section_5}
\subsection{Evaluation of predictions at Earth according to DTW}

The results from the evaluation of the solar wind velocity predictions at Earth based on the default and the three optimal WSA setups identified in Fig.~\ref{Fig:Comp_of_all_distributions_1stway}, are summarized in Table~\ref{Table:DTW_Earth}. From there we conclude that the default WSA configuration is the best to reconstruct the solar wind velocities at Earth for all cases, except encounter 5. During the latter encounter, the WSA setup of $[w, \beta] = [0.026 \text{ rad}, 0.50]$ (red line in Fig.~\ref{Fig:TimeseriesEarth_1stway}e) performs the best compared to all other configurations. We also notice that for all other encounters, the second best WSA configuration is $[w, \beta] = [0.026 \text{ rad}, 0.50]$ which always performs better compared to $[w, \beta] = [0.031 \text{ rad}, 0.50]$ and $[w, \beta] = [0.035 \text{ rad}, 0.50]$. 

\begin{sidewaysfigure}[ht]
   \centering
   \includegraphics [width=1.0\textwidth, keepaspectratio, angle=0]{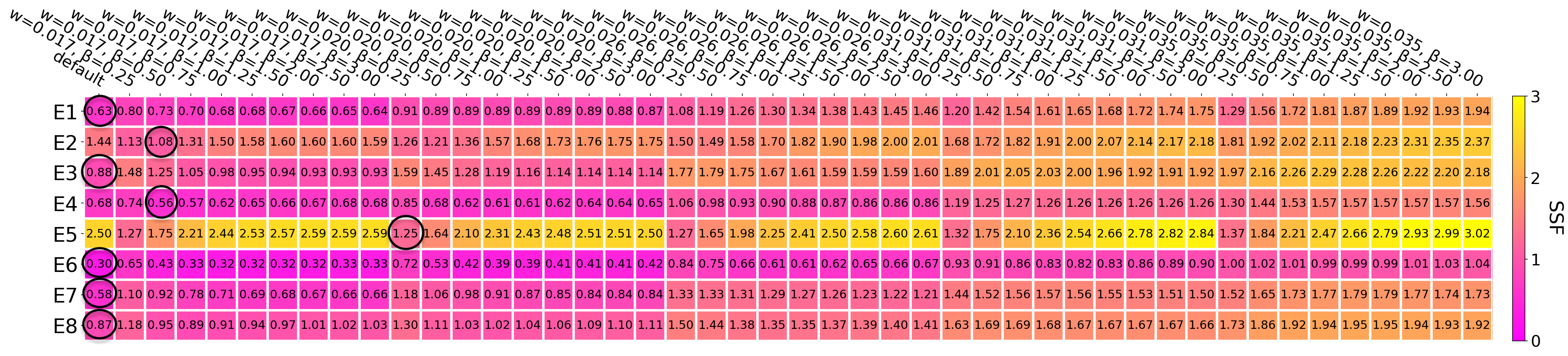}
  \caption{Summarized table of SSF skill scores as calculated by DTW. The x-axis represents all the different WSA configurations while the y-axis shows the individual PSP encounters. The color bar extends from the best possible prediction (SSF = 0; magenta) to the worst one occurred in our dataset (SSF = 3; yellow). For every encounter and WSA configuration, the exact SSF value appears in black at the center of each box. Circled boxes indicate the lowest SSF value that corresponds to the best prediction.}
  \label{Fig:Total_SSF_table}
\end{sidewaysfigure}

For a more global assessment of our results, we present in the table of Fig.~\ref{Fig:Total_SSF_table} the complete array of DTW evaluation results at Earth for all performed runs and encounters. We then ask the opposite question than before, namely, for which pair of $[w, \beta]$ do we get the best results at Earth? Is it any of the pairs we selected from Fig.~\ref{Fig:Comp_of_all_distributions_1stway} based on the calibration close to the Sun, or is it a different one? The first thing we notice is that the default WSA setup performs optimally for 5 out of 8 encounters (from now on the encounters are abbreviated as ``E"), namely, for E1, E3, E6, E7, E8 (i.e., the SSF factor is the lowest among all performed runs, see circled numbers in Fig.~\ref{Fig:Total_SSF_table}). For the three remaining encounters, the best results are given from the combination $[w, \beta] = [0.017 \text{ rad}, 0.50]$ for E2 and E4, and $[w, \beta] = [0.020\text{ rad}, 0.25]$ for E5. We also notice that as we move to regions of higher $w$ (right part of the table in Fig.~\ref{Fig:Total_SSF_table}), the performance of the time series becomes generally worst (higher SSF) for all encounters except E5. This is because, for the same $\beta$, the solar wind becomes increasingly slower. A comparison of time series of different $w$ and same $\beta$ is presented in Fig.~\ref{Fig:Appendix_sameb}, for E1. There, we notice a systematic shift towards lower velocities as $w$ grows. This finding is directly associated to the discussion in section~\ref{subsection:Calibration_w_beta}, in which Fig.~\ref{Fig:same_b_different_w}a, b, c showed that as $w$ grows bigger, the transition from slow to fast solar wind migrates deeper into the CH, causing the intermediate solar wind velocities and transition to fast to occur deeper inside the CH.
% Larger values of $w$ would decrease even more the amount of modeled velocities in the fast regime, creating an increasingly high skeweness towards low speeds. On the other hand, larger values of $\beta$, would create an increasingly bimodal distribution and probably improve our predictions in case the observations had a bimodal shape as well. 
Another important piece of information that Fig.~\ref{Fig:Total_SSF_table} provides, is that the best solar wind predictions for E1, E3, E6, E7 based on the parametric study we performed (ignoring for a moment the default WSA configuration), is 
given by a WSA setup with the lowest $w$ (0.017 rad) and high $\beta$ values (e.g., mostly for $\beta$ = 2.0, 2.5, 3.0). This is because, in these four encounters the arriving HSSs were characterized by very high amplitudes (see, e.g., Fig.~\ref{Fig:TimeseriesEarth_1stway}a, c, f, g). Higher $\beta$ values invoke sharper and bigger changes between slow and fast solar wind which would be necessary to capture more correctly the fast flows during these periods. To visualize the influence of $\beta$ better, we present in Fig.~\ref{Fig:Appendix_samew} how the predicted time series at Earth change for the same $w$ but different $\beta$ values, during E1. We notice that the background level of slow solar wind remains approximately the same, however, the changes between slow and fast solar wind are very different each time. Additional information from the table of Fig.~\ref{Fig:Total_SSF_table} include the fact that the best performed runs for encounters E2 and E4 originate from setups with the lowest $w$ (0.017 rad) and quite low $\beta$ (e.g., $\beta$ = 0.50). This is because the peaks of HSSs are generally overestimated by the default WSA setup during these periods (see, e.g., Fig.~\ref{Fig:TimeseriesEarth_1stway}b, d) so a smaller $\beta$ would be needed to capture better the fast flows. For E8 the default WSA setup provides the best simulations at Earth, while the second best is given by the combination $[w, \beta] = [0.017\text{ rad}, 0.50]$. Lastly, E5 is an exception among all other encounters because of the overall slower solar wind recorded. Only a few HSSs arrive at Earth during this period with amplitudes that do not exceed 500 km/s. The best WSA configuration for E5 is given by a different combination, i.e., $[w, \beta] = [0.020\text{ rad}, 0.25]$. It is interesting to point out that even if we had found an ideal $w, \beta$ pair that would provide the best results for all other encounters, it would not describe well the solar wind velocities for E5 because of their much lower amplitudes. This result shows that WSA equation should be globally modified when coupled with MHD models beyond just choosing a unique pair of $[w, \beta]$ to describe both the slow and fast solar wind.

\subsection{Adding heating to the system}

Since the supposably optimal WSA configurations (magenta circled boxes in Fig.~\ref{Fig:Comp_of_all_distributions_1stway}) led to systematically underestimated solar wind forecasts at Earth (see Fig.~\ref{Fig:TimeseriesEarth_1stway} and Table~\ref{Table:DTW_Earth}), we checked what is the effect of additional heating to the heliospheric part of EUHFORIA which extends from 0.1~au onwards. Could this provide more acceleration to the system and improve the results? The heliospheric part of EUHFORIA is based on a polytropic MHD approach with a polytropic index ($\gamma$) equal to 1.5 \citep[][]{pomoell18}{}{}. To simulate additional heating and thus additional acceleration, we decreased $\gamma$ from 1.5 to 1.1. We also tested the output for $\gamma$ = 5/3 (adiabatic case).

\begin{figure}[ht]
\centering
\gridline{\fig{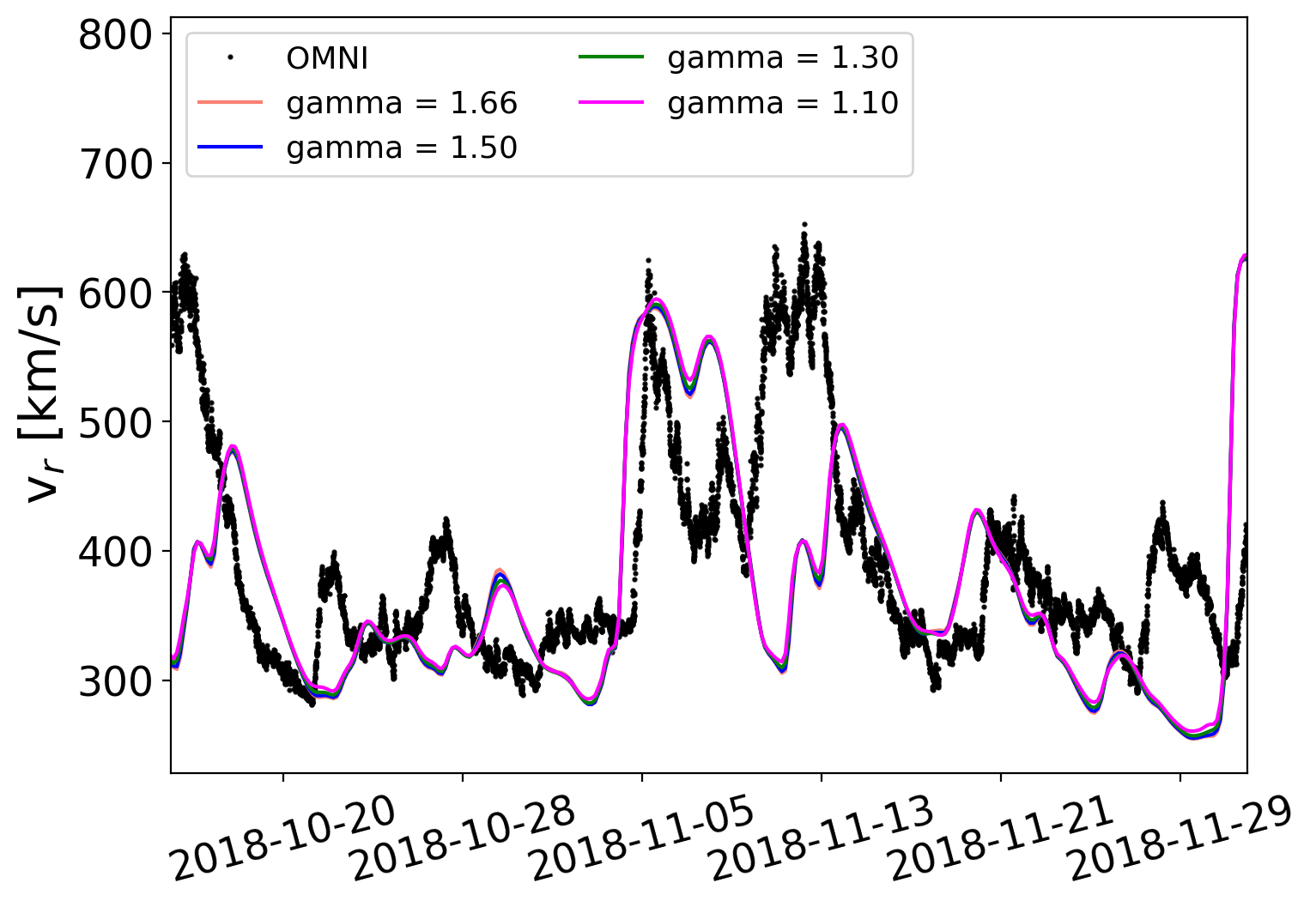}        {0.48\textwidth}{(a)}
            \fig{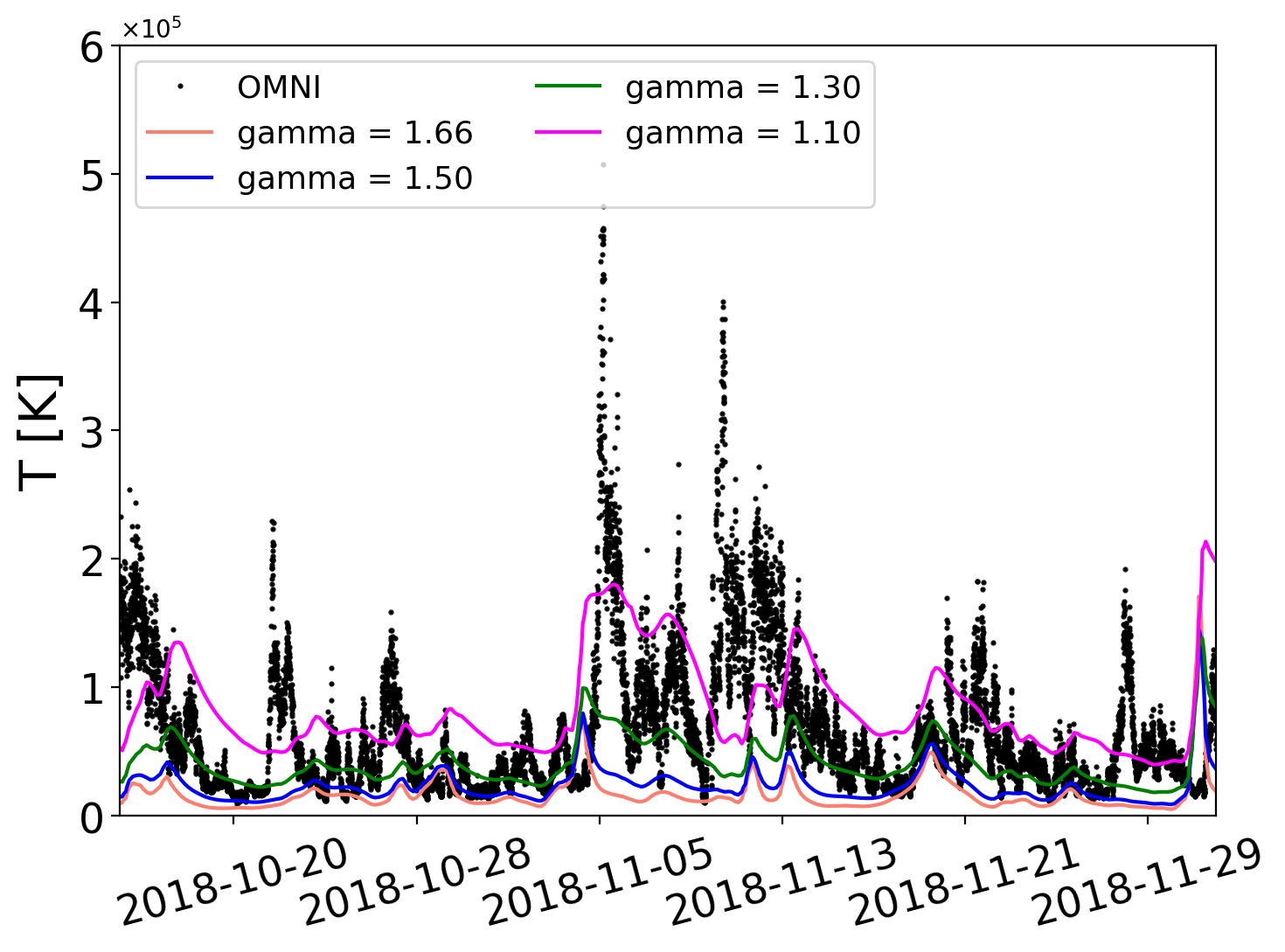}{0.46\textwidth}{(b)}}
            
\caption{Comparison between OMNI data and EUHFORIA simulations. Colorful solid lines correspond to heliospheric simulations made with different $\gamma$. Panel (a): velocity signatures. Panel (b): temperature signatures.}
\label{Fig:Temp_dens_vr}
\end{figure}

\begin{figure}
\centering
\gridline{\fig{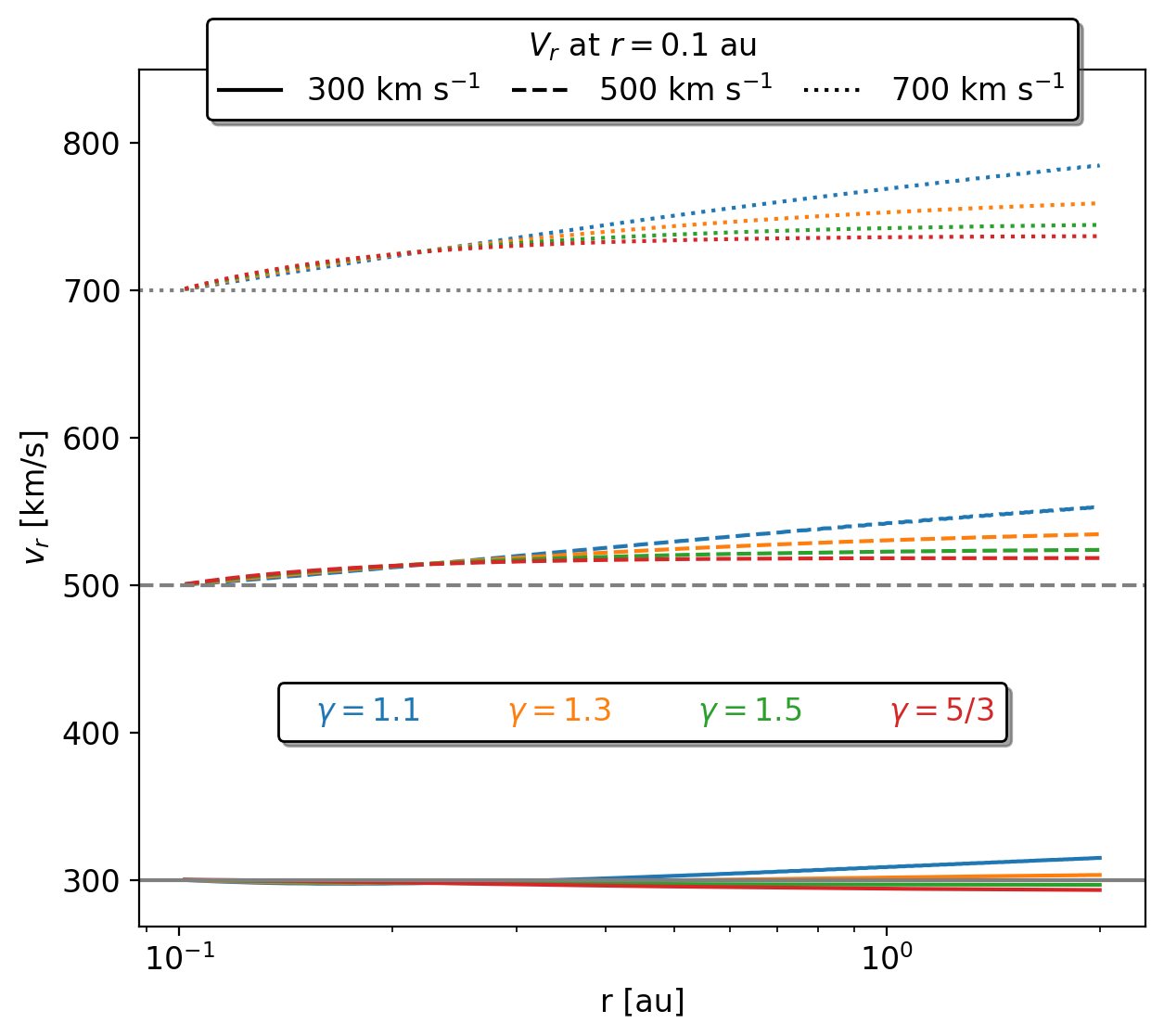}{0.47\textwidth}{(a)}
            \fig{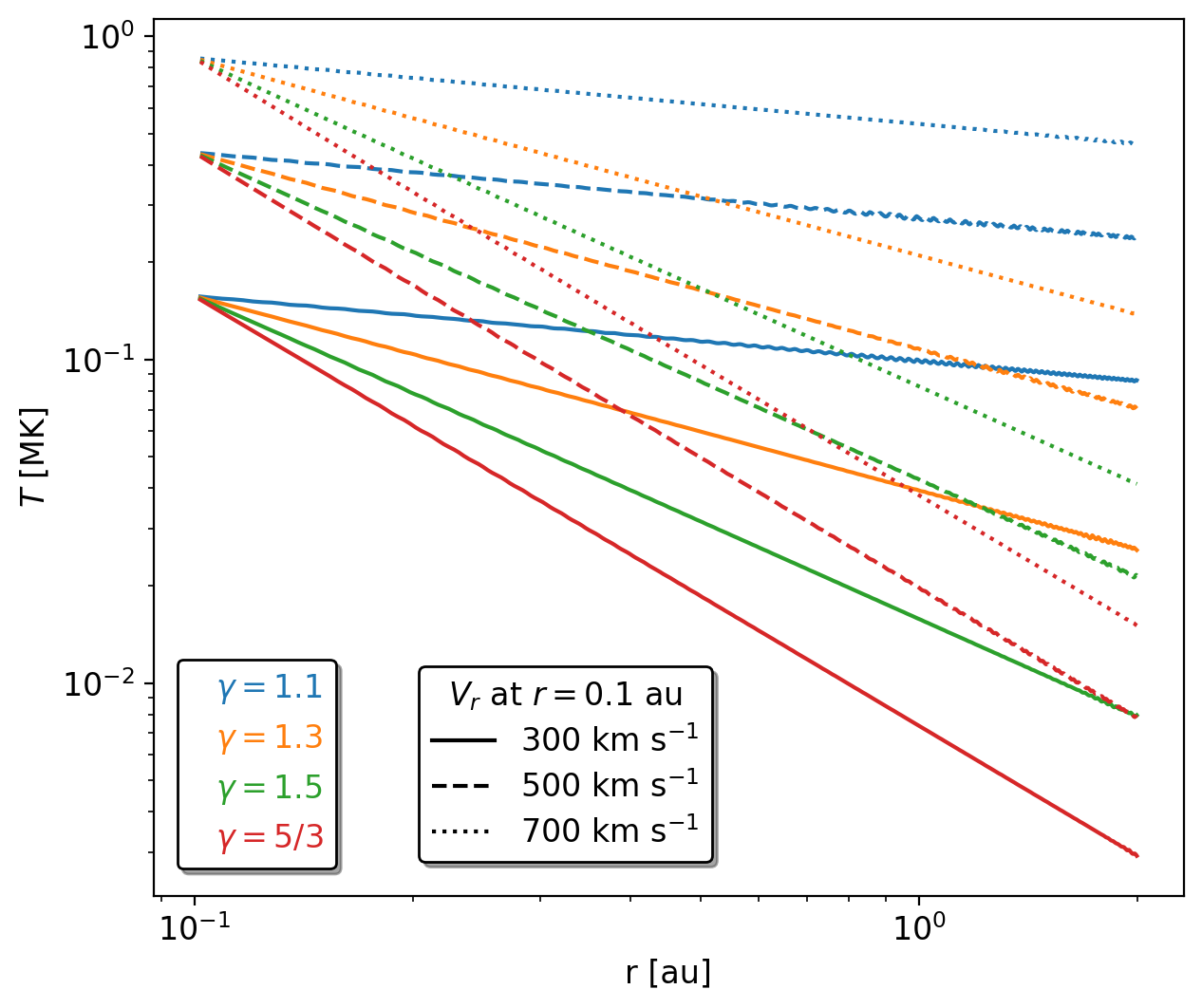}{0.47\textwidth}{(b)}}

\caption{Velocity and temperature variations as a function of heliocentric distance for three synthetic solar wind streams that have initial velocities of 300, 500 and 700 km/s at the inner boundary of EUHFORIA's heliospheric domain (0.1~au).}
\label{Fig:Dakeyo}
\end{figure}

Our results show that the change in the modeling output at Earth based on the different $\gamma$ is noticeable for the temperature and mildly noticeable for velocity (Fig.~\ref{Fig:Temp_dens_vr}a,b). This behaviour is expected, and can be predicted from the analysis of the heliospheric part of Parker's polytropic solar wind solutions. Figure \ref{Fig:Dakeyo}a shows how the velocity varies as a function of heliocentric distance for three synthetic solar wind stream cases with an initial velocity of 300, 500, and 700 km/s at 0.1~au. Different colors correspond to different $\gamma$ values. Figure \ref{Fig:Dakeyo}b presents how the temperature varies as a power law with heliocentric distance, for the same solar wind streams and $\gamma$ values. When we change the $\gamma$ at 0.1~au we practically change the temperature slope. Therefore, at any heliocentric distance the temperature will differ depending on the $\gamma$ employed and the different slope that this change caused. However, this effect is only slightly apparent for velocities.
All curves after $\approx$ 0.25~au have very similar slopes. As we are varying $\gamma$ only from 0.1~au upwards, the effect on the velocity is minimal because the flows are already well supersonic and approaching their asymptotic values. The radial decay of temperature, on the other hand, depends directly on $\gamma$, and differences can be significant at 1\;au even for a modification of $\gamma$ from 0.1\;au upwards only. Therefore, adding more heating to the system by lowering $\gamma$ does not expect to lead to better velocity predictions at Earth. 

\section{Summary and conclusions}
\label{section:Section_6}

In this study we investigated how the calibration of the WSA model, as implemented in EUHFORIA, can be performed based on single-point PSP observations. We employed PSP data between 0.1 -- 0.4~au from the first 8 encounters during the solar minimum period of solar cycle 24 -- 25 so that the influence from solar transients is minimal. We focused on the calibration of four constants in the WSA relationship shown in eq.~\ref{WSA_vr}, namely, those that could be fine-tuned based on observations and past studies according to the methodology of \citet[][]{mcgregor11}{}{}. These are the parameters $V_{0}$ and $V_{1}$, which correspond to the lowest velocity observed at the heliocentric distance of interest, and the maximum velocity range that solar wind can vary above $V_{0}$. Also, the parameter $w$ which indicates the angular distance from the CH boundary at which the transition from slow to fast solar wind takes place, and the parameter $\beta$ which shows how abrupt this transition is.

Calibrating the WSA model at 0.1~au based on PSP observations did not lead to better predictions at Earth. Even though we achieved good resemblance between WSA velocities and PSP observations close to the Sun by finding appropriate values of $V_{0}, V_{1}, w, \beta$, the predictions at Earth usually ended up worse than the ones produced by the default WSA model. Therefore, we approached the problem backwards, namely, by looking for the best set of parameters (specifically $w, \beta$) that give the best predictions at Earth and verifying whether they could be applied to resemble PSP observations close to the Sun. It appears that the Earth-bound set of parameters do not agree with the optimal ones for low heliocentric distances. By evaluating all performed runs and encounters based on the DTW technique, we saw that the optimal predictions at Earth were achieved by a pair of $[w, \beta]$ values other than those displayed in Fig.~\ref{Fig:Comp_of_all_distributions_1stway} and shown in Fig.~\ref{Fig:Distributions}b, c, d. Even though the latter pairs of $[w, \beta]$ resulted in a good resemblance between the WSA model and PSP observations close to the Sun, they led to poor predictions at Earth. Our results with DTW showed that the only way to optimize the predictions at Earth was to calibrate the WSA model with higher velocities closer to the Sun, something that was in contradiction with the very slow velocities recorded by PSP. Depending on the case, higher modeled velocities close to the Sun can be achieved either by reducing the value of $w$ which increases the region covered by the fast solar wind regime, or by raising the value of $\beta$ which eventually creates an increasingly bimodal distribution with one peak in the slow and another peak in the fast velocity regime, respectively (see examples of how the WSA distribution changes for different values of $w, \beta$ in Fig.~\ref{Fig:Examples_of_distr_closeToTheSun_appendix} and Fig.~\ref{Fig:Examples_of_distr_closeToTheSun_appendix2}). The smaller the $w$ value is close to the Sun, the higher levels of background solar wind at Earth will provide, and vice versa (see Fig.~\ref{Fig:Appendix_sameb}). Additionally, the smaller the $\beta$ value is, the smoother and slower the transition between the slow and fast solar wind becomes diminishing the ability of the model to capture fast solar wind flows (see Fig.~\ref{Fig:Appendix_samew}). This is indeed what we saw at Earth. Ignoring the default WSA configuration for a moment, the $w$ value that provided good predictions for all encounters, was the lowest considered value (0.017~rad), except for E5 which was characterized by overall slower velocities and fewer HSSs (for this case, the best $w$ value was 0.020 rad). On the contrary, the $w$ values closer to the Sun that made the WSA modeling distribution to be in good agreement with the PSP one, were 0.026, 0.031 and 0.035 rad (see Fig.~\ref{Fig:Comp_of_all_distributions_1stway} and Fig.~\ref{Fig:Distributions}b, c, d). However, those values under predicted the solar wind velocity at Earth as discussed in section~\ref{section:Section_4} and shown in Fig.~\ref{Fig:TimeseriesEarth_1stway}. Our results show that:

\begin{itemize}

    \item There should be some processes that still accelerate the solar wind from distances close to the Sun (i.e., 0.1 -- 0.4~au) to Earth, that cannot be captured efficiently (or, at all), by EUHFORIA. Such processes could include residual acceleration of the solar wind \citep[see, e.g., ][]{schwenn90, rileyAndLionelo2011, marino2023}{}{}, as well as interactions between neighboring wind streams, co-rotating interaction regions, and CMEs \citep[see, e.g.,][who state that the slowest of all streams detected by Helios disappear with height, presumably due to this kind of interactions]{sanchez2016}{}{}.

    \item Even when the calibration of the WSA equation is performed solely during a period of low solar activity, unique values of $V_{0}, V_{1}, w, \beta$ may still not be enough to properly describe the solar wind at a specific heliocentric distance for different Carrington rotations (see, e.g., the difference between solar wind velocities during E5 and the rest of the encounters). In their work, \citet[][]{kumarAndSrivastava2022}{}{} and \citet[][]{riley2015}{}{} showed, for example, that the optimal WSA parameters can vary greatly from one Carrington rotation to another. Can this problem be overcome? To answer this question, efforts towards reforming the WSA equation in a way that would better reflect the physics and dynamics of slow and fast solar wind in association to their source regions on the Sun, should be undertaken.

    \item The optimal set of WSA parameters (especially $w, \beta$) necessary to correctly approximate PSP observations close to the Sun, is different than those identified to give the best predictions at Earth after running the MHD part of EUHFORIA. Because of this discrepancy, and to understand better how the WSA should be modified when coupled with the current MHD model, an efficient uncertainty quantification study of the free parameters of the whole chain of models (WSA+EUHFORIA), would be useful. In their recent study, \citet[][]{issan2023}{}{} performed a sensitivity analysis of all the free parameters in the PFSS+WSA+HuX chain. They found five most influential parameters that are all related to the WSA model and vary from one Carrington rotation to another. In a future study, it would be interesting to apply their uncertainty quantification framework for the more complex WSA+EUHFORIA pipeline to understand the sensitivity of the different parameters for both modeling parts and get a better understanding of its efficiency.
    
\end{itemize}

\begin{figure}
\centering
\gridline{\fig{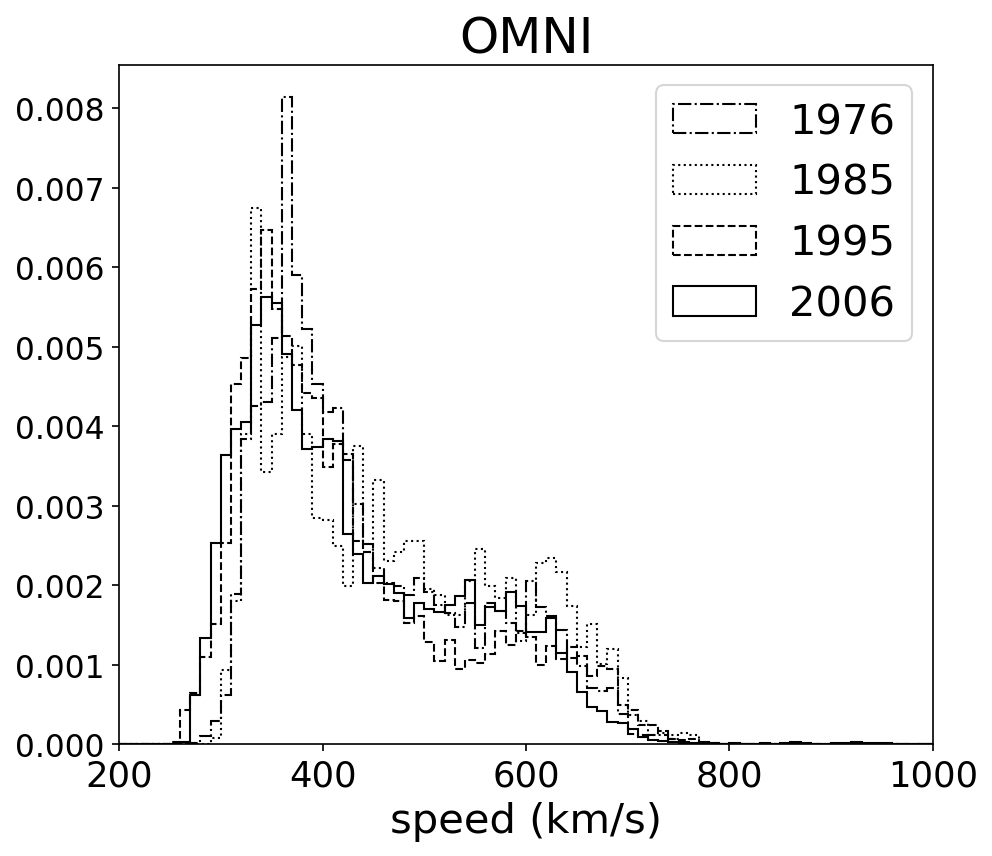}{0.46\textwidth}{(a)}
          \fig{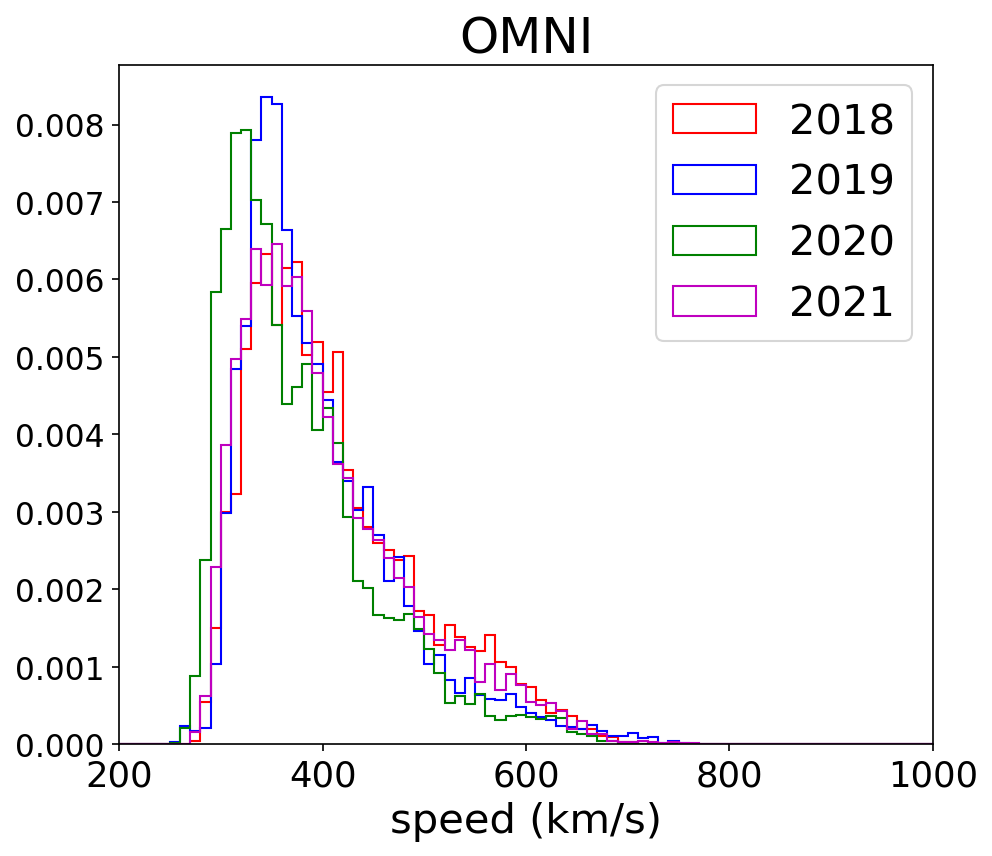}{0.46\textwidth}{(b)}}

\caption{Speed distributions at Earth based on OMNI data for years during different solar minimum periods. Panel (a): yearly speed distributions during past consecutive solar minima. Panel (b): yearly speed distributions during the latest solar minimum.}
\label{Fig:Distributions_at_Earth}
\end{figure}

\begin{figure}
\centering
\gridline{\fig{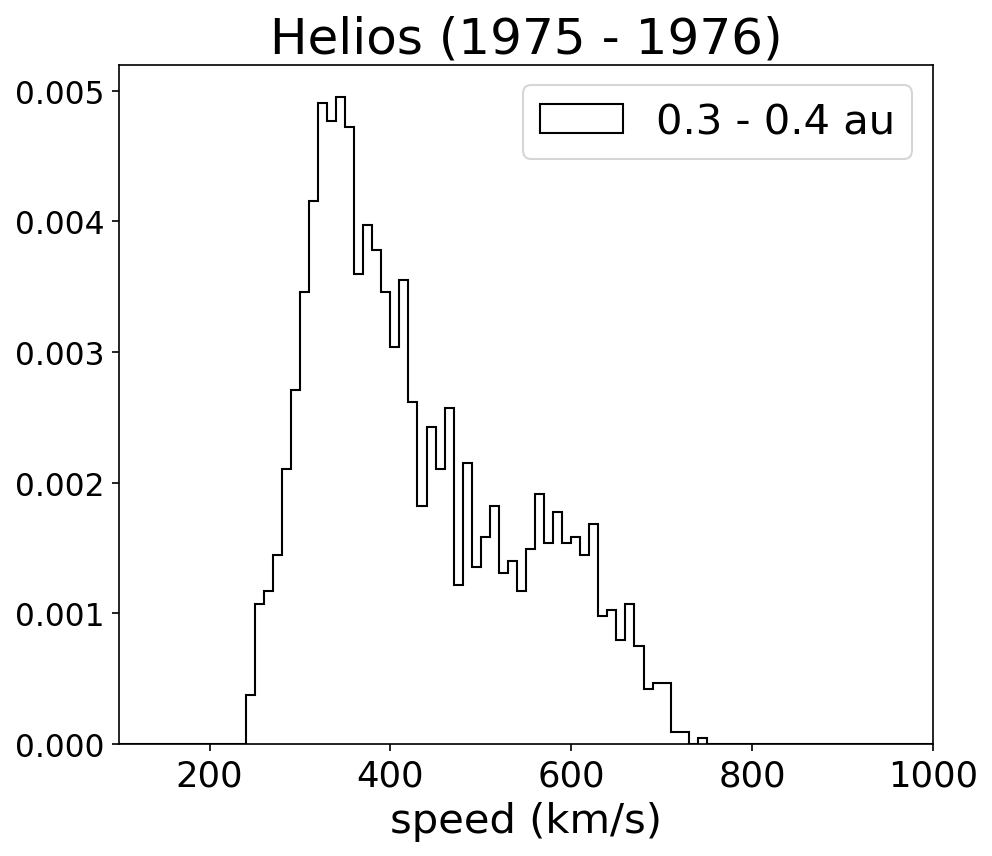}{0.46\textwidth}{(a)}
          \fig{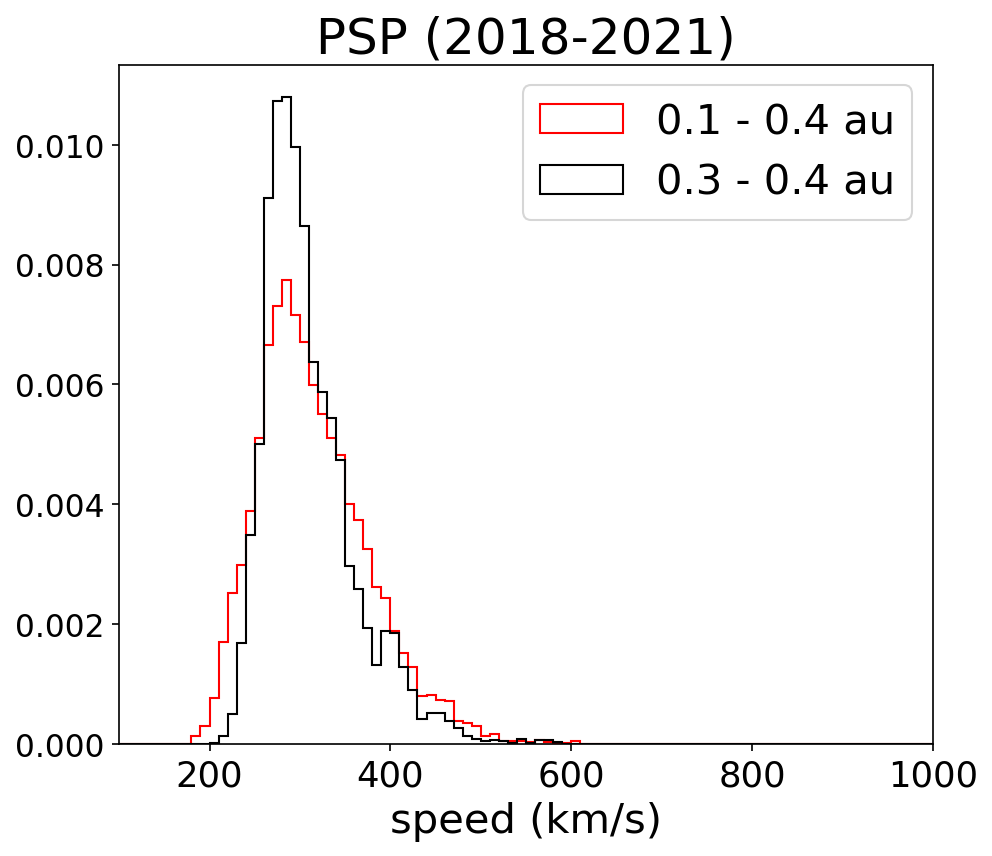}{0.46\textwidth}{(b)}}

`\caption{Speed distributions closer to the Sun. Panel (a): speed distributions as measured by Helios between \mbox{0.3 - 0.4 au} during the solar minimum of 1975-1976. Panel (b): speed distributions as measured by PSP (first eight encounters) between \mbox{0.1 - 0.4 au} during the latest solar minimum. }
\label{Fig:Distributions_close_to_the_Sun}
\end{figure}

Our findings are different from those of \citet[][]{mcgregor11} who, by following a similar approach, improved solar wind forecasts at 1\;au during the years 1996 - 2003 and at Ulysses during 1994 - 1995 by calibrating the WSA model with Helios data. The outcome of our work raises questions as to why this difference between the two studies occurs. First, we were expecting a direct improvement of the solar wind predictions with EUHFORIA because we employed observations much closer to the Sun than \citet[][]{mcgregor11} did, for the calibration of WSA. This means that the fine-tuning of WSA boundary output based on PSP data was more consistent in our case. Second, the magnetograms we used to produce the WSA output at 0.1\;au referred exactly to the same period as the PSP observations employed for the calibration (i.e., 2018 - 2021, during the first eight PSP encounters). In their work, \citet[][]{mcgregor11} exploited Helios measurements from the solar minimum period 1974 - 1976 to calibrate the WSA output more than 20 years later (1996 - 2003). To justify their choice, they showed that a number of consecutive solar minimum periods behaved the same by displaying the similarity of speed distributions at \mbox{1 au}. We show these distributions in Fig.~\ref{Fig:Distributions_at_Earth}a for the years that \citet[][]{mcgregor11} present in their study. We notice that all speed distributions at Earth during 1976, 1985, 1995 and 2006 show a bimodal shape with a high peak at slow speeds and a low peak at high speeds. This, however, does not seem to be the case for the latest solar minimum during which PSP operated (Fig.~\ref{Fig:Distributions_at_Earth}b). In the latter plot, only a single peak in slow speeds is visible for all four years that were taken into account (2018 - 2021). By comparing similar distributions closer to the Sun, the same pattern is revealed when we set side by side observations from past solar minimum periods and the latest one. Namely, a bimodal pattern is apparent in the speed distribution from Helios measurements between 0.3 - 0.4\;au during the years 1975-1976 which were part of solar cycle 21-22 minimum. This is similar to the distribution that \citet[][]{mcgregor11} used to calibrate the WSA model and it is presented in Fig.~\ref{Fig:Distributions_close_to_the_Sun}a. However, as seen in Fig.~\ref{Fig:Distributions_close_to_the_Sun}b, a bimodal distribution is missing from PSP observations. The red distribution in the latter figure represents speeds within the range \mbox{0.1 - 0.4\;au}. The black distribution depicts speeds within \mbox{0.3 - 0.4\;au} which can be directly
compared to Fig.~\ref{Fig:Distributions_close_to_the_Sun}a. We note that the integrated area under each of the histograms in Fig.~\ref{Fig:Distributions_at_Earth}, \ref{Fig:Distributions_close_to_the_Sun} is equal to one.

From the observational point of view, we believe that the scarcity of fast solar wind velocities recorded by PSP close to the Sun during the latest solar minimum period is another reason why this study did not yield improved solar wind predictions after the WSA calibration. Even though the observations and input data used to perform this analysis were more accurate compared to the case of \citet[][]{mcgregor11}{}{}, the PSP velocities themselves- based on which the whole study was relied on- were by far too low to lead to a reasonable calibration of the model. We calibrated WSA according to unexpectedly slow velocities and, consequently, not only we got a general underestimation of the predictions but could also not improve the dynamics between the slow and the fast solar wind throughout the heliosphere since there was no fast solar wind recorded by PSP to account for. In a future study, it would be interesting to explore why this scarcity of fast solar wind velocities happens during the latest solar minimum compared to previous ones. However, PSP observations should not be the only factor to blame for not achieving better predictions at Earth after the calibration. They currently consist of the only in situ ground truth for distances close to the Sun, and this is why we relied on them. Our results showed that the WSA configurations which resembled optimally the PSP observations close to the Sun, were different than the ones needed to provide better predictions at Earth. Therefore, missing physical processes from the heliospheric part of EUHFORIA can be one factor contributing to this discrepancy. Another factor can be the fact that the currently employed WSA relationship coupled to a heliospheric MHD domain may need to be globally reformed beyond of just updating the four $V_{0}, V_{1}, w, \beta$ constant factors in order to capture better the physics taking place. Other factors that should be mentioned but we believe they do not drastically influence the results, is the date, type of magnetograms used, the observatory that magnetograms came from, as well as the disadvantages of single-point observations compared to multi-point ones. In a future study, we aim to use multi-point observations during different periods of solar activity to calibrate the boundary conditions from WSA in an effort to decrease the uncertainties coming from single-point observations but also extend this analysis for both solar minimum and maximum intervals.

The last part of this study was devoted to better understanding the influence of the acceleration from the heliospheric MHD part of EUHFORIA on the solar wind boundary velocities provided by the WSA model at 0.1~au. The goal was to verify if we can correct for the systematic underestimation of the solar wind seen in the predictions at \mbox{1 au} by decreasing the polytropic index ($\gamma$) of the MHD part of EUHFORIA. With that, we aimed to provide more heating and thus, more acceleration to the system. However, this was not eventually the case as a change in $\gamma$ at 0.1\;au mainly affects the temperatures of the system, but not so much the velocities.

\begin{acknowledgments}
E.S. research was supported by an appointment to the NASA Postdoctoral Program at the NASA Goddard Space Flight Center, administered by Oak Ridge Associated Universities under contract with NASA. C.N.A is supported by the NASA competed Heliophysics Internal Scientist Funding Model (ISFM). 
J.M. acknowledges funding by the BRAIN-be project SWiM (Solar Wind Modelling
with EUHFORIA for the new heliospheric missions) and
funding from the FED-tWIN project PERIHELION. EUHFORIA was created as a joint effort between KU Leuven and the University of Helsinki and is being developed further by the the project EUHFORIA 2.0, a European Union’s Horizon 2020 research and innovation programs under grant agreement No 870405. The ROB team thanks the Belgian Federal Science Policy Office (BELSPO) for the provision of financial support in the framework of the PRODEX Programme of the European Space Agency (ESA) under contract number 4000134474.
N.W.\ acknowledges support from the Research Foundation - Flanders (FWO-Vlaanderen, fellowship no.\ 1184319N).
\end{acknowledgments}

\bibliography{bibliography}{}

\begin{thebibliography}{}
\expandafter\ifx\csname natexlab\endcsname\relax\def\natexlab#1{#1}\fi
\providecommand{\url}[1]{\href{#1}{#1}}
\providecommand{\dodoi}[1]{doi:~\href{http://doi.org/#1}{\nolinkurl{#1}}}
\providecommand{\doeprint}[1]{\href{http://ascl.net/#1}{\nolinkurl{http://ascl.net/#1}}}
\providecommand{\doarXiv}[1]{\href{https://arxiv.org/abs/#1}{\nolinkurl{https://arxiv.org/abs/#1}}}

\bibitem[{{Altschuler} \& {Newkirk}(1969)}]{altschuler69}
{Altschuler}, M.~D., \& {Newkirk}, G. 1969, \solphys, 9, 131,
  \dodoi{10.1007/BF00145734}

\bibitem[{{Arge} {et~al.}(2004){Arge}, {Luhmann}, {Odstrcil}, {Schrijver}, \&
  {Li}}]{Arge04}
{Arge}, C.~N., {Luhmann}, J.~G., {Odstrcil}, D., {Schrijver}, C.~J., \& {Li},
  Y. 2004, Journal of Atmospheric and Solar-Terrestrial Physics, 66, 1295,
  \dodoi{10.1016/j.jastp.2004.03.018}

\bibitem[{{Arge} {et~al.}(2003){Arge}, {Odstrcil}, {Pizzo}, \&
  {Mayer}}]{arge03}
{Arge}, C.~N., {Odstrcil}, D., {Pizzo}, V.~J., \& {Mayer}, L.~R. 2003, in Am
  Inst Phys CS, Vol. 679, Solar Wind Ten, ed. M.~{Velli}, R.~{Bruno},
  F.~{Malara}, \& B.~{Bucci}, 190--193, \dodoi{10.1063/1.1618574}

\bibitem[{Case {et~al.}(2020)Case, Kasper, Stevens, Korreck, Paulson, Daigneau,
  Caldwell, Freeman, Henry, Klingensmith, {et~al.}}]{case2020}
Case, A.~W., Kasper, J.~C., Stevens, M.~L., {et~al.} 2020, The Astrophysical
  Journal Supplement Series, 246, 43

\bibitem[{Fox {et~al.}(2016)Fox, Velli, Bale, Decker, Driesman, Howard, Kasper,
  Kinnison, Kusterer, Lario, {et~al.}}]{fox2016}
Fox, N., Velli, M., Bale, S., {et~al.} 2016, Space Science Reviews, 204, 7

\bibitem[{Green \& Baker(2015)}]{green2015coronal}
Green, L., \& Baker, D. 2015, Weather, 70, 31

\bibitem[{Hapgood(2011)}]{hapgood2011towards}
Hapgood, M. 2011, Advances in Space Research, 47, 2059

\bibitem[{{Hinterreiter} {et~al.}(2019){Hinterreiter}, {Magdalenic}, {Temmer},
  {Verbeke}, {Jebaraj}, {Samara}, {Asvestari}, {Poedts}, {Pomoell}, {Kilpua},
  {Rodriguez}, {Scolini}, \& {Isavnin}}]{Hinterreiter19}
{Hinterreiter}, J., {Magdalenic}, J., {Temmer}, M., {et~al.} 2019, \solphys,
  294, 170, \dodoi{10.1007/s11207-019-1558-8}

\bibitem[{Issan {et~al.}(2023)Issan, Riley, Camporeale, \& Kramer}]{issan2023}
Issan, O., Riley, P., Camporeale, E., \& Kramer, B. 2023, arXiv preprint
  arXiv:2305.08009

\bibitem[{Kim {et~al.}(2014)Kim, Pogorelov, Borovikov, Jackson, Yu, \&
  Tokumaru}]{kim2014}
Kim, T., Pogorelov, N., Borovikov, S., {et~al.} 2014, Journal of Geophysical
  Research: Space Physics, 119, 7981

\bibitem[{Kumar \& Srivastava(2022)}]{kumarAndSrivastava2022}
Kumar, S., \& Srivastava, N. 2022, Space Weather, 20, e2022SW003069

\bibitem[{{MacNeice} {et~al.}(2018){MacNeice}, {Jian}, {Antiochos}, {Arge},
  {Bussy-Virat}, {DeRosa}, {Jackson}, {Linker}, {Mikic}, {Owens}, {Ridley},
  {Riley}, {Savani}, \& {Sokolov}}]{MacNeice2018}
{MacNeice}, P., {Jian}, L.~K., {Antiochos}, S.~K., {et~al.} 2018, Space
  Weather, 16, 1644, \dodoi{10.1029/2018SW002040}

\bibitem[{Marino \& Sorriso-Valvo(2023)}]{marino2023}
Marino, R., \& Sorriso-Valvo, L. 2023, Physics Reports, 1006, 1

\bibitem[{{McGregor} {et~al.}(2008){McGregor}, {Hughes}, {Arge}, \&
  {Owens}}]{mcgregor08}
{McGregor}, S.~L., {Hughes}, W.~J., {Arge}, C.~N., \& {Owens}, M.~J. 2008, J
  Geophys Res (Space Phys), 113, A08112, \dodoi{10.1029/2007JA012330}

\bibitem[{{McGregor} {et~al.}(2011){McGregor}, {Hughes}, {Arge}, {Owens}, \&
  {Odstrcil}}]{mcgregor11}
{McGregor}, S.~L., {Hughes}, W.~J., {Arge}, C.~N., {Owens}, M.~J., \&
  {Odstrcil}, D. 2011, J Geophys Res (Space Phys), 116, A03101,
  \dodoi{10.1029/2010JA015881}

\bibitem[{M{\"o}stl {et~al.}(2017)M{\"o}stl, Isavnin, Boakes, Kilpua, Davies,
  Harrison, Barnes, Krupar, Eastwood, Good, {et~al.}}]{mostl2017}
M{\"o}stl, C., Isavnin, A., Boakes, P., {et~al.} 2017, Space Weather, 15, 955

\bibitem[{M{\"o}stl {et~al.}(2020)M{\"o}stl, Weiss, Bailey, Reiss, Amerstorfer,
  Hinterreiter, Bauer, McIntosh, Lugaz, \& Stansby}]{mostl2020}
M{\"o}stl, C., Weiss, A.~J., Bailey, R.~L., {et~al.} 2020, The Astrophysical
  Journal, 903, 92

\bibitem[{{Owens} {et~al.}(2008){Owens}, {Spence}, {McGregor}, {Hughes},
  {Quinn}, {Arge}, {Riley}, {Linker}, \& {Odstrcil}}]{owens08}
{Owens}, M.~J., {Spence}, H.~E., {McGregor}, S., {et~al.} 2008, Space Weather,
  6, S08001, \dodoi{10.1029/2007SW000380}

\bibitem[{Parker(2020)}]{PSPmission1}
Parker, E.~N. 2020, Nature Astronomy, 4, 19

\bibitem[{{Pomoell} \& {Poedts}(2018)}]{pomoell18}
{Pomoell}, J., \& {Poedts}, S. 2018, J Space Weather Space Climate, 8, A35,
  \dodoi{10.1051/swsc/2018020}

\bibitem[{Riley {et~al.}(2015)Riley, Linker, \& Arge}]{riley2015}
Riley, P., Linker, J.~A., \& Arge, C.~N. 2015, Space Weather, 13, 154

\bibitem[{Riley \& Lionello(2011)}]{rileyAndLionelo2011}
Riley, P., \& Lionello, R. 2011, Solar Physics, 270, 575

\bibitem[{Riley {et~al.}(2003)Riley, Mikic, Linker, \& Zurbuchen}]{riley2003}
Riley, P., Mikic, Z., Linker, J., \& Zurbuchen, T.~H. 2003in , American
  Institute of Physics, 79--82

\bibitem[{Samara {et~al.}(2022)Samara, Laperre, Kieokaew, Temmer, Verbeke,
  Rodriguez, Magdaleni{\'c}, \& Poedts}]{Samara2022DTW}
Samara, E., Laperre, B., Kieokaew, R., {et~al.} 2022, The Astrophysical
  Journal, 927, 187

\bibitem[{Samara {et~al.}(2021)Samara, Pinto, Magdaleni{\'c}, Wijsen,
  Jer{\v{c}}i{\'c}, Scolini, Jebaraj, Rodriguez, \& Poedts}]{Samara2021MVP}
Samara, E., Pinto, R.~F., Magdaleni{\'c}, J., {et~al.} 2021, Astronomy \&
  Astrophysics, 648, A35

\bibitem[{Sanchez-Diaz {et~al.}(2016)Sanchez-Diaz, Rouillard, Lavraud, Segura,
  Tao, Pinto, Sheeley~Jr, \& Plotnikov}]{sanchez2016}
Sanchez-Diaz, E., Rouillard, A.~P., Lavraud, B., {et~al.} 2016, Journal of
  Geophysical Research: Space Physics, 121, 2830

\bibitem[{Schatten(1971)}]{schatten1971current_npp}
Schatten, K.~H. 1971, Current sheet magnetic model for the solar corona, Tech.
  rep.

\bibitem[{{Schatten} {et~al.}(1969){Schatten}, {Wilcox}, \&
  {Ness}}]{schatten69}
{Schatten}, K.~H., {Wilcox}, J.~M., \& {Ness}, N.~F. 1969, \solphys, 6, 442,
  \dodoi{10.1007/BF00146478}

\bibitem[{{Schrijver} {et~al.}(2015){Schrijver}, {Kauristie}, {Aylward},
  {Denardini}, {Gibson}, {Glover}, {Gopalswamy}, {Grande}, {Hapgood},
  {Heynderickx}, {Jakowski}, {Kalegaev}, {Lapenta}, {Linker}, {Liu},
  {Mandrini}, {Mann}, {Nagatsuma}, {Nandy}, {Obara}, {Paul O'Brien}, {Onsager},
  {Opgenoorth}, {Terkildsen}, {Valladares}, \& {Vilmer}}]{schrijver15}
{Schrijver}, C.~J., {Kauristie}, K., {Aylward}, A.~D., {et~al.} 2015, Advances
  in Space Research, 55, 2745, \dodoi{10.1016/j.asr.2015.03.023}

\bibitem[{Schwadron {et~al.}(2005)Schwadron, McComas, Elliott, Gloeckler,
  Geiss, \& Von~Steiger}]{schwadron2005}
Schwadron, N., McComas, D., Elliott, H., {et~al.} 2005, Journal of Geophysical
  Research: Space Physics, 110

\bibitem[{{Schwenn}(1990)}]{schwenn90}
{Schwenn}, R. 1990, Physics of the Inner Heliosphere: 1. Large-Scale Phenomena,
  99

\bibitem[{Szabo(2018)}]{PSPmission2}
Szabo, A. 2018, Nature astronomy, 2, 829

\bibitem[{{van der Holst} {et~al.}(2010){van der Holst}, {Manchester},
  {Frazin}, {V{\'a}squez}, {T{\'o}th}, \& {Gombosi}}]{vanderHolst10}
{van der Holst}, B., {Manchester}, W.~B., I., {Frazin}, R.~A., {et~al.} 2010,
  \apj, 725, 1373, \dodoi{10.1088/0004-637X/725/1/1373}

\bibitem[{Wang \& Sheeley~Jr(1997)}]{wang1997}
Wang, Y.-M., \& Sheeley~Jr, N. 1997, Geophysical research letters, 24, 3141

\bibitem[{Whittlesey {et~al.}(2020)Whittlesey, Larson, Kasper, Halekas,
  Abatcha, Abiad, Berthomier, Case, Chen, Curtis, {et~al.}}]{whittlesey2020}
Whittlesey, P.~L., Larson, D.~E., Kasper, J.~C., {et~al.} 2020, The
  Astrophysical Journal Supplement Series, 246, 74

\bibitem[{{Wiegelmann} {et~al.}(2017){Wiegelmann}, {Petrie}, \&
  {Riley}}]{Wiegelmann2017}
{Wiegelmann}, T., {Petrie}, G. J.~D., \& {Riley}, P. 2017, \ssr, 210, 249,
  \dodoi{10.1007/s11214-015-0178-3}

\end{thebibliography}
\bibliographystyle{aasjournal}

%% This command is needed to show the entire author+affiliation list when the collaboration and author truncation commands are used.  It has to go at the end of the manuscript.


\clearpage
\appendix
\section{Influence of different $w$ and $\beta$ at Earth and 0.1~au}

\begin{figure}[h]
\centering
\gridline{\fig{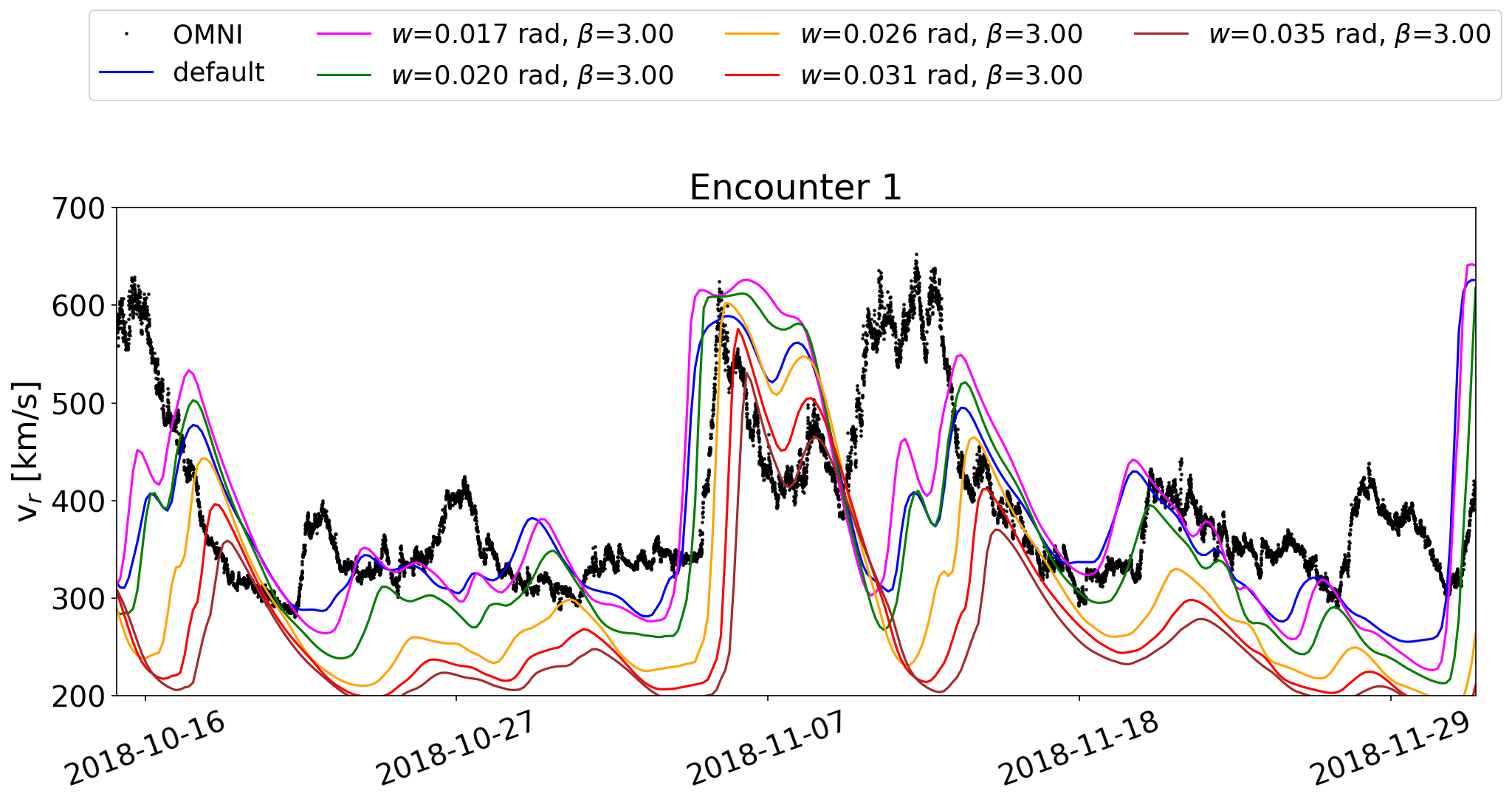}{0.7\textwidth}{}}
\caption{Comparison of predicted solar wind velocity time series at Earth for the same $\beta$ and different values of $w$ during PSP encounter 1.}
\label{Fig:Appendix_sameb}
\end{figure}

\begin{figure}[h]
\centering
\gridline{\fig{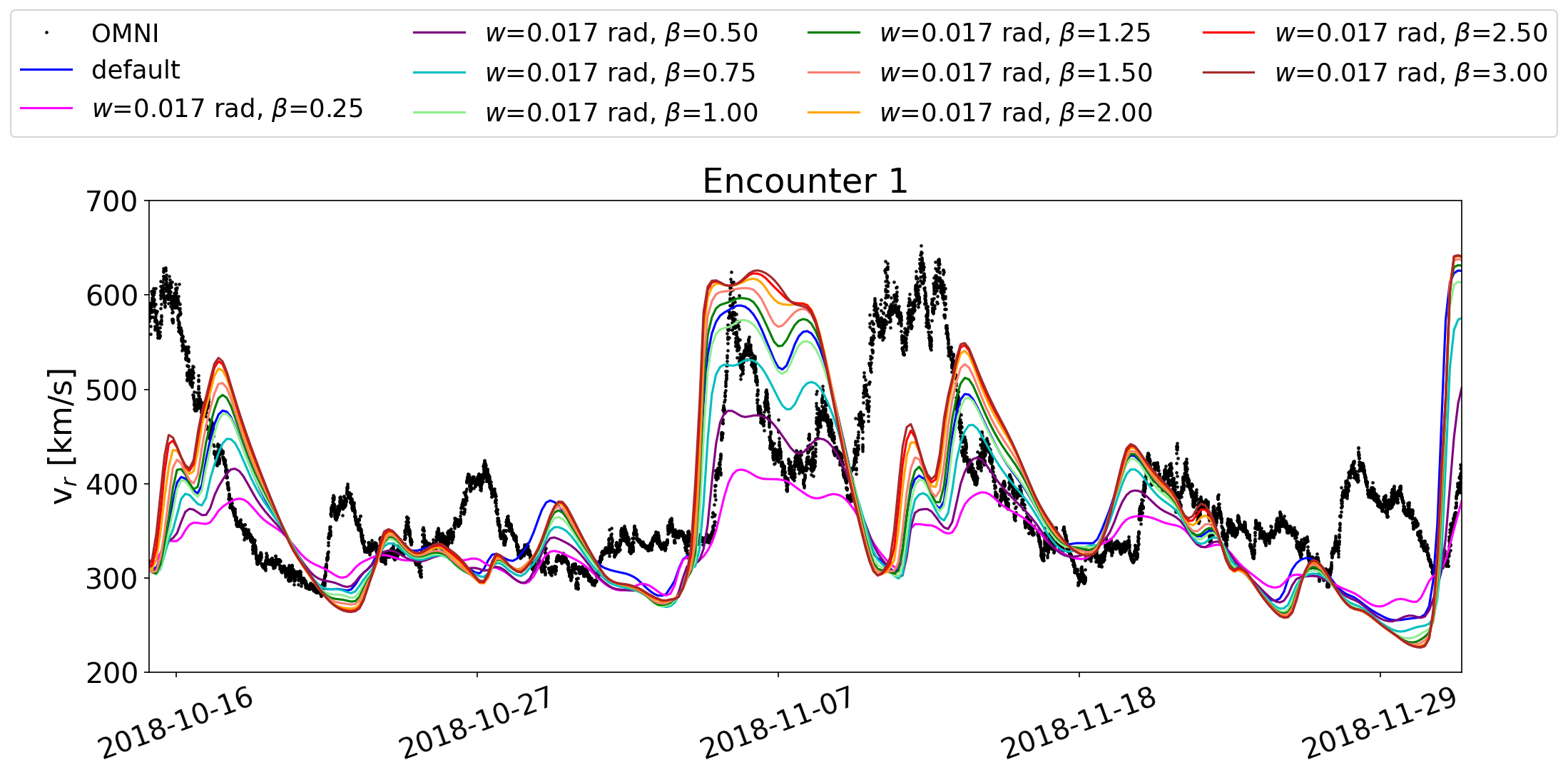}{0.7\textwidth}{}}
\caption{Same as \ref{Fig:Appendix_sameb} but for the same $w$ and different values of $\beta$.}
\label{Fig:Appendix_samew}
\end{figure}

Figures~\ref{Fig:Appendix_sameb} and \ref{Fig:Appendix_samew} show how the predictions at Earth during the period of PSP encounter 1 vary for the same $w$ and different values of $\beta$, as well as for the same $\beta$ and different values of $w$, respectively. In Fig.~\ref{Fig:Appendix_sameb} we notice a systematic shift towards lower velocities as $w$ grows. This is because, according to Fig.~\ref{Fig:same_b_different_w}a, b, c, as $w$ grows bigger, the transition from slow to fast solar wind mitigates deeper into the CH, causing the intermediate solar wind velocities to become increasingly slower. In Fig.~\ref{Fig:Appendix_samew} we notice that higher $\beta$ values invoke sharper and bigger changes between slow and fast solar wind. Figure~\ref{Fig:Examples_of_distr_closeToTheSun_appendix} and ~\ref{Fig:Examples_of_distr_closeToTheSun_appendix2} show how the velocity distributions close to the Sun vary, for different values of $w$ and $\beta$, compared to the same PSP velocity distribution between 0.1 -- 0.4~au.

 \begin{figure}
 \centering
 \gridline{\fig{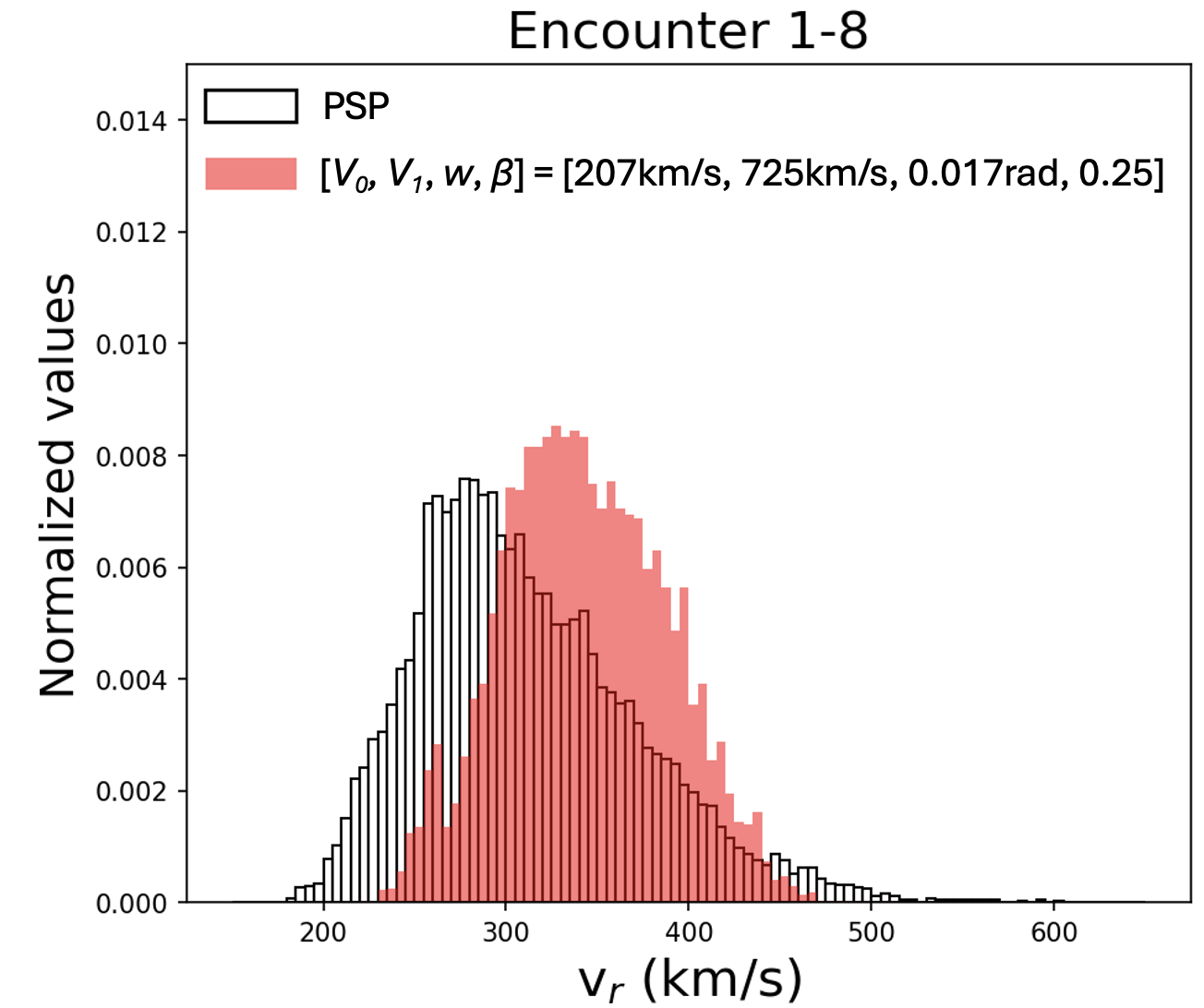}{0.25\textwidth}{(a)}
             \fig{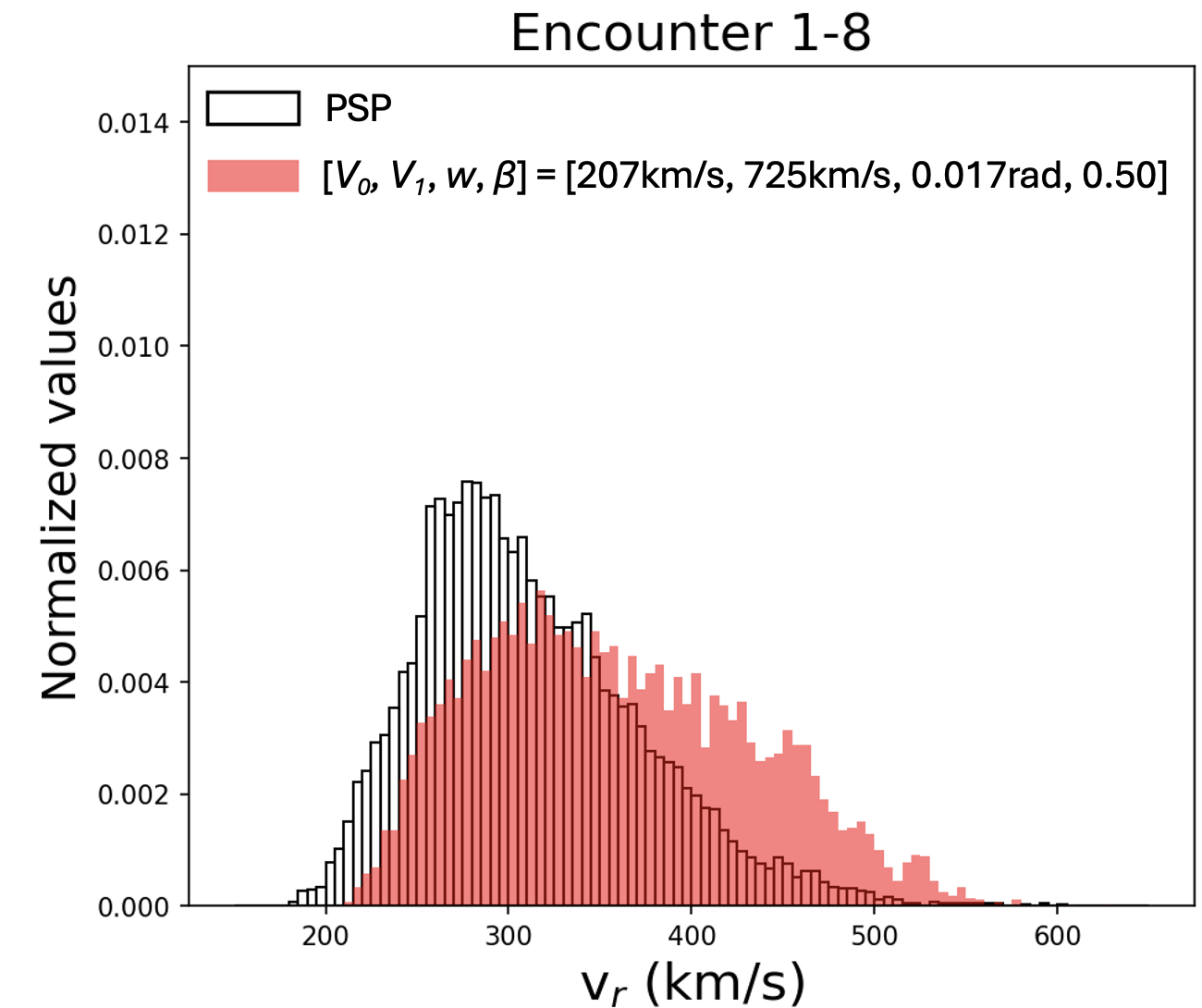}{0.25\textwidth}{(b)}
             \fig{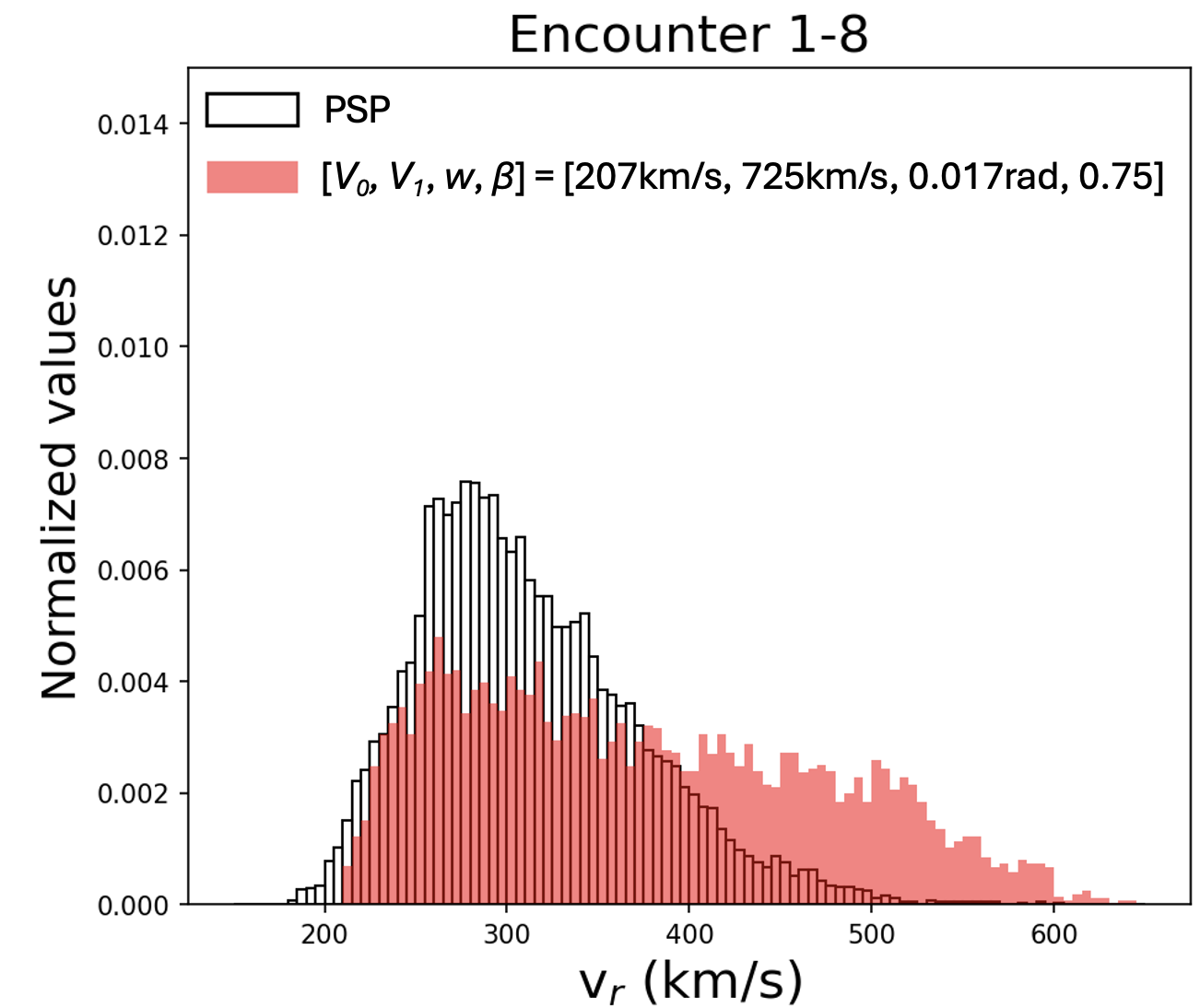}{0.25\textwidth}{(c)}
             \fig{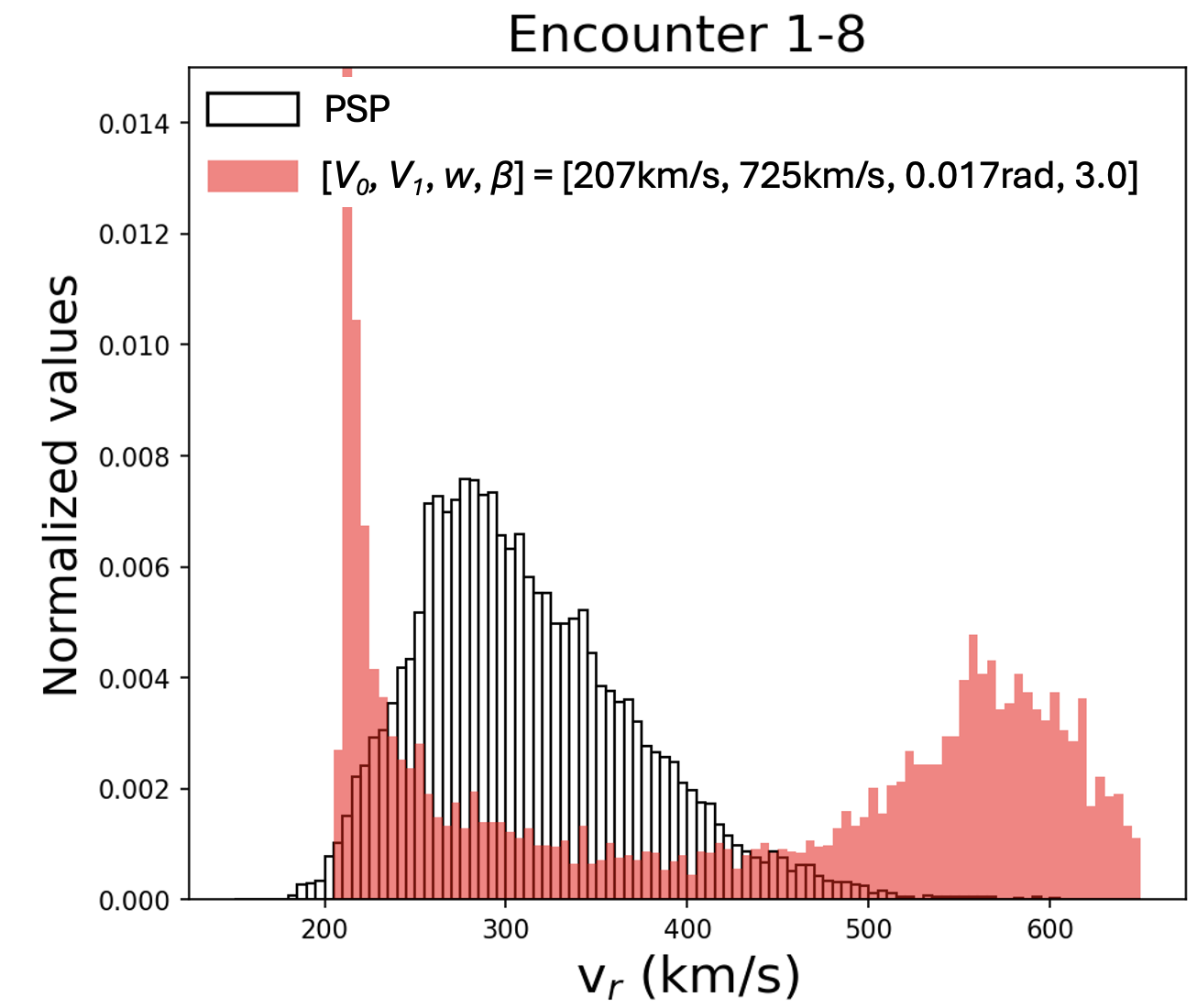}{0.25\textwidth}{(c)}}

 \gridline{\fig{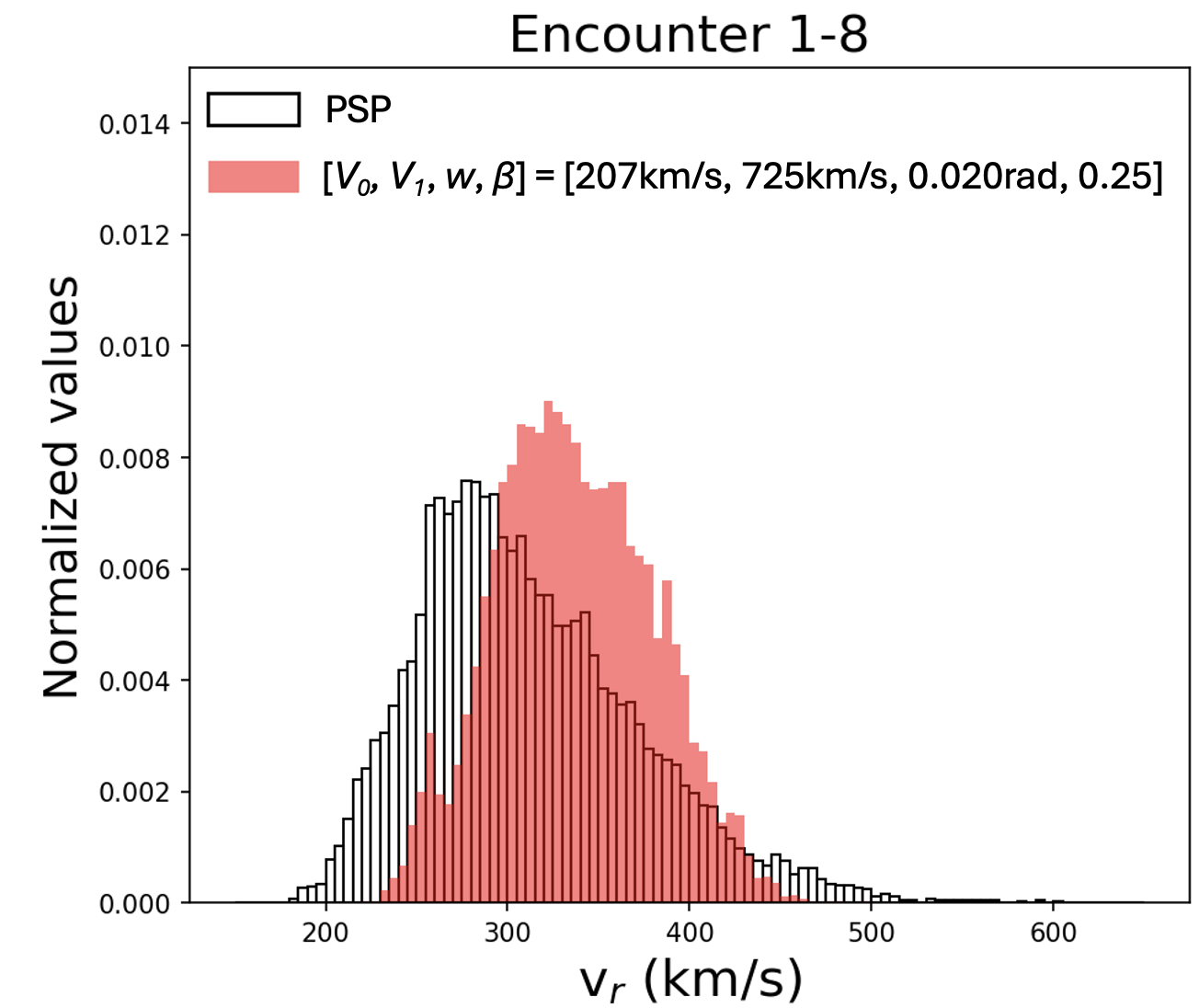}{0.25\textwidth}{(a)}
             \fig{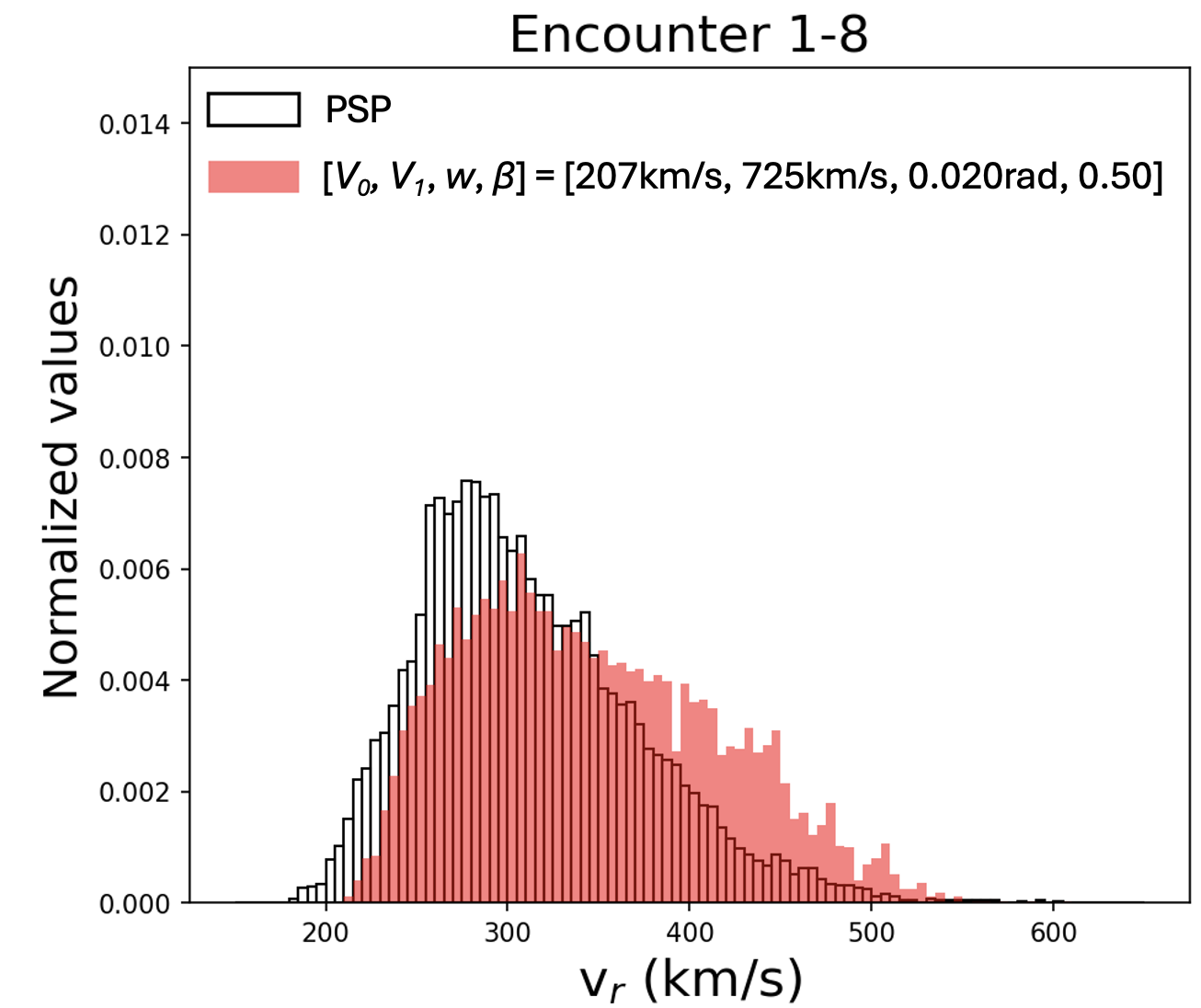}{0.25\textwidth}{(b)}
             \fig{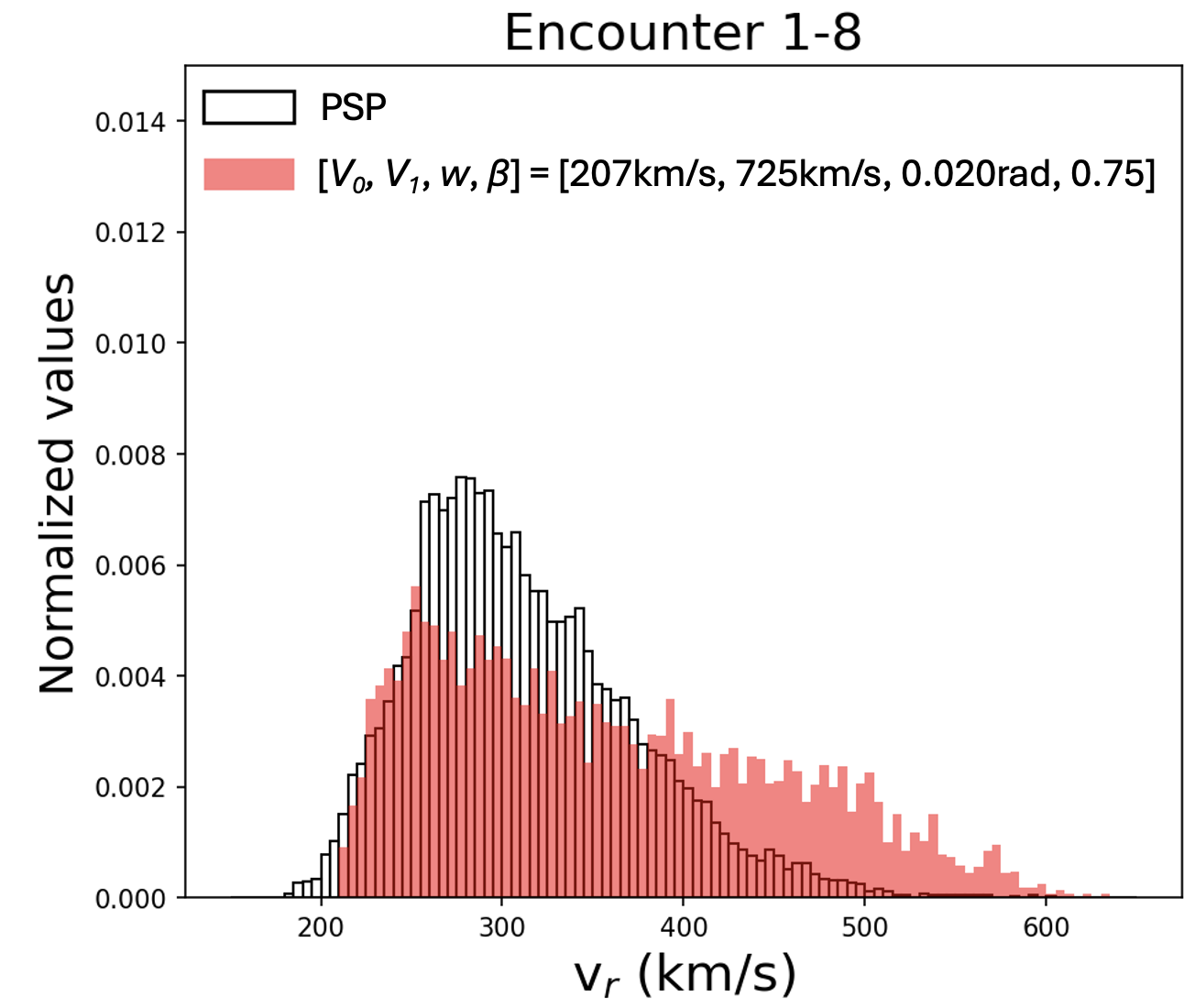}{0.25\textwidth}{(c)}
             \fig{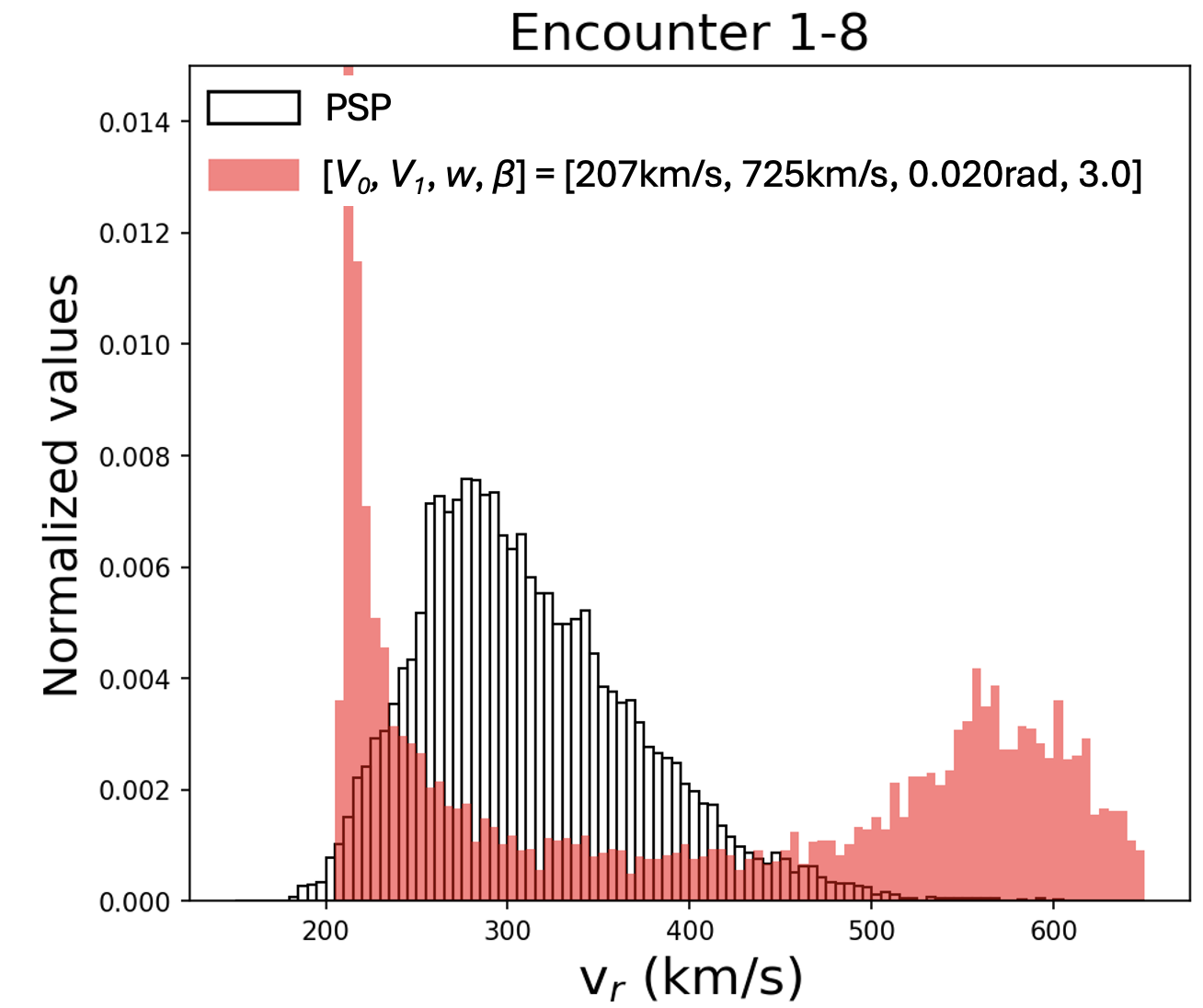}{0.25\textwidth}{(c)}}

 \gridline{\fig{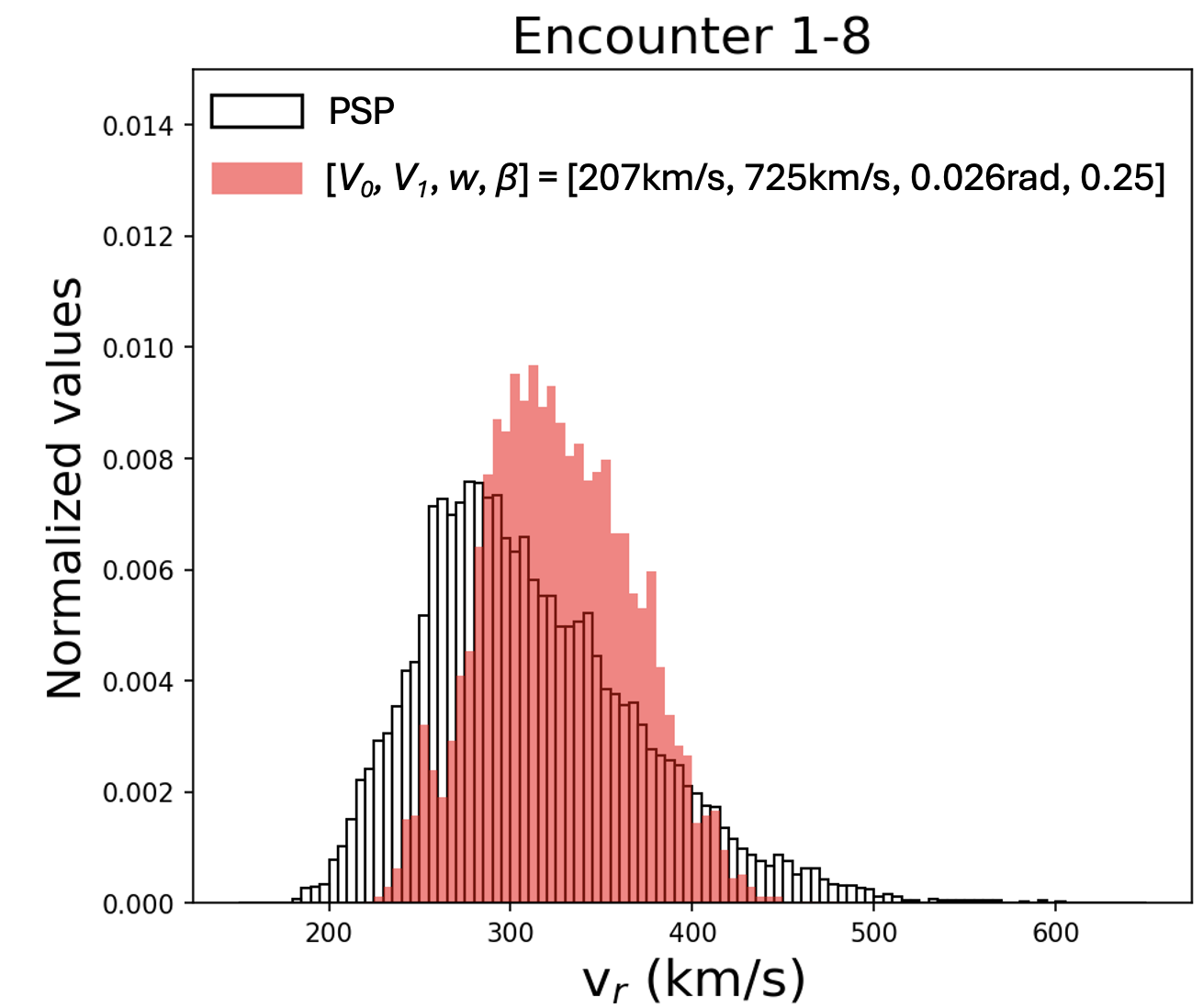}{0.25\textwidth}{(a)}
             \fig{Figures/E1toE8_minvr_207_w0_026_beta0_50_vmax725_noadhoc_editted.png}{0.25\textwidth}{(b)}
             \fig{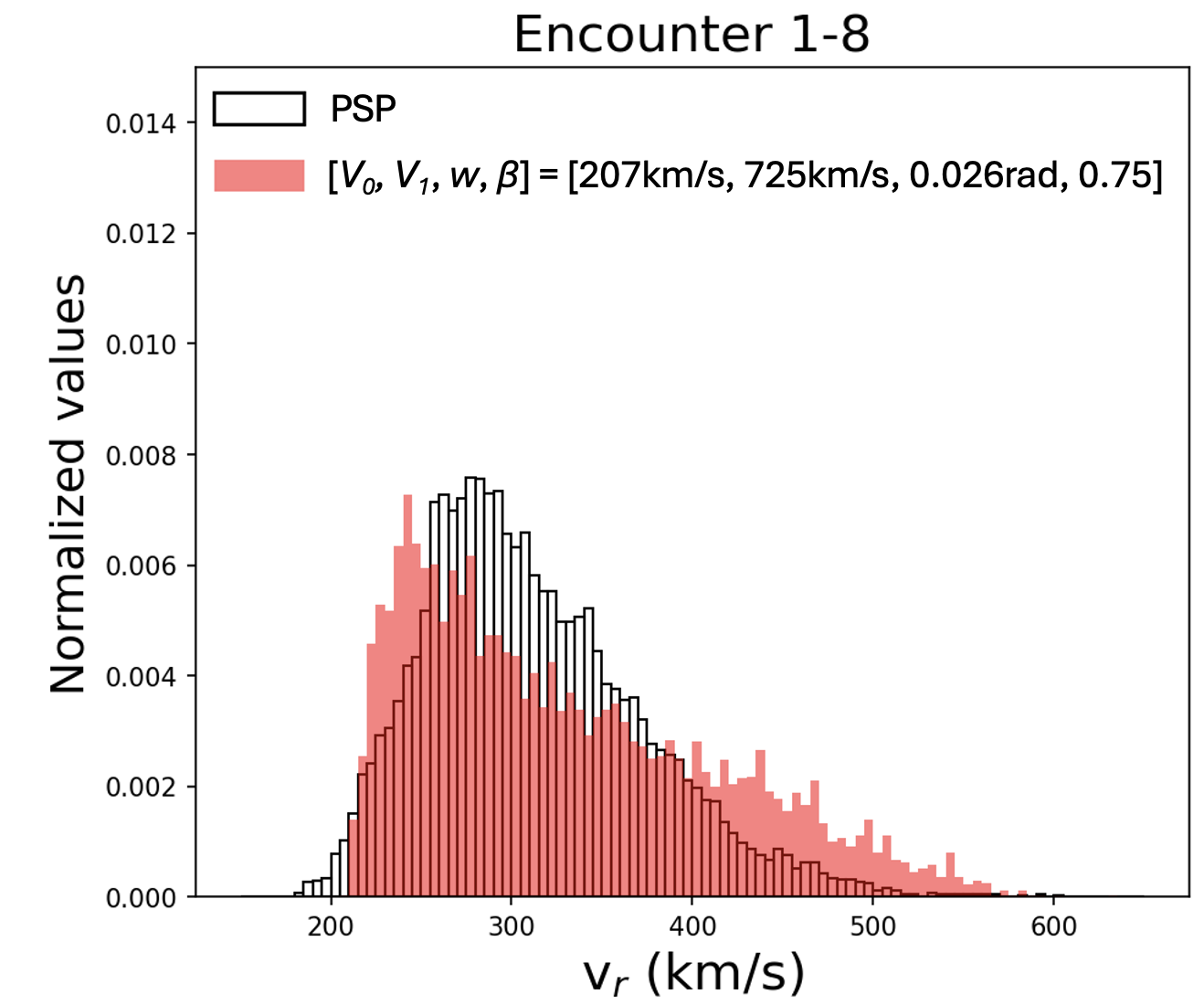}{0.25\textwidth}{(c)}
             \fig{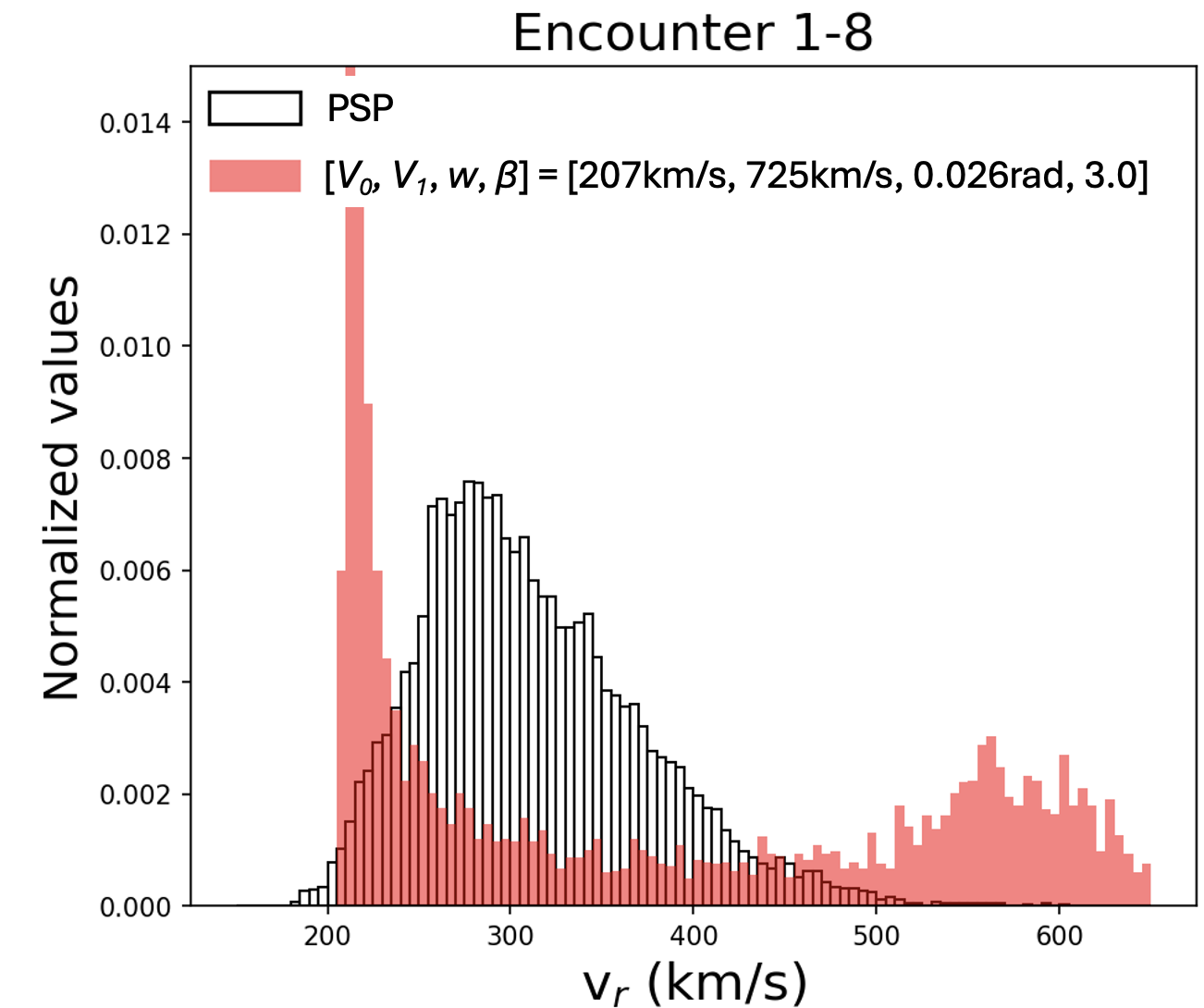}{0.25\textwidth}{(c)}}

 \caption{Examples of how the WSA modeled distributions at 0.1~au vary for $w$ = 0.017 rad, 0.020 rad and 0.026 rad and $\beta$ = 0.25, 0.50, 0.75 and 3.0.}
 \label{Fig:Examples_of_distr_closeToTheSun_appendix}
 \end{figure}

 \begin{figure}
 \centering

 \gridline{\fig{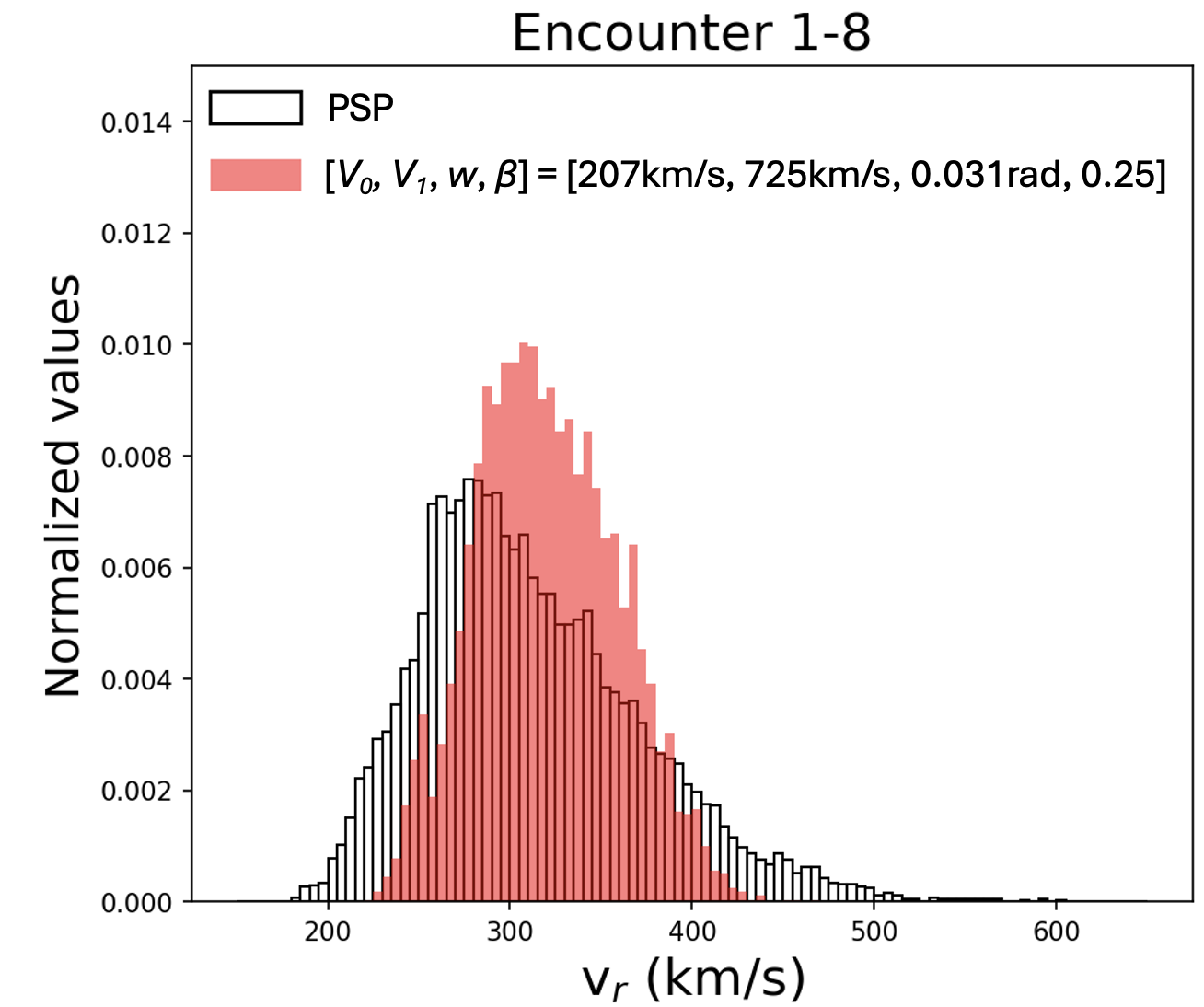}{0.25\textwidth}{(a)}
             \fig{Figures/E1toE8_minvr_207_w0_031_beta0_50_vmax725_noadhoc_editted.png}{0.25\textwidth}{(b)}
             \fig{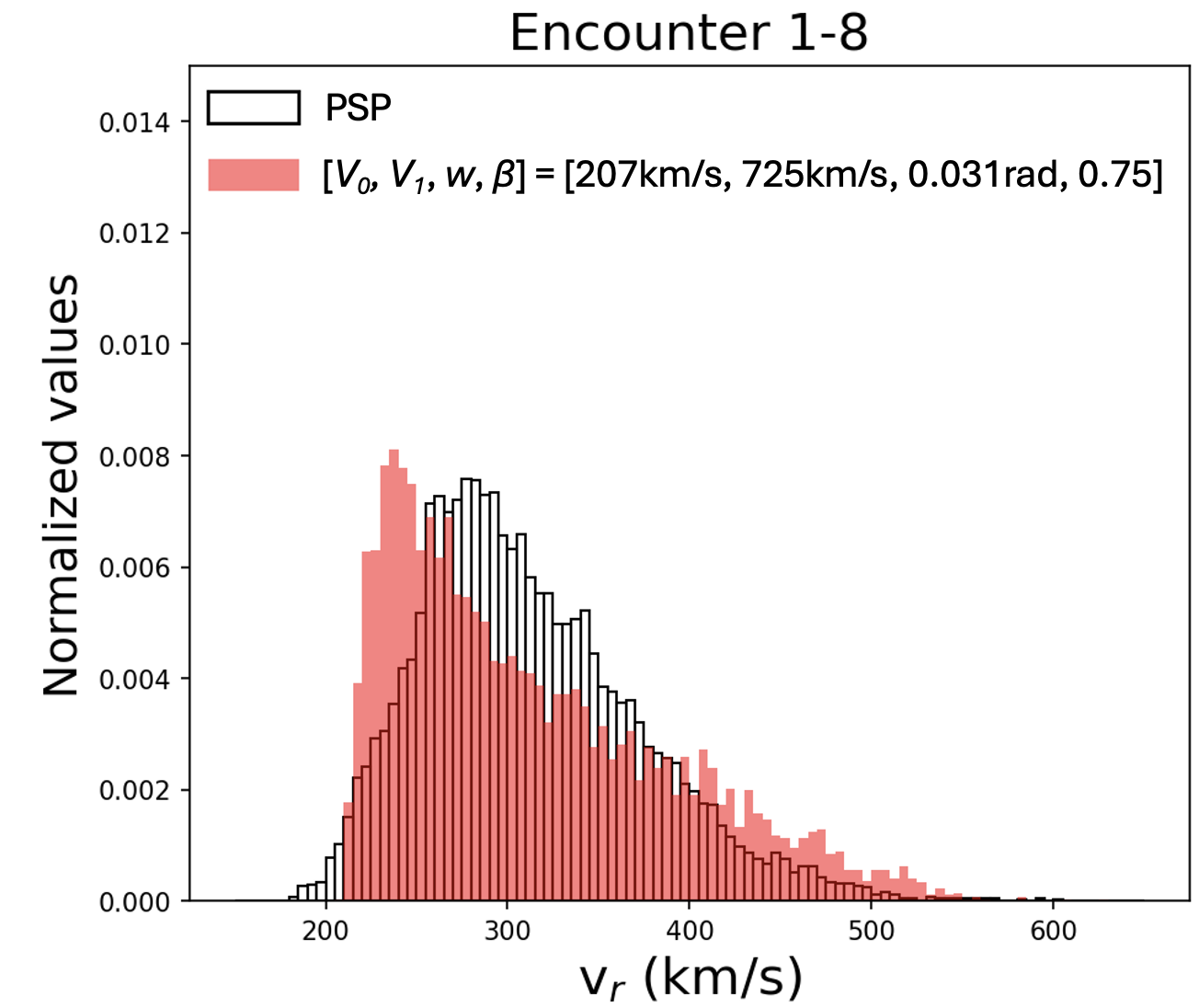}{0.25\textwidth}{(c)}
             \fig{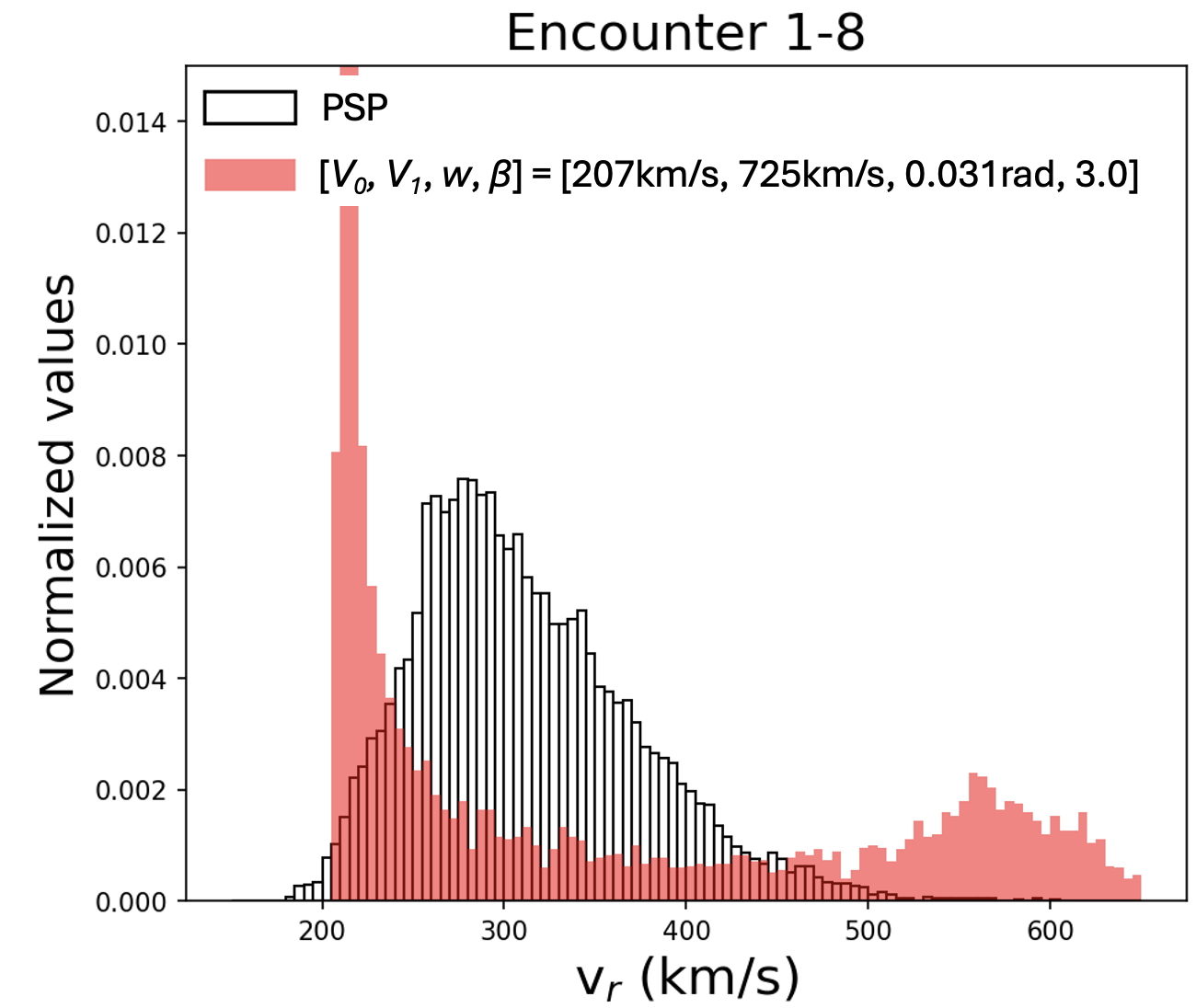}{0.25\textwidth}{(c)}}

 \gridline{\fig{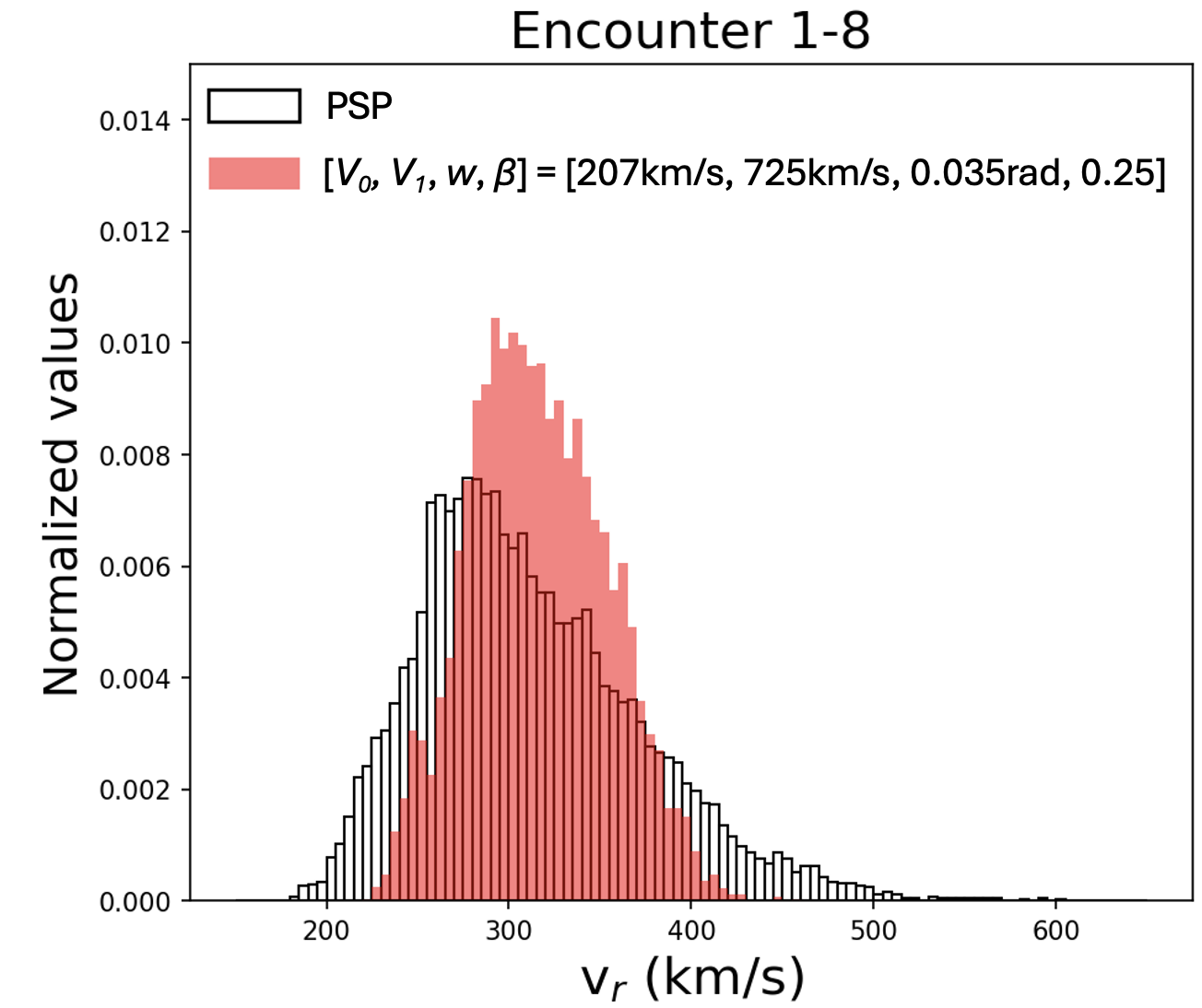}{0.25\textwidth}{(a)}
             \fig{Figures/E1toE8_minvr_207_w0_035_beta0_50_vmax725_noadhoc_editted.png}{0.25\textwidth}{(b)}
             \fig{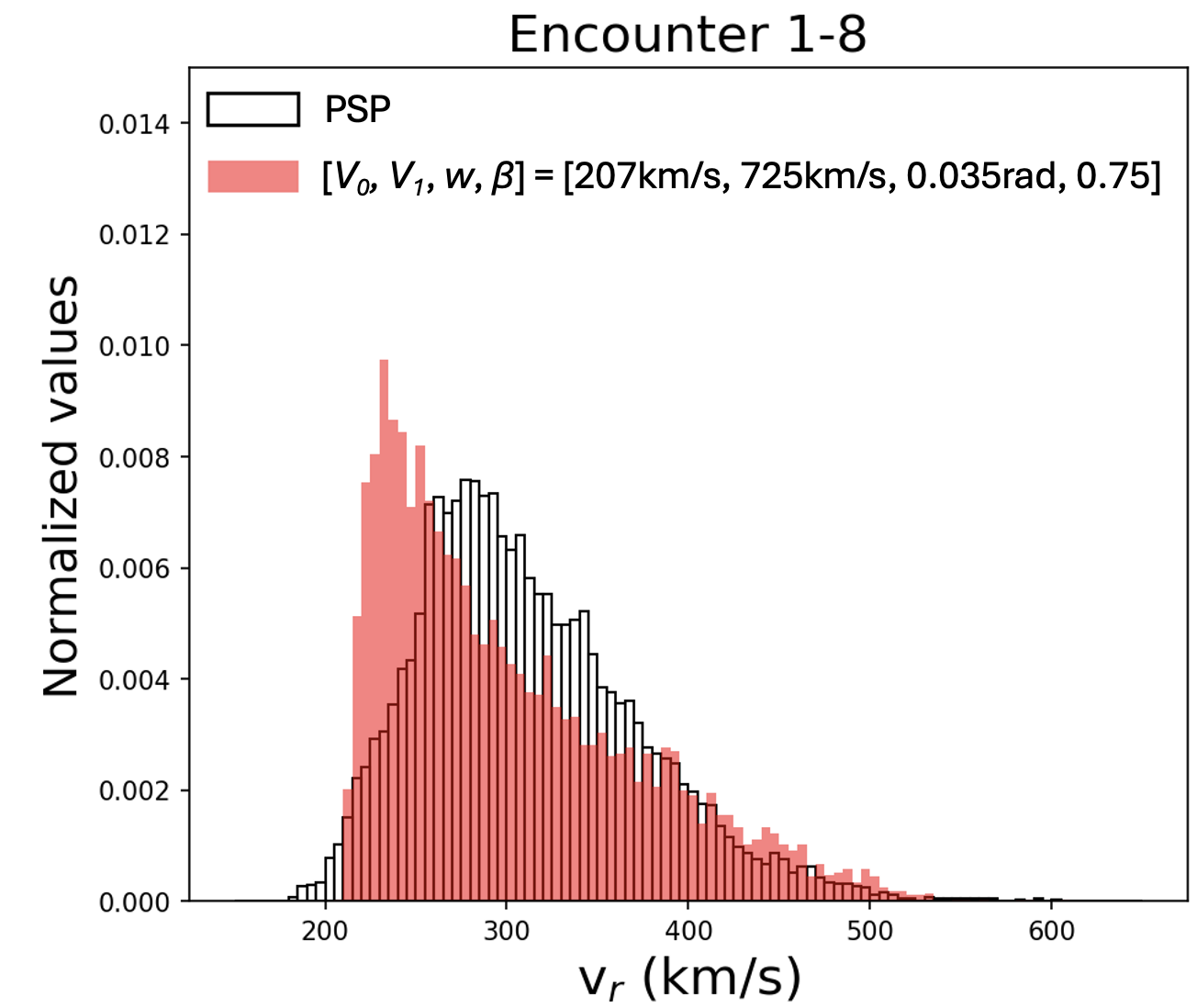}{0.25\textwidth}{(c)}
             \fig{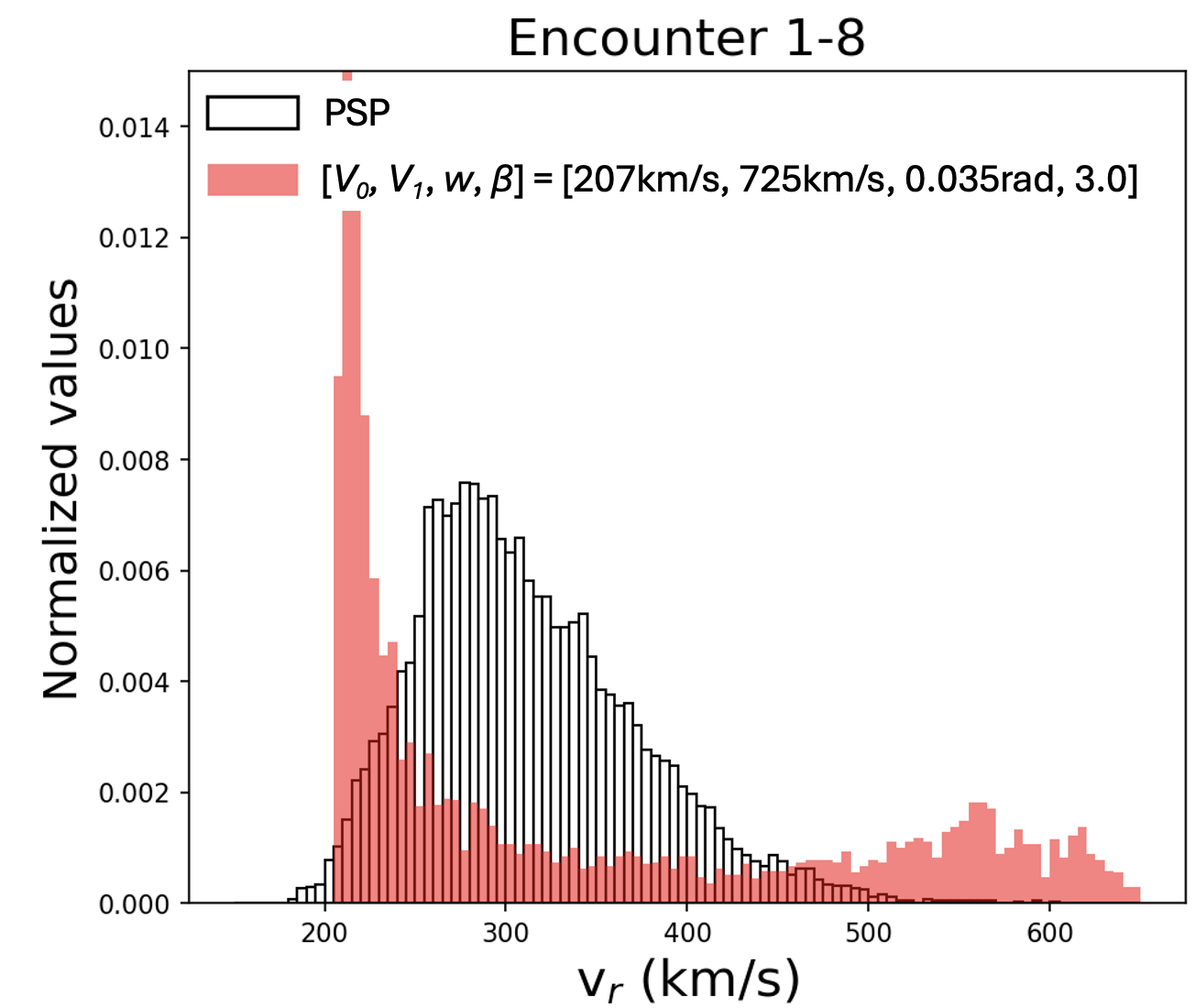}{0.25\textwidth}{(c)}}
            
 \caption{Continuation from Fig.~\ref{Fig:Examples_of_distr_closeToTheSun_appendix} for $w$ = 0.031 rad and 0.035 rad and same $\beta$ values.}
 \label{Fig:Examples_of_distr_closeToTheSun_appendix2}
 \end{figure}

\end{document}